\documentclass[12pt]{article}
\usepackage{mathrsfs}
\makeatletter \@addtoreset{equation}{section} \makeatother
\renewcommand{\theequation}{\thesection.\arabic{equation}}
\newcommand{\ba}{\begin{array}}
\newcommand{\ea}{\end{array}}
\newcommand{\beq}{\begin{equation}}
\newcommand{\eeq}{\end{equation}}
\newcommand{\bea}{\begin{eqnarray}}
\newcommand{\eea}{\end{eqnarray}}

\def\bce{\begin{center}}
\def\ece{\end{center}}

\def\nonu{\nonumber}

\def\pa{\partial}
\def\al{\alpha}
\def\be{\beta}
\def\ga{\gamma}

\def\de{\delta}

\def\si{\sigma}

\def\eps6{{\displaystyle \mathop{\epsilon}^{6}}{}}
\def\g6{{\displaystyle \mathop{g}^{6}}{}}

\def\nab6{{\displaystyle \mathop{\nabla}^{6}}{}}

\def\0{{\sst{(0)}}}
\def\1{{\sst{(1)}}}
\def\2{{\sst{(2)}}}
\def\3{{\sst{(3)}}}
\def\4{{\sst{(4)}}}
\def\5{{\sst{(5)}}}
\def\6{{\sst{(6)}}}
\def\7{{\sst{(7)}}}
\def\8{{\sst{(8)}}}

\def\ba{\begin{array}}
\def\ea{\end{array}}
\def\beq{\begin{equation}}
\def\eeq{\end{equation}}
\def\be{\begin{equation}}
\def\ee{\end{equation}}

\def\eps{\epsilon}

\def\ba{\begin{array}}
\def\ea{\end{array}}
\def\beq{\begin{equation}}
\def\eeq{\end{equation}}
\def\be{\begin{equation}}
\def\ee{\end{equation}}

\def\eps{\epsilon}

\def\eps6{{\displaystyle \mathop{\epsilon}^{6}}{}}

\def\nab6{{\displaystyle \mathop{\nabla}^{6}}{}}

\newcommand{\bean}{\begin{eqnarray*}}
\newcommand{\eean}{\end{eqnarray*}}

\begin{document}
\thispagestyle{empty} \addtocounter{page}{-1}
   \begin{flushright}
\end{flushright}

\vspace*{1.3cm}
  
\centerline{ \large \bf
  Adding Complex Fermions  to
  the Grassmannian-like  Coset Model
}
\vspace*{1.5cm}
\centerline{ {\bf  Changhyun Ahn}
} 
\vspace*{1.0cm} 
\centerline{\it 
 Department of Physics, Kyungpook National University, Taegu
41566, Korea} 
\vspace*{0.5cm}
\centerline{\tt ahn@knu.ac.kr
} 
\vskip2cm

\centerline{\bf Abstract}
\vspace*{0.5cm}

In the ${\cal N}=2$ supersymmetric coset model,
$\frac{SU(N+M)_k \times SO(2 N M)_1}{
  SU(N)_{k+M} \times U(1)_{ N M (N+M)(k+N+M)}}$,
we construct the $SU(M)$ nonsinglet ${\cal N}=2$ multiplet
of spins $(1, \frac{3}{2}, \frac{3}{2}, 2)$ in terms of coset fields.
The next $SU(M)$ singlet and nonsinglet
 ${\cal N}=2$ multiplets
of spins $(2, \frac{5}{2}, \frac{5}{2}, 3)$
are determined by applying the ${\cal N}=2$ supersymmetry
currents of spin $\frac{3}{2}$ to the bosonic
singlet and nonsinglet currents of spin $3$ in the bosonic coset model.
We also obtain the operator product expansions(OPEs) between
the currents of the ${\cal N}=2$ superconformal algebra
and above three kinds of ${\cal N}=2$ multiplets.
These currents in two dimensions
play the role of the asymptotic symmetry, as the generators of
${\cal N}=2$ ``rectangular $W$-algebra'',
  of the $M \times M$  matrix
  generalization of  ${\cal N}=2$
  $AdS_3$ higher spin theory in the bulk.
The structure constants in the right hand sides of these OPEs
are dependent on the three parameters $k, N$ and $M$ explicitly.

Moreover, the OPEs between 
$SU(M)$ nonsinglet
${\cal N}=2$ multiplet of spins $(1, \frac{3}{2}, \frac{3}{2}, 2)$
and itself are analyzed in detail.
The complete OPE between the lowest component of the $SU(M)$
singlet  ${\cal N}=2$ multiplet
of spins $(2, \frac{5}{2}, \frac{5}{2}, 3)$ and itself
is described.
In particular, when $M=2$,
it is known that the above ${\cal N}=2$
supersymmetric coset model provides the
realization of the extension of the large ${\cal N}=4$ nonlinear
superconformal algebra. 
We determine the currents of  the large ${\cal N}=4$ nonlinear
superconformal algebra and the higher spin-$\frac{3}{2}, 2$
currents of the lowest ${\cal N}=4$ multiplet
for generic $k$ and $N$ in terms of the coset fields.
For the remaining higher spin-$\frac{5}{2},3$ currents
 of the lowest ${\cal N}=4$ multiplet,
some of the results are given.


\baselineskip=18pt
\newpage
\renewcommand{\theequation}
{\arabic{section}\mbox{.}\arabic{equation}}

\tableofcontents

\section{ Introduction}

It is known that the most general coset
in the bosonic theory
can be described by \cite{CHR1306}
\bea
\frac{SU(N+M)_k }{
  SU(N)_{k} \times U(1)_{ N M (N+M)k}},
\label{coset1}
\eea
where the three parameters, $k, M$ and $N$ are present
\footnote{The possibility
  of four parameters
  in the different coset model is studied in \cite{EP2006}.}.
According to the observation of 
\cite{CH1812},
this coset model is dual to $M \times M$ matrix generalization
of $AdS_3$ Vasiliev higher spin theory \cite{PV1,PV2}
by taking the appropriate limit on these parameters.
Note that for $M=1$ case,  the Gaberdiel-Gopakumar
conjecture \cite{GG1011} can be seen and
see also \cite{GG1205,GG1207,AGKP} for the relevant works.
At the particular value of
the level $k$ with generic $N$ and $M$,
the operator product expansion (OPE)
between the charged spin-$2$ current and itself leads to
the one of the ``rectangular'' $W$-algebra \cite{AM}
with $SU(M)$ symmetry
which is the asymptotic symmetry of $AdS_3$ higher spin theory
\cite{CH1812} and see also \cite{EP1}.
For generic $k, N$ and $M$, this OPE is further studied in
\cite{Ahn2011} and see also \cite{Ahn1111,AK1308}
\footnote{There is a previous work on the
Grassmannian coset model in \cite{BK1990}.}.

Why do we add the complex fermions into the above coset model
(\ref{coset1})? This is one of the ways to
make the bosonic theory to be the supersymmetric theory.
In other words, by using the fermionic operators
of spin-$\frac{1}{2}$, we can construct
the half integer currents including the supersymmetry generators
of spin-$\frac{3}{2}$ explicitly \footnote{
Without introducing the fermions,
we can have the supersymmetric version of (\ref{coset1}) by considering
the special value of the level $k$.
This is because
  we can realize the various spin-$1$ currents
  in terms of fermions. For example, see also \cite{Douglas,BBSScon,GS88,
  GKO,HR,ASS,SS,Ahn1211,Ahn1305,BFK,GHKSS,Ahn1604,Ahn1701,AP1902}.}.
We would like to discuss about the supersymmetric version of
(\ref{coset1}) and it is given by \cite{CHR1306,CH1906}
\bea
\frac{SU(N+M)_k \times SO(2 N M)_1}{
  SU(N)_{k+M} \times U(1)_{ N M (N+M)(k+N+M)}}.
\label{coset2}
\eea
Note that
there exists an $ SO(2 N M)_1$ factor
associated with complex fermions in the numerator
and the levels in the denominator are changed appropriately.
For $M=1$, by dividing (\ref{coset2}) out the $SU(M=1)_{k+N}$
further,
the ${\cal N}=2$ $AdS_3$ higher spin gravity is related to
the Kazama-Suzuki model \cite{KSnpb,KSplb},
according to \cite{CHR1111}. See also the relevant works in
\cite{CG1203,Henneauxetal,HP,Ahn1206,CG,Ahn1208,CHR1211,GK1,
Ahn1604,GK2,Dattaetal,Eberhardtetal,Castroetal}.
For $M=2$, the above coset arises in the context of
the large
${\cal N}=4$ holography  \cite{GG1305}.
See also previous works on this holography
\cite{Ahn1311,GP1403,BCG,GG1406,Ahn1408,AP1410,GG1501,
Ahn1504,AK1506,AK1509,AKP1510,GG1512,AK1607,AHK2106,Ferreira,AKK1703,
Eberhardtetal1,Ahn1711,EGR,Ahn1805,EG1904,AKP1904,AKK1910,EP2006,
AGK2004,AK2009,GG2105}.

For generic $M > 2$,
in \cite{CH1906}, the extension of ${\cal N}=2$ superconformal algbra
is described. The generators of ${\cal N}=2$ superconformal algebra
are given by the spin-$1, \frac{3}{2}, \frac{3}{2}, 2$ currents
and denoted by $K, G^+, G^-, T$ respectively \footnote{
  The OPE between $G^+$ and $G^-$ does not produce the
  standard ${\cal N}=2$ superconformal algebra for generic $M >1$ because
  the first order pole of this OPE has the additional terms.
  However, we are using the same
  terminology ``${\cal N}=2$ superconformal
  algebra'' in this paper.}.
The extra generators of $SU(M)$ nonsinglets
are classified by the spin-$1$ currents,
spin-$\frac{3}{2}$ currents, and the spin-$2$ currents.
Their numbers are given by $(M^2-1)$, $2(M^2-1)$ and $(M^2-1)$
respectively.
If we increase the spin by one, then
there are  the spin-$2$ currents,
spin-$\frac{5}{2}$ currents, and the spin-$3$ currents
of $SU(M)$ singlets and nonsinglets.
Their numbers are given by $ M^2$, $2M^2$ and $M^2$
respectively.
As we increase further, then the general features
of $SU(M)$ singlets and nonsinglets
go through: there are the $M^2$ spin-$s$ currents,
the $2M^2$ spin-$(s+\frac{1}{2})$ currents, and the
$M^2$
spin-$(s+1)$ currents \footnote{The ${\cal N}=2$ ``rectangular''
  $W$-algebra appears as the asymptotic symmetry
  of the $M \times M$  matrix
  generalization of  ${\cal N}=2$
  $AdS_3$ higher spin theory \cite{CH1812}.  }.

We present them here, in the notation of ${\cal N}=2$
multiplet \footnote{Strictly speaking,
  when we are saying about ``${\cal N}=2$ multiplet'' in this paper,
  the four component currents do not satisfy
  the standard ${\cal N}=2$ primary conditions with the
  currents $(K,G^+,G^-,T)$: there are some higher order poles appear
and some structure constants appear differently.}, with the assignment of
spin and the $SU(M)$ index $a=1,2, \cdots, (M^2-1)$
as follows:
\bea
&& (W^{-(1), 0}, G^{+(\frac{3}{2}), 0},
G^{-(\frac{3}{2}), 0}, W^{+(2), 0}) \equiv (K, G^+, G^-, T),
\nonu\\
&& (W^{-(1), a}, G^{+(\frac{3}{2}), a},
G^{-(\frac{3}{2}), a}, W^{+(2), a}),
\nonu \\
&& (W^{-(2), 0}, G^{+(\frac{5}{2}), 0},  G^{-(\frac{5}{2}), 0}, W^{+(3), 0}),
\nonu\\
&& (W^{-(2), a}, G^{+(\frac{5}{2}), a},  G^{-(\frac{5}{2}), a}, W^{+(3), a}),
\nonu \\
&& (W^{-(3), 0}, G^{+(\frac{7}{2}), 0},  G^{-(\frac{7}{2}), 0}, W^{+(4), 0}),
\nonu\\
&& (W^{-(3), a}, G^{+(\frac{7}{2}), a},  G^{-(\frac{7}{2}), a}, W^{+(4), a}),
\qquad \cdots \qquad, 
\nonu \\
&& (W^{-(s), 0}, G^{+(s+\frac{1}{2}), 0},  G^{-(s+\frac{1}{2}), 0}, W^{+(s+1), 0}),
\nonu \\
&& (W^{-(s), a}, G^{+(s+\frac{1}{2}), a},  G^{-(s+\frac{1}{2}), a}, W^{+(s+1), a}),
\qquad \cdots \qquad .
\label{Ws}
\eea
In (\ref{Ws}), the generators of ${\cal N}=2$ superconformal algebra
are denoted by $(K, G^+, G^-, T)$.
Each current of 
half integer spins appears in the same ${\cal N}=2$ multiplet
while each current of integer spins appears
in the different ${\cal N}=2$ multiplets.
The coset realization on the above currents
is done for the generators of $(K,G^+, G^-, T)$ and
the next three kinds of $ (W^{-(1), a}, G^{+(\frac{3}{2}), a},
G^{-(\frac{3}{2}), a})$  so far in \cite{CH1906} \footnote{We are using the
  notation \cite{CH1906}
  of  ${\cal N}=2$ ``rectangular'' $W$-algebra in the bulk
  for its dual CFT
  currents in (\ref{Ws}).}.

\begin{itemize}
  \item[]
In this paper, we would like to
construct the coset realization for the following
currents
\bea
&& (\mbox{known}, \mbox{known},
\mbox{known}, W^{+(2), a}),
\nonu \\
&& (W^{-(2), 0}, G^{+(\frac{5}{2}), 0},  G^{-(\frac{5}{2}), 0}, W^{+(3), 0}),
\nonu\\
&& (W^{-(2), a}, G^{+(\frac{5}{2}), a},  G^{-(\frac{5}{2}), a}, W^{+(3), a}).
\label{findings}
\eea
Here, the first three components of the first ${\cal N}=2$
multiplet (\ref{findings})
are known in terms of coset fields.
\end{itemize}
Moreover, we describe some of the OPEs
between the currents for low spins.
Then how we can construct the currents
in (\ref{findings}) in terms of coset fields explicitly?
For the spin-$2$ current appearing in the first line of (\ref{findings}),
we can use both the previous known currents which belong to the same
${\cal N}=2$ multiplet and the generators of $(K,G^+,G^-,T)$.
For the currents belonging to the second and third ${\cal N}=2$ multiplets
in (\ref{findings}), we can use the previous known currents of spin-$2,3$
currents in the bosonic coset model (\ref{coset1}).
We figure out that the singlet spin-$3$ current $W^{(3)}$ will appear
in the last component of the second ${\cal N}=2$ multiplet
while  the nonsinglet spin-$3$ current $P^a$ will appear
in the last component of the third ${\cal N}=2$ multiplet
and  the nonsinglet spin-$2$ current $K^a$ will appear
in the first component of the third ${\cal N}=2$ multiplet, as the
${\cal N}=2$ supersymmetric versions.
This is because the OPE between the spin-$2$ current $K^a$
and the spin-$2$ current $K^b$ leads to the spin-$3$ current
$P^c$ in the bosonic coset model.
Furthermore,
when we fix $M=1$, the singlet spin-$3$ current
arises in the last component of the corresponding
${\cal N}=2$ multiplet (For example,
 \cite{Ahn1206,Ahn1208,Ahn1604})
and therefore it is natural to think about the above description
for general $M$ case.

\begin{itemize}
  \item[]
One way to determine the ${\cal N}=2$ superpartners of
the spin-$2,3$ currents found in the bosonic coset model is to
consider, as a first step,
the OPEs between the ${\cal N}=2$ supersymmetry
generators $G^{\pm}$ and the spin-$3$ currents $W^{(3)}$ and
$P^a$ of the bosonic theory.
\end{itemize}
Then we will obtain
the intermediate
spin-$\frac{5}{2}$ currents in the specific poles which depend on
the complex fermions as well as the bosonic currents.
Further computations for the OPEs  between the
above supersymmetry generators $G^{\pm}$ and these intermediate
spin-$\frac{5}{2}$ currents
obtained at the previous stage can be performed. Then we will obtain
the spin-$2$ currents which will contain both the complex fermions
and the bosonic currents as before.
Then we have singlet spin-$2$ current $W^{-(2),0}$
and the nonsinglet spin-$2$ currents $W^{-(2),a}$ as in (\ref{findings}).
By calculating the OPEs between the spin-$\frac{3}{2}$ currents
$G^{\pm}$ and these spin-$2$ currents obtained newly, we can determine
the singlet spin-$\frac{5}{2}$ currents $W^{\pm (\frac{5}{2}),0}$
and the nonsinglet spin-$\frac{5}{2}$ currents $W^{\pm (\frac{5}{2}),a}$.
In general,
they are different from the above intermediate spin-$\frac{5}{2}$
currents.
Finally, after further action of supersymmetry generators $G^{\pm}$
on these spin-$\frac{5}{2}$ currents, the singlet spin-$3$ current
$W^{+(3),0}$ and the nonsinglet spin-$3$ currents $W^{+(3),a}$ can be
determined completely.

So far, the OPEs we are considering are
the ones between the currents of the ${\cal N}=2$
superconformal algebra (the first line of (\ref{Ws}))
and the currents of the
singlet and nonsinglet ${\cal N}=2$ multiplets (\ref{findings}).
Therefore, the right hand sides of these OPEs look similar to the
behavior of standard ${\cal N}=2$ primary currents.
The difference is that
we observe that there appear some other additional singular terms
and other type of currents living in other ${\cal N}=2$ multiplet.

\begin{itemize}
  \item[]
    The next things we should do, for the simplest
    cases, is to compute 1) the OPEs between the
currents in the first ${\cal N}=2$ multiplet in (\ref{findings})
and 2) the OPE between $W^{-(2),0}$ and itself.
\end{itemize}
In this case
we do expect that the right hand sides of these OPEs will produce
other ${\cal N}=2$ multiplets nontrivially.
For the OPEs between the first nonsinglet
${\cal N}=2$ multiplet in (\ref{findings}),
there exist new primary spin-$\frac{5}{2}$ currents having two free indices
and primary spin-$3$ current also having two free indices.
For the OPE between the lowest component and itself of the second
singlet
${\cal N}=2$ multiplet in (\ref{findings}), there is no new primary current.
Although we consider the OPE between the singlet spin-$2$ current and itself,
we do expect that the OPEs between the currents of the
singlet ${\cal N}=2$ multiplets $ (W^{-(s), 0}, G^{+(s+\frac{1}{2}), 0},
G^{-(s+\frac{1}{2}), 0}, W^{+(s+1), 0})$ do not contain the currents from
the nonsinglet ${\cal N}=2$ multiplets, in the presence of the
currents of the ${\cal N}=2$ superconformal algebra with modified
stress energy tensor.

\begin{itemize}
  \item[]
Furthermore, we can study the extension of the large
${\cal N}=4$ nonlinear superconformal algebra in the present context
because we have some information on the coset fields via
(\ref{findings}).
The question is how to reorganize the
currents of ${\cal N}=4$ multiplets in
terms of the currents of ${\cal N}=2$ multiplets
and the coset fields.
\end{itemize}
In this case, the parameter $M$ is fixed by $2$.
We can try to obtain the relevant currents of the above extension
by recalling the defining OPEs in \cite{AK1411}
between them starting from the low spins.
The nontrivial part is to obtain
the currents of  the
the large
${\cal N}=4$ nonlinear superconformal algebra
because the stress energy tensor appears very nontrivially
(For example, \cite{cqg1989})
and we should fix
the normalizations for the spin-$1$ currents and the spin-$\frac{3}{2}$
currents correctly.
Once we identify the currents of the
the large
${\cal N}=4$ nonlinear superconformal algebra, then it is rather straightforward
to determine the currents 
of the lowest ${\cal N}=4$ multiplet by using the defining relations in the
OPEs in \cite{AK1411}.

In section $2$,
we review how we can add the complex fermions into the bosonic coset
model and the coset realization for the
${\cal N}=2$ superconformal algebra is reviewed.
We describe the known currents in (\ref{findings}).
In section $3$, after finding the nonsinglet spin-$2$ currents
$W^{+(2),a}$, we compute the OPEs between the first ${\cal N}=2$
multiplet in the second line of (\ref{Ws})
and the ${\cal N}=2$ stress energy tensor in the first line of
(\ref{Ws}).
In section $4$, by considering the OPEs between the spin-$\frac{3}{2}$
currents $G^{\pm}$ and the spin-$3$ current $W^{(3)}$ and
analyzing the OPEs
 between the spin-$\frac{3}{2}$
currents $G^{\pm}$ and the spin-$\frac{5}{2}$ current found newly
successively,
the lowest current of the second ${\cal N}=2$ multiplet in the third
of (\ref{Ws}) can be obtained. Moreover, the remaining other three
components of this ${\cal N}=2$ multiplet can be determined.
As before,
 the OPEs between the second ${\cal N}=2$
multiplet in the third line of (\ref{Ws})
and the ${\cal N}=2$ stress energy tensor in the first line of
(\ref{Ws}) are determined.
In section $5$, we do this section by considering the
third ${\cal N}=2$ multiplet in the fourth line of (\ref{Ws})
by following the procedure done in section $4$.
In section $6$,
after analyzing the OPEs between the first ${\cal N}=2$ multiplet,
the arising of the new primary currents of spin-$\frac{5}{2}, 3$
having the free indices $a b$ of $SU(M)$ is described. 
In section $7$,
some of the OPEs
between the second ${\cal N}=2$ multiplet are analyzed.
In section $8$,
for the $M=2$ case,
the coset realization gives us an extension of the
large ${\cal N}=4$ nonlinear superconformal algebra
\cite{GS,cqg1989,npb1989,GK}
and some of the currents of the lowest ${\cal N}=4$
multiplet are given explicitly.
In section $9$,
we conclude our work and the future directions are given.
The various Appendices are for the details in previous sections
we describe.

\section{ Review}

Some known results are reviewed in this section.

\subsection{The singlet
  currents of spins $(1, \frac{3}{2}, \frac{3}{2}, 2)$}

\subsubsection{The role of complex fermions}

In the coset 
\bea
\frac{SU(N+M)_k \times SO(2 N M)_1}{
  SU(N)_{k+M} \times U(1)_{ N M (N+M)(k+N+M)}},
\label{coset}
\eea
the decomposition of $SU(N+M)$ 
into the $SU(N) \times SU(M)$
in the numerator can be performed as in the bosonic case \cite{CH1812}.
We use the generators $(t^{\alpha}, t^a, t^{u(1)}, t^{(\rho \bar{i})},
  t^{(\bar{\si} j)})$ with the normalized metric.
  The index $\alpha$ runs over $\alpha=1,2, \cdots, (N^2-1)$
  while the index $a$ runs over $a=1,2, \cdots, (M^2-1)$.
  The fundamental indices $\rho$ and $j$ run over
  $\rho =1,2, \cdots, N$ and $j=1,2,\cdots, M$
  while the antifundamental indices 
  $\bar{\si}$ and $\bar{i}$ run over
  $\bar{\si}=1,2, \cdots, N$ and $\bar{i}=1,2, \cdots, M$.
  The $f$ and $d$ symbols can be obtained from the above generators.
  The $SU(N+M)$ currents of spin-$1$ in the numerator satisfy the 
standard OPEs.
  
The $N M$ complex fermions of spin-$\frac{1}{2}$
in the numerator satisfy the following OPE
\bea
\psi^{(\rho \bar{i})}(z) \, \psi^{(\bar{\si} j)}(w) = \frac{1}{(z-w)} \,
\de^{\rho \bar{\si}} \, \de^{j \bar{i}} + \cdots.
\label{psipsi}
\eea
We introduce the $SU(N)$ currents, the $SU(M)$ currents
and $U(1)$ current from these complex fermions living in
the $SO(2 N M)_1$ factor as follows \cite{CH1906}:
\bea
J^{\al}_f \equiv t^{\al}_{\rho \bar{\si}} \, \de_{j \bar{i}} \,
\psi^{(\rho \bar{i})} \, \psi^{(\bar{\si} j)}, \qquad
J^{a}_f \equiv -t^{a}_{j \bar{i}} \, \de_{\rho \bar{\si}}  \,
\psi^{(\rho \bar{i})} \, \psi^{(\bar{\si} j)}, \qquad
J^{u(1)}_f \equiv \de_{\rho \bar{\si}} \, \de_{j \bar{i}} \,
\psi^{(\rho \bar{i})} \, \psi^{(\bar{\si} j)}.
\label{threeJf}
\eea
The appropriate contractions between the indices are taken.
We can also choose the second currents with positive sign.
Then the currents in the denominator of the coset (\ref{coset})
are given by
\bea
J^{\alpha} + J^{\alpha}_f, \qquad
\sqrt{M N (M+N)} \, J^{u(1)} + (M+N) \, J_f^{u(1)},
\label{denom}
\eea
together with (\ref{threeJf}).
Note that the level for the $J^{\alpha}_f$ is given by
$M$ and the level for the $J_f^{u(1)}$ is given by
$M N$ by using (\ref{psipsi}).
See also Appendix $A.2$.
Compared with the bosonic coset (\ref{coset1}),
the levels in the denominator
are increased by $M$ and $N M (N+M)^2$ respectively.
Note that the level for $J^{\alpha}$ and $J^{u(1)}$ is given by $k$.

The role of $J_f^a$ will appear in the next subsection.
Note that the OPEs between any two different currents in
(\ref{threeJf}) are regular \footnote{It is obvious that
the OPEs between the bosonic currents
$(J^{\alpha}, J^a, J^{u(1)}, J^{(\rho \bar{i})},J^{(\bar{\si} j)})$
and the currents in (\ref{threeJf}) are regular. }.
See also Appendix $A.2$.
   With the help of complex fermions, we can construct the currents
   of half-integer spin which are necessary to obtain the
   supersymmetric theory.

\subsubsection{The ${\cal N}=2$ superconformal algebra}

The stress energy tensor
by Sugawara construction is the difference between 
the stress energy tensor in the numerator
and the one in the denominator and 
is given by  
 \bea
  T  & =  &\frac{1}{2(k+M+N)} \Bigg[ J^{\al}  J^{\al} + J^a  J^a +    \de_{\rho \bar{\si}} \de_{j \bar{i}} \,
    J^{(\rho \bar{i})}  J^{(\bar{\si} j)} + \de_{\rho \bar{\si}} \de_{j \bar{i}} 
   J^{(\bar{\si} j)}  J^{(\rho \bar{i})} + J^{u(1)}  J^{u(1)} 
    \Bigg]
  \nonu \\
&-& \frac{1}{2} \Bigg[\de_{\rho \bar{\si}} \, \de_{j \bar{i}} \,
  \psi^{(\rho \bar{i})} \, \pa\, \psi^{(\bar{\si} j)} -
\de_{\rho \bar{\si}} \, \de_{j \bar{i}} \,
\pa \, \psi^{(\rho \bar{i})} \, \psi^{(\bar{\si} j)}
\Bigg]
- \frac{1}{2(k+M+N)} \,  \Bigg[ (J^{\al}+ J_f^{\al}) \,
  (J^{\al} + J_f^{\al}) \nonu \\
  & + &
  ( J^{u(1)} + \sqrt{\frac{M+N}{M N}}\, J^{u(1)}_f ) \, (J^{u(1)}+
   \sqrt{\frac{M+N}{M N}} J^{u(1)}_f) \Bigg].
  \label{T}
\eea
Compared with the bosonic case \cite{CH1812} \footnote{
\label{bosonT} The bosonic stress energy tensor is given by $
  T_{boson}   =  \frac{1}{2(k+N+M)} \Bigg[ J^{\al}  J^{\al} + J^a  J^a +
    \de_{\rho \bar{\si}} \de_{j \bar{i}} \,
    J^{(\rho \bar{i})}  J^{(\bar{\si} j)} + \de_{\rho \bar{\si}} \de_{j \bar{i}} 
   J^{(\bar{\si} j)}  J^{(\rho \bar{i})} + J^{u(1)}  J^{u(1)} 
    \Bigg]
  - \frac{1}{2(k+N)} \,  J^{\al} \, J^{\al} - \frac{1}{2k} \,
  J^{u(1)} \, J^{u(1)}$.}, due to the presence of the
$SO(2 N M)_1$ factor in the numerator, its contribution
appears in the second line of (\ref{T}).
The quantity $(k+M+N)$ in the first line
can be interpreted as the sum of the level $k$ and $(N+M)$
of $SU(N+M)$.
There exists a common factor $\frac{1}{2(k+M+N)}$ in the
stress energy tensor of the denominator.
For the $SU(N)$ current in (\ref{denom}), the quantity
$(k+M+N)$ can be regarded as the sum of the level $(k+M)$
and the $N$ of $SU(N)$.
When we multiply $M N (M+N)$ in the third line 
of (\ref{T}) and divide it, then the corresponding level provides the
overall factor 
$\frac{1}{2 N M (M+N)(k+M+N)}$ for the $U(1)$ current (\ref{denom}) in the
denominator (\ref{denom}).
In other words, the level can be identified with
$ N M (M+N)(k+M+N)$.

Then it is straightforward to compute the central charge
for the supersymmetric coset model
\bea
c= \frac{3 M N k}{(k+M+N)} + \frac{(k+N)(M^2-1)}{(k+M+N)}=
\frac{(k M^2+3 k M N-k+M^2 N-N)}{(k+M+N)}.
\label{central}
\eea
When we have further $SU(M)_{k+N}$ factor in the denominator
of (\ref{coset2}),
then the central charge is given by
the first contribution of the first relation in
(\ref{central}).
According to the observation of \cite{KSnpb},
this can be obtained
from the level $k$ of $SU(N+M)$, the dual Coxeter number $(N+M)$ of
$SU(N+M)$ and the dimension $2 N M$ of
$\frac{SU(N+M)}{SU(N) \times SU(M) \times U(1)}$.
See also \cite{Ahn1206,Ahn1208}.
The corresponding current of
$SU(M)_{k+N}$ factor is given by $(J^a + J^a_f)$
and the level of $J_f^a$ is $N$. Therefore, the
total level is $(k+N)$.
By adding  $SU(M)_{k+N}$ factor in the previous coset,
we obtain the above coset (\ref{coset}). Then
the central charge
for the $SU(M)_{k+N}$ factor is given by the
second contribution 
of the first relation in
(\ref{central}) \footnote{We can compute the central charges
  for the group $G=SU(N+M)_k \times SO(2NM)_1$ and for the
  group $H=SU(N)_{k+M} \times U(1)_{NM(N+M)(k+M+N)}$ directly.
  For the former, we have the contributions from each factor
  $c_G=\frac{k ((M+N)^2-1)}{k +(M+N)}+
  \frac{1}{1+(2NM-2)} \frac{1}{2} 2 N M (2N M-1)$. For the
  latter $c_H=\frac{(k+M)(N^2-1)}{(k+M)+N}+\frac{NM(N+M)(k+M+N)}{
  NM(N+M)(k+M+N)+0}$ from each contribution. Then we obtain
 that the central charge (\ref{central}) is given by their
 difference as follows: $c=c_G-c_H$. By using the property of
 $T(z) \, T_H(w) = 0 +\cdots$ where $T=T_G-T_H$, we confirm that
 the OPE $T(z) \, T(w)$ can be reduced to the OPE
of $T_G(z) \, T_G(w)-T_H(z) \, T_H(w)$ \cite{BBSS1,BS}.}.

By taking the product of the above complex fermions and
the spin $1$ currents transforming $({\bf N},
\overline{\bf {M}})$
or $(\overline{\bf {N}},{\bf M})$ with the appropriate
contractions between the indices, we
obtain the spin-$\frac{3}{2}$ currents \cite{CH1906}
of ${\cal N}=2$ superconformal algebra  as follows: 
\bea
G^{+} \equiv  \de_{\rho \bar{\si}} \, \de_{j \bar{i}} \,
J^{(\rho \bar{i})} \, \psi^{(\bar{\si} j)}, \qquad
G^{-} \equiv  \de_{\rho \bar{\si}} \, \de_{j \bar{i}} \,
\psi^{(\rho \bar{i})} \, J^{(\bar{\si} j)}.
\label{gpm}
\eea

It turns out that
the remaining spin $1$ current
of ${\cal N}=2$ superconformal algebra
is described by
\bea
K \equiv \frac{1}{(k+M+N)} \, \Bigg(  \sqrt{ \frac{M+N}{M N}}
\, M N \, J^{u(1)}
- k \, J_f^{u(1)} \Bigg).
\label{K}
\eea

Note that the currents (\ref{T}), (\ref{gpm}) and (\ref{K})
of the ${\cal N}=2$ superconformal algebra
\bea
(K, G^{+}, G^{-}, T)
\label{kggt}
\eea
are regular in  the OPEs between these and the currents
(\ref{denom})
of the denominator of the coset. 

In Appendix $A$, we present some OPEs between the
spin $\frac{1}{2}$ operators and the spin $1$ operators,
some OPEs between the spin $1$ operators, the OPEs
between the spin $\frac{1}{2}, 1$ operators and the
currents (\ref{kggt}),  and the complete OPEs of the
${\cal N}=2$ superconformal algebra.
As noted by \cite{CH1906}, there is a difference between the
${\cal N}=2$ superconformal algebra we are using in this paper
and the standard ${\cal N}=2$ superconformal algebra for general
$M (\neq 1)$:
i) the normalizations in the spin-$1,\frac{3}{2}$ currents
are different and ii) in the OPE of $G^{+}(z) \, G^{-}(w)$,
the stress energy tensor
term has an additional term, $(J^a + J_f^a)^2$. For $M=1$ case,
this additional term vanishes and we obtain the standard
${\cal N}=2$ superconformal algebra.

\subsubsection{The so far known currents}

In the bosonic coset model,
the currents contain the $J^a$ of spin-$1$, the $K^a$ of spin-$2$,
the $P^a$ of spin-$3$ and the $W^{(3)}$ of spin-$3$.
They are all primary under the
corresponding stress energy tensor appearing in the footnote
\ref{bosonT}.
The question is whether they are primary under the above
stress energy tensor (\ref{T}) or not.
Recall that the bosonic stress energy tensor
consists of the first line of (\ref{T}) in addition to
$J^{\alpha}\, J^{\alpha}$ and $J^{u(1)} \, J^{u(1)}$ terms.
By construction, the above primary currents do not have any singular
terms in the OPEs between them and these quadratic terms.
Then we can regard  the first line of (\ref{T}) as the
bosonic stress energy tensor. Let us look at the second and third
lines of (\ref{T}). They are either purely complex fermions dependent
terms, $J^{\alpha}$ dependent term, $J^{u(1)}$ dependent term,
$J^{\alpha}$ with complex fermions, or $J^{u(1)}$ with complex fermions.
It is clear to observe that the above primary currents consisting of
purely bosonic operators do not produce any nontrivial singular terms
when we calculate the OPEs between them and
 the second and third
 lines of (\ref{T}). This implies that
 the above primary currents with bosonic
 stress energy tensor are primary also under the above
 stress energy tensor (\ref{T}).

 Furthermore, because the above primary currents do not contain
 the complex fermions, 
 it is obvious that
 the regular conditions for these currents with the operators
 (\ref{denom}) in the denominator hold.
 Then all the (quasi)primary currents in the bosonic
 coset model can play the role of primary operators under the
 stress energy tensor (\ref{T}).
 However, we should check whether they transform under the
 spin-$\frac{3}{2}$ currents in (\ref{gpm}) and
 the spin-$1$ current (\ref{K}) or not.

 So far, there exist the following currents
 \cite{CH1812,Ahn2011,CH1906}
\bea
 && \mbox{spin-1}: J^a, \qquad \mbox{spin-2}: K^a, \qquad
 \mbox{spin-3}: P^a, \qquad \cdots,
 \nonu \\
 && \mbox{spin-1}: K, \qquad \mbox{spin-$\frac{3}{2}$}: G^{+}, 
 G^{-}, \qquad \mbox{spin-2}: T, \qquad
 \mbox{spin-3}: W^{(3)}, \qquad \cdots .
 \label{fields}
 \eea
 We would like to construct the additional higher spin currents
 of integer or half-integer spins in terms of the coset operators
 $(J^{\alpha}, J^a, J^{u(1)}, J^{(\rho \bar{i})},J^{(\bar{\si} j)})$ and
 $(\psi^{(\rho \bar{i})}, \psi^{(\bar{\si} j)})$. Moreover,
 by using their OPEs, we determine the possible new higher spin
 currents.
 We will obtain ${\cal N}=2$ supersymmetric currents
 corresponding to the above $K^a, W^{(3)}$ and $P^a$.
 Although these currents do not belong to the extension of
 ${\cal N}=2$ superconformal algebra we are describing in this paper,
 they can provide how we can construct their ${\cal N}=2$
 supersymmetric versions explicitly by using the currents of
 ${\cal N}=2$ superconformal algebra.
 
\subsection{The nonsinglet
  currents of spins $(1, \frac{3}{2}, \frac{3}{2}, 2)$}


In \cite{CH1906}, the three currents of this multiplet
were obtained.
We can consider the $SU(M)$ current in (\ref{threeJf})
as the lowest component of this multiplet. This current
has the adjoint index
$a$. In order to have the spin-$\frac{3}{2}$ current which has an index
$a$, we use the generator of $SU(M)$ by making the product of
spin-$1$ current and spin-$\frac{1}{2}$ operator with the appropriate
contractions of the indices. 
It turns out that in \cite{CH1906} there exist the nonsinglet
spin-$\frac{3}{2}$
currents 
\bea
G^{+,a} \equiv  -t^a_{j \bar{i}} \, \de_{\rho \bar{\si}}  \,
J^{(\rho \bar{i})} \, \psi^{(\bar{\si} j)}, \qquad
G^{-,a} \equiv  -t^a_{j \bar{i}}\, \de_{\rho \bar{\si}}  \,
\psi^{(\rho \bar{i})} \, J^{(\bar{\si} j)}.
\label{gpma}
\eea

Then it is natural to describe the following${\cal N}=2$ multiplet
\bea
(J^a_f, G^{+\, a}, G^{-\, a}, ?).
\label{spin1mul}
\eea
In next section, we will determine the last component of this multiplet.

\subsection{The regular condition}

By construction, all the currents should satisfy
the regular conditions in the OPEs between them and the currents
in the denominator of the coset (\ref{coset}).
All the bosonic currents in the coset model of \cite{CH1812}
satisfy automatically because the additional currents in (\ref{threeJf})
come from the complex fermions.
For example, the spin-$1$ current in (\ref{fields})
satisfies the trivial OPEs 
\bea
J^a(z) \,  ( J^{u(1)} + \sqrt{\frac{M+N}{M N}}\, J^{u(1)}_f )(w) =
0 + \cdots,
\nonu \\
J^a(z) \, (J^{\alpha} + J_f^{\alpha})(w) =0 +\cdots.
\label{reg}
\eea
If we determine the currents from the coset fields from the beginning
by taking the multiple product between them and introducing the
arbitrary coefficients to be fixed, then it is necessary to check
the regular conditions like as (\ref{reg}) explicitly. However,
when we are calculating some OPEs between the known currents
which satisfy the regular conditions already, we do not need to check
them at each step because some composite operators
appearing in the right hand sides of these OPEs
satisfy these regular conditions automatically. At the final stage,
it is better to check these conditions for consistency checks.

Therefore, the lowest singlet and nonsinglet
${\cal N}=2$ multiplets are classified by
(\ref{kggt}) and (\ref{spin1mul}). The former is the ${\cal N}=2$ supersymmetric
extension of the stress energy tensor of the bosonic coset model.

\section{The nonsinglet  multiplet of spins
  $(1,\frac{3}{2},\frac{3}{2},2)$}

After identifying the last component of the ${\cal N}=2$ multiplet
correctly, we present the OPEs between the generators of
${\cal N}=2$ superconformal algebra and this ${\cal N}=2$ multiplet.
Moreover, the OPEs between the supersymmetry generators and the
spin-$2$ current $K^a$ obtained in the bosonic coset model are given.

\subsection{The spin-$2$ current}

The simplest ${\cal N}=2$ multiplet is described in (\ref{spin1mul}).
In this section, we would like to obtain the last component of this
multiplet in terms of coset fields explicitly. Later we will also
consider their OPEs and determine their algebraic structures
in section $6$.

How do we obtain the last component of the multiplet (\ref{spin1mul})?
According to the primary condition of the ${\cal N}=2$ multiplet under
the multiplet of (\ref{kggt}) in the ${\cal N}=2$ superconformal
algebra, we can use the OPEs between $G^{\pm}(z)$ and $G^{\mp,a}(w)$
and look at the first order pole. See Appendix $B$.
Based on the explicit expressions of (\ref{gpm}) and (\ref{gpma}),
we can compute the OPE  between
$G^{+}(z)$ and $G^{-,a}(w)$ explicitly.

After subtracting the descendant terms from the first order pole
of the OPE $G^{+}(z)\, G^{-,a}(w)$,
the primary spin-$2$ current is
\footnote{
  After extracting the bosonic spin-$2$ current from
  this spin-$2$ current, we obtain the following
  spin-$2$ current $
 \hat{W}^{+(2),a}  = 
  k \,t^a_{j \bar{i}} \, \de_{\rho \bar{\si}} \,
\psi^{(\rho \bar{i})}\,
\pa \psi^{(\bar{\si} j)}-\sqrt{\frac{M+N}{M N}} \, J^{u(1)}\, J_f^a
+  t^{\al}_{\rho \bar{\si}} \, t^a_{j \bar{i}} \, J^{\al} \,
 \psi^{(\rho \bar{i})}\,
 \psi^{(\bar{\si} j)} -\frac{1}{M} \, J^a \, J_f^{u(1)} +
 \frac{1}{2} (i f +d)^{b a c} \, J^b \, J_f^c
 +    \frac{k}{2}\, \pa \, J_f^a + \Big( - \frac{N}{2(2k+M)}\,
 d^{a b c} \, J^b \, J^c + \frac{N}{k}\, \sqrt{\frac{M+N}{M N}}\,
 J^a \, J^{u(1)} \Big)$ which is a
 primary. In section $6$, we will use the last
 relation
 of (\ref{nonspin2}) in order to apply the results in \cite{Ahn2011}.}
\bea
W^{+(2),a}& = &  -t^a_{j \bar{i}} \, \de_{\rho \bar{\si}} \, J^{(\rho \bar{i})}\,
J^{(\bar{\si} j)} + k \,t^a_{j \bar{i}} \, \de_{\rho \bar{\si}} \,
\psi^{(\rho \bar{i})}\,
\pa \psi^{(\bar{\si} j)}-\sqrt{\frac{M+N}{M N}} \, J^{u(1)}\, J_f^a
\nonu \\
&+ & t^{\al}_{\rho \bar{\si}} \, t^a_{j \bar{i}} \, J^{\al} \,
 \psi^{(\rho \bar{i})}\,
 \psi^{(\bar{\si} j)} -\frac{1}{M} \, J^a \, J_f^{u(1)} +
 \frac{1}{2} (i f +d)^{b a c} \, J^b \, J_f^c
 \nonu \\
 &- & \frac{N}{2} \, \pa \, J^a + \frac{k}{2}\, \pa \, J_f^a
 \equiv  -\frac{1}{2} \, K^a + \hat{W}^{+(2),a},
 \label{nonspin2}
 \eea
 where the spin-$2$ current obtained in the
 bosonic coset model is given by \cite{CH1812}
\bea
K^a & = & \de_{\rho \bar{\si}} \,
t^a_{j\bar{i}} \, (J^{(\rho \bar{i})} \,  J^{(\bar{\si} j)} +
J^{(\bar{\si} j)} \,  J^{(\rho \bar{i})})  
 -\frac{N}{(M+2k)} \, d^{abc} \, J^b\, J^c
 \nonu \\
 & + & \frac{2N}{k} \sqrt{\frac{M+N}{M N}} \, J^a  \,  J^{u(1)}.
\label{spin2Ka}
 \eea
 We realize that
 the field contents in (\ref{nonspin2})
 cannot be written in terms of the known currents in
 (\ref{fields}) and (\ref{spin1mul})
 due to the first five terms.
 In particular, the fourth term of (\ref{nonspin2})
 looks like the first term of spin-$3$ current $P^a$
 appearing in (\ref{fields}) in the sense that
 $J^{(\rho \bar{i})}\,
 J^{(\bar{\si} j)}$ is replaced by
$\psi^{(\rho \bar{i})}\,
 \psi^{(\bar{\si} j)}$ and other factors remain unchanged.
 The reason why we write down the above spin-$2$ current
 in terms of the known previous spin-$2$ current and other piece is
 that when we calculate some OPEs including this current, we can use
 the previous result found in \cite{Ahn2011}.
 For example, in the OPE between $W^{+(2),a}(z)$ and $W^{+(2),b}(w)$,
 it is rather complicated to determine the OPE between the first term
 of (\ref{nonspin2}) and itself. Instead, we can change those first term
 by using (\ref{spin2Ka}) in terms of $J^a, J^{u(1)}$ and $K^a$.
 We will see the details in section $6$.

 \subsection{ The OPEs with the currents of
 ${\cal N}=2$ superconformal algebra}

 Once we have determined the four currents in the multiplet,
 \bea
(J_f^a, G^{+,a}, G^{-,a}, W^{+(2),a}),
 \label{simplest}
 \eea
 we need to check whether they are really the components of
 ${\cal N}=2$ multiplet or not.
 It is straightforward to calculate the following OPEs
 with the help of (\ref{threeJf}),
 (\ref{K}), (\ref{gpma}) and (\ref{nonspin2})
\bea
K(z) \, J_f^a(w) & = & 0 +\cdots,
\nonu \\
K(z) \, G^{\pm,a}(w) & = & \pm \frac{1}{(z-w)}\, G^{\pm,a}(w) +\cdots,
\nonu \\
K(z) \, W^{+(2),a}(w) & = & \frac{1}{(z-w)^2}\, \Bigg[ N\, J^a -k \,
  J_f^a \Bigg](w)  +\cdots.
\label{class1}
\eea
When we compare the ones in Appendix $B$ with these OPEs (\ref{class1}),
the last OPE contains the additional term $J^a$ in the second order pole
of the right hand side. There are two ways to fix this inconsistency
by considering the above spin-$2$ current with some modifications
in order to have $J^a_f(w)$ in the above second order pole 
or by taking the spin-$1$ current as $(N\, J^a -k \,
  J_f^a)$ rather than the previous $J_f^a$.
  Recall that the OPE between $K(z)$ and $J^a(w)$ is regular because
  it is obvious to see that the OPE between the spin-$1$ current $J^a$
  and the spin-$1$ $J_f^{u(1)}$ having complex fermions
does not have any singular terms
  and moreover
  it is known that $J^a(z) \, J^{u(1)}(w) = 0 + \cdots$ from
  the observation of \cite{CH1812}.
  This implies that the above linear combination of spin-$1$ current
  satisfies the corresponding OPE which is the first one in
  (\ref{class1}). Then we will have the right property in Appendix $B$.

However, we will take the lowest component of the multiplet as
  $J^a_f$ rather than $(N\, J^a -k \,J_f^a)$ as in (\ref{simplest}).
  One of the reasons why we do not consider this linear combination
  as the lowest component of the above multiplet is that it is better
  to treat $J^a$ and $J_f^a$ separately when we decide which composite
  operators are written in terms of the known currents or not.
  Otherwise we should replace all the $J^a_f$ term by adding
  $J^a$ term in order to preserve the above linear combination
  in all the composite operators \footnote{
  Note that all the ${\cal N}=2$ multiplets in this paper
  do not satisfy the properties in Appendix $B$. In other words,
  some of the coefficients appearing in Appendix $B$ appear
  differently and the vanishing coefficients in the higher order
  poles in Appendix $B$ appear nontrivially.}.
  We observe that the combination of $(N\, J^a -k \,J_f^a)$
  will appear at many places of the OPEs we will calculate later.
  
  Let us describe the OPEs between the spin-$\frac{3}{2}$ current
  of ${\cal N}=2$ superconformal algebra 
  and the currents of the multiplet in (\ref{simplest})
  \footnote{
\label{iden}
    For convenience, we present the following OPEs
    $J^a(z) \, G^{\pm}(w)=\pm \, \frac{1}{(z-w)}\, G^{\pm,a}(w) +
    \cdots$, $J^{\al}(z) \, G^+(w)= \frac{1}{(z-w)}\, t^{\al}_{\rho
      \bar{\si}}\, \de_{j \bar{k}}\,
    \psi^{(\bar{\si} j)}\, J^{(\rho \bar{k})}(w) + \cdots$
    and  $J^{\al}(z) \, G^-(w)= -\frac{1}{(z-w)}\, t^{\al}_{\rho
      \bar{\nu}}\,\de_{m \bar{m}} \, \psi^{(\rho \bar{m})}\,
    J^{( \bar{\nu} m)}(w) + \cdots$ which will be used in the
    last OPE of (\ref{class2}).
  }
\bea
G^+(z) \, J_f^a(w) & = & \frac{1}{(z-w)}\, G^{+,a}(w) + \cdots,
\nonu \\
G^{+}(z) \, G^{+,a}(w) & = & 0 +\cdots,
\nonu\\
G^{+}(z) \, G^{-,a}(w) & = & \frac{1}{(z-w)^2} \, \Bigg[ N\, J^a -k \,
  J_f^a \Bigg](w) \nonu \\
& + & \frac{1}{(z-w)} \Bigg[\frac{1}{2}\,
  \Bigg( N\, \pa \, J^a -k \,
  \pa \, J_f^a \Bigg) +W^{+(2),a} \Bigg](w) + \cdots,
\nonu \\
G^{+}(z) \, W^{+(2),a}(w) & = & \frac{1}{(z-w)^2}\, \frac{3}{2}\, (k+N)\,
G^{+,a}(w)\nonu \\
& + &  \frac{1}{(z-w)}\, \frac{1}{2}\, (k+N)\,
\pa \, G^{+,a}(w) + \cdots.
\label{class2}
\eea
By changing the sign of $J_f^a(w)$ (and the remaining currents are the
same) we obtain the standard OPE
of $G^{+}(z) \, J_f^a(w)$ which leads to the minus sign of the
right hand side of this OPE. See also Appendix $B$.
Note that there is a term of
$J^a(w)$ in the second order pole of $G^{+}(z) \, G^{-,a}(w)$
as before.
From this  OPE, we have obtained the nonsinglet
spin-$2$ current we mentioned
before.
In the last OPE, the numerical value in the right hand side
has an extra factor $(k+N)$ when we compare with the one in Appendix
$B$ \footnote{In the last OPE, we use the following identity
  $t^a_{j \bar{i}} \, \de_{k \bar{l}}\, \de_{\rho \bar{\mu}}\, \de_{\tau \bar{\si}}\,
  \psi^{(\bar{\si} j)}\, \psi^{(\rho \bar{i})}\, \psi^{(\bar{\mu} k)}\, J^{(\tau
    \bar{l})}= \frac{1}{2} (i \, f + d)^{b a c}\, G^{+,b} \, J_f^c -
  \frac{1}{M} \, G^+ \, J_f^a -\frac{1}{M}\, G^{+,a}\, J_f^{u(1)}$
  by using the rearrangement lemma in \cite{BBSS,BS}.
  Here the $J_f^{u(1)}$ term is canceled by other term in the first order
  pole of the last OPE. Then we are left with the known currents
  finally.}.

Similarly, we obtain the following OPEs
\bea
G^-(z) \, J_f^a(w) & = & -\frac{1}{(z-w)}\, G^{-,a}(w) + \cdots,
\nonu \\
G^{-}(z) \, G^{+,a}(w) & = &  \frac{1}{(z-w)^2} \, \Bigg[ -N\, J^a +k \,
  J_f^a \Bigg](w) \nonu \\
& + & \frac{1}{(z-w)} \Bigg[\frac{1}{2}\,
  \Bigg( -N\, \pa \, J^a + k \,
  \pa \, J_f^a \Bigg) +W^{+(2),a}  + i \, f^{a b c}\, J^b \, J_f^c
  \Bigg](w) + \cdots,
\nonu \\
G^{-}(z) \, G^{-,a}(w) & = & 0 +\cdots,
\nonu \\
G^{-}(z) \, W^{+(2),a}(w) & = & \frac{1}{(z-w)^2}\,
\frac{1}{2}\, (3k+2M+3N)\, G^{-,a}(w) \nonu \\
& + &
\frac{1}{(z-w)} \, \Bigg[ \frac{1}{3} \,
  \frac{1}{2}\, (3k+2M+3N)\, \pa \, G^{-,a}
  + i \, f^{a b c}\, J^b \, G^{-,c}- i \, f^{a b c}\, G^{-,b}\, J_f^c
  \nonu \\
  & - & \frac{ M}{3} \, \pa \, G^{-,a} \Bigg](w) + \cdots.
\label{class3}
\eea
By changing the sign of $J_f^a(w)$ as before,
we obtain the standard OPE
of $G^{-}(z) \, J_f^a(w)$ in Appendix $B$.
There exists a term of
$J^a(w)$ in the second order pole of $G^{-}(z) \, G^{+,a}(w)$.
In the last OPE of (\ref{class3}), we do not combine
with the first and the fourth terms in the first order pole
\footnote{In all the OPEs in this paper, we intentionally put the exact
  coefficients coming from the descendant terms in the right hand sides
  of the OPEs without simplifying them.}.
The last three terms of the last OPE (which is primary
under the stress energy tensor)
are written in terms of the
known currents in (\ref{fields}) \footnote{
  In the last OPE, the following identity is used
  $t^a_{j \bar{i}} \, \de_{\rho \bar{\mu}}\, \de_{\tau \bar{\si}}\, \de_{k \bar{l}}
  \, \psi^{(\rho \bar{i})}\, \psi^{(\bar{\si} j)}\, \psi^{(\tau \bar{l})}\,
  J^{(\bar{\mu} k)}= \frac{1}{M}\, J_f^a \, G^- + \frac{1}{M}\, J_f^{u(1)}\,
  G^{-,a}-\frac{1}{2}(i \, f+ d)^{a b c}\, J_f^c \, G^{-,b}$
  with the help of \cite{BBSS,BS} again. The $J_f^{u(1)}$
  term we do not want to have can be canceled by other term appearing in
  the
first order pole of the last OPE of (\ref{class3}).}.

We can check that each current in the multiplet
we are considering in this subsection
satisfies the primary condition under the stress energy tensor (\ref{T})
as follows:
\bea
T(z) \, J_f^a(w) & = & \frac{1}{(z-w)^2} \, J_f^a(w) +
\frac{1}{(z-w)} \, \pa \, J_f^a(w) + \cdots,
\nonu \\
T(z) \, G^{\pm,a}(w) & = & \frac{1}{(z-w)^2} \, \frac{3}{2} \,
G^{\pm,a}(w) +
\frac{1}{(z-w)} \, \pa \, G^{\pm,a}(w) + \cdots,
\nonu \\
T(z) \, W^{+(2),a}(w) & = & \frac{1}{(z-w)^2} \, 2 
\, W^{+(2),a}(w) +
\frac{1}{(z-w)} \, \pa \, W^{+(2),a}(w) + \cdots.
\label{class4}
\eea
Of course, it is rather nontrivial to check these OPEs
(\ref{class4})
for generic $k,N$ and $M$ by hands but we can check them
for fixed $N$ and $M$ by using the Thielemans package
\cite{Thielemans}.
When we notice the presence of higher order terms where
the pole is greater than or equal to three, then we can try to
do this checking by hands on the higher order terms.
Usually, it is not necessary to
check the first and second order poles of the OPEs.
For the quasi primary operators, it is nontrivial to check the
fourth order poles by hands.

As a consistency check, we can check that
the currents satisfy the following regular conditions
\bea
(J_f^a, G^{+,a}, G^{-,a}, W^{+(2),a})(z) \,
( J^{u(1)} + \sqrt{\frac{M+N}{M N}}\, J^{u(1)}_f )(w) =
0 + \cdots,
\nonu \\
(J_f^a, G^{+,a}, G^{-,a}, W^{+(2),a})(z) \, (J^{\alpha} + J_f^{\alpha})(w) =0 +\cdots.
\label{REG}
\eea
For example, the relative coefficients in (\ref{nonspin2})
are fixed from these constraints and the primary
condition under the stress energy tensor (\ref{T}) (and its
${\cal N}=2$ extension).

\subsection{ Some OPEs with the spin-$2$ current $K^a$}

It is clear that the OPE between $K(z)$ and the spin-$2$ current
$K^a(w)$
does not contain the singular terms because
the OPE between the $J^{u(1)}(z)$ and $K^a(w)$
is regular and the $K^a(w)$ does not have the complex fermions.
Then the question is what are the OPEs between $G^{\pm}(z)$
and $K^a(w)$ \footnote{The OPE between $K(z)$ and $K^a(w)$ is regular
  and the OPE between  $T(z)$ and $K^a(w)$ satisfies the standard OPE for
a primary field mentioned before.}.

By using the explicit expressions of (\ref{gpm}) and
(\ref{spin2Ka})
we obtain
\bea
G^{+}(z) \, K^{a}(w) & = & -\frac{1}{(z-w)^2} \,
\frac{2(k^2-1)(2k+M+N)}{k(2k+M)} \, G^{+,a}(w) \nonu \\
& + & 
\frac{1}{(z-w)} \, \Bigg[ -\frac{1}{3}
\frac{2(k^2-1)(2k+M+N)}{k(2k+M)} \, \pa \, G^{+,a}
+ V^{+(\frac{5}{2}),a} \Bigg](w) \nonu \\
& + & \cdots.
\label{gkplus}
\eea
Note that the first order pole provides
the spin-$\frac{5}{2}$ current
which cannot be written in terms of the known currents
(\ref{fields}) and (\ref{nonspin2}).
The primary spin-$\frac{5}{2}$ current under the stress
energy tensor (\ref{T}), after subtracting the descendant term
in the first order pole is 
\bea
V^{+(\frac{5}{2}),a} & = &
-2 (k+N) \, t^a_{j \bar{i}} \, \de_{\rho \bar{\si}} \, \pa \,
J^{(\rho \bar{i})}\,
\psi^{(\bar{\si} j)}
-\frac{2(k+N)}{k} \,\sqrt{\frac{M+N}{M N}} \, J^{u(1)}\, G^{+,a}
\nonu \\
&-& \frac{2}{k M} \, (k+M+N) \, J^a \, G^{+}
-(i f - \frac{2k+M+2N}{2k+M} \, d)^{a b c} \, J^b \, G^{+,c}
\nonu \\
& + & 2 \, t^{\al}_{\rho \bar{\si}} \, t^a_{j \bar{i}} \, J^{\al} \,
 J^{(\rho \bar{i})}\,
 \psi^{(\bar{\si} j)} -\frac{2}{3}\,
 \frac{2(k^2-1)(2k+M+N)}{k(2k+M)} \, \pa \, G^{+,a}.
\label{5halfone}
\eea
Note that the first, second and fifth terms are the reasons why
we cannot write down this current in terms of the known
currents.
Again, the field content of the fifth term looks like
the fourth term of (\ref{nonspin2}) in the sense that
the spin-$\frac{1}{2}$ operator is replaced by the spin-$1$
current.

Similarly, we can determine the following OPE
\bea
G^{-}(z) \, K^{a}(w) & = &
 -\frac{1}{(z-w)^2} \,
\frac{2(k^2-1)(2k+M+N)}{k(2k+M)} \, G^{-,a}(w) \nonu \\
& + & 
\frac{1}{(z-w)} \, \Bigg[ -\frac{1}{3}
\frac{2(k^2-1)(2k+M+N)}{k(2k+M)} \, \pa \, G^{-,a}
+ V^{-(\frac{5}{2}),a} \Bigg](w) \nonu \\
& + & \cdots,
\label{gkminus}
\eea
where 
the primary spin-$\frac{5}{2}$ current is 
\bea
 V^{-(\frac{5}{2}),a} & = &
-2 (k+N) \, t^a_{j \bar{i}} \, \de_{\rho \bar{\si}} \, \pa \,
J^{(\bar{\si} j)}\,
\psi^{(\rho \bar{i})}
+\frac{2(k+N)}{k} \,\sqrt{\frac{M+N}{M N}} \, J^{u(1)}\, G^{-,a}
\nonu \\
&+& \frac{2}{k M} \, (k+M+N) \, J^a \, G^{-}
-(i f + \frac{2k+M+2N}{2k+M} \, d)^{a b c} \, J^b \, G^{-,c}
\nonu \\
& - & 2 \, t^{\al}_{\rho \bar{\si}} \, t^a_{j \bar{i}} \, J^{\al} \,
 J^{(\bar{\si} j)}\,
 \psi^{(\rho \bar{i})} -\frac{2}{3}\,
 \frac{2(k^2-1)(2k+M+N)}{k(2k+M)} \, \pa \, G^{-,a}.
 \label{5halftwo}
\eea
The field contents of (\ref{5halftwo})
look similar to the one in (\ref{5halfone}).
They will appear at many places in the OPEs we will consider later.
Due to the presence of the new primary fields (\ref{5halfone}) and
(\ref{5halftwo}) for given OPEs, the bosonic spin-$2$ current
$K^a$ by itself
cannot play the role of the last component of the multiplet
in (\ref{spin1mul}). In other words, the right spin-$2$
current in (\ref{nonspin2}) contains  both fermionic dependent terms and
bosonic current dependent terms in addition to $K^a$.
If we remember the OPEs in Appendix $B$,
the OPEs in (\ref{gkplus}) and (\ref{gkminus})
imply that the spin-$2$ current $K^a$ can be a candidate for the
lowest component of other nonsinglet
${\cal N}=2$ multiplet which will be explained
later if we succeed to eliminate the second order poles of
(\ref{gkplus}) and (\ref{gkminus}) \footnote{Of course,
  if we add some terms to $K^a$ and can remove
  $V^{\pm (\frac{5}{2}), a}$ terms properly, then this modified
  spin-$2$ current can be the last component of the ${\cal N}=2$
  multiplet of this section. See the last OPEs in (\ref{class2})
and (\ref{class3}).}.

In summary, the equations (\ref{class1}), (\ref{class2}),
(\ref{class3}) and (\ref{class4})
imply that the right hand sides of these OPEs
contain the currents (\ref{simplest}), the
nonsinglet spin-$1$ current
$J^a$ and their composite operators.
As emphasized before, these OPEs do not satisfy Appendix $B$.

\section{ The singlet
   multiplet of spins $(2,\frac{5}{2},\frac{5}{2},3)$}

We would like to construct this multiplet
starting from the singlet spin-$3$ current found in the
bosonic coset model \cite{CH1812}. 

\subsection{Construction of lowest component}

Once we know any component of this multiplet, then
the other three components can be determined, in principle, by using
the ${\cal N}=2$ primary conditions described in Appendix $B$.
Let us first consider how we obtain the lowest component
of this ${\cal N}=2$ multiplet \footnote{
  Although we have realized that
the spin-$2$ current $K^a$ will participate in the lowest component
from the previous section,
it is not clear how we can continue to calculate the additional terms
by adding the possible composite operators of spin-$2$ to the $K^a$
with an appropriate contraction in the indices.}.

There exists a singlet spin-$3$ current $W^{(3)}$
in the bosonic coset model and its expression
in terms of coset fields is described by \cite{CH1812}
 \bea
W^{(3)} & = &
b_1 \, d^{\alpha \beta \gamma} \, J^{\alpha} J^{\beta} J^{\gamma}
+ b_2 \, d^{a b c} \, J^a J^b J^c+
b_3 \, J^{u(1)}  J^{u(1)}  J^{u(1)} + b_4 \, J^{\al} J^{\al} J^{u(1)}
\nonu \\
&+& b_5 \, J^a \, J^a \, J^{u(1)} + b_6 \, 
t^{\al}_{\rho \bar{\si}} \, \de_{j\bar{i}}\,
J^{\al}   \, (J^{(\rho \bar{i})}  \, J^{(\bar{\si} j)} +
J^{(\bar{\si} j)} \,   J^{(\rho \bar{i})}) 
\nonu \\
&+& b_7  \, \de_{\rho \bar{\si}} \,
t^a_{j\bar{i}} \, J^a  (J^{(\rho \bar{i})}  J^{(\bar{\si} j)} +
J^{(\bar{\si} j)}  J^{(\rho \bar{i})})  
+b_8 \, \de_{\rho \bar{\si}} \,
\de_{j\bar{i}} \, J^{u(1)}  (J^{(\rho \bar{i})}  J^{(\bar{\si} j)} +
J^{(\bar{\si} j)}  J^{(\rho \bar{i})})   
\nonu \\
&+& b_{12} \, 
\de_{\rho \bar{\si}} \,
\de_{j\bar{i}} \, \pa  \, J^{(\rho \bar{i})}  \,J^{(\bar{\si} j)}
+ b_{13} \,
\, \de_{\rho \bar{\si}} \,
\de_{j\bar{i}}  \, \pa  \, J^{(\bar{\si} j)} \,  J^{(\rho \bar{i})}
+ b_{14} \, \pa^2 \, J^{u(1)}.
\label{W}
\eea
Here the relative coefficients are functions of $k, N$ and $M$
as follows:
\bea
b_2 & = & -\frac{ N (k+N) (k+2 N)}{M (k+M) (k+2 M)} \, b_1,
\qquad
b_3 = \sqrt{\frac{M+N}{M N}} \,
\frac{ 2  (k+N) (k+2 N) (M+N)}{k^2 M} \, b_1,
\nonu \\
b_4 & = & \sqrt{\frac{M+N}{M N}} \, \frac{ 6  (k+N)}{k} \, b_1,
\qquad
b_5 = \sqrt{\frac{ M+N}{M N}} \,
\frac{ 6 N (k+N) (k+2 N)}{k M (k+2 M)}\,
b_1,
\nonu \\
  b_6 & = & -\frac{3  (k+N)}{M} \, b_1,\qquad
  b_7 = \frac{3  (k+N) (k+2 N)}{M (k+2 M)} \, b_1,
  \nonu \\
  b_8 & = & -\sqrt{\frac{M+N}{M N}}\,
  \frac{ 3  (k+N) (k+2 N)}{k M}\, b_1,
  \qquad
b_{12} = \frac{3  (k+N) (k+2 N)}{M}\, b_1,
\nonu \\
b_{13} & = & -\frac{3  (k+N) (k+2 N)}{M} \, b_1,
\qquad b_{14} = - \sqrt{\frac{ M+N}{M N}}\,
N (k+N) (k+2 N) \, b_1.
\label{bvalues}
\eea

Although we add the spin-$1$ currents coming from the complex
fermions to the ones in the bosonic coset model studied in
\cite{CH1812}, the regular conditions with the coset fields
in the coset (\ref{coset}) hold as follows:
\bea
W^{(3)}(z) \,  ( J^{u(1)} + \sqrt{\frac{M+N}{M N}}\, J^{u(1)}_f )(w) =
0 + \cdots,
\nonu \\
W^{(3)}(z) \, (J^{\alpha} + J_f^{\alpha})(w) =0 +\cdots,
\label{REG1}
\eea
because the OPEs between the spin-$3$ current $W^{(3)}$
(consisting of purely bosonic fields)
and these spin-$1$ currents (consisting of purely fermionic fields)
do not have any singular terms as in (\ref{REG}). 

We need to calculate the OPEs between the supersymmetry generators
of the ${\cal N}=2$ superconformal algebra
and the above singlet spin-$3$ current in order to determine
its superpartners. Because the spin-$\frac{3}{2}$ currents consist of
the spin-$1, \frac{1}{2}$ operators transforming as
$({\bf N}, \overline{{\bf M}})$ or $(\overline{{\bf N}},
{\bf M})$, we
should first calculate the OPEs between the spin-$1$ currents and
the spin-$3$ current $W^{(3)}$. Note that it is obvious to see
that the OPEs between the spin-$\frac{1}{2}$ operator
and $W^{(3)}$ are regular. In Appendix $C$, we present
some relevant OPEs between the spin-$1$ current and the singlet spin-$3$
current.

It turns out that after extracting
the OPE $W^{(3)}(z) \, G^{\pm}(w)$ and then
changing the order of the two currents we have
\bea
&& G^{\pm}(z) \, W^{(3)}(w)  =\nonu \\
&&
\mp \frac{1}{(z-w)^3} \,
\frac{ (k^2-1) (k^2-4) (k+M+N) (2 k+M+N) (3 k+2 M+2 N)}{
  k^2 M (k+M) (k+2 M)} \, b_1 \, G^{\pm}(w)\nonu \\
&& \pm
\frac{1}{(z-w)^2} \,
G^{\pm (\frac{5}{2}),0}(w) + {\cal O} (\frac{1}{(z-w)}).
\label{gw3}
\eea
We do not specify the first order poles of (\ref{gw3}) here.
They can be written explicitly from the results of Appendix $C$.
Because there are no descendant terms
associated with the spin-$\frac{3}{2}$ currents
at the second order pole, the primary singlet
spin-$\frac{5}{2}$ currents
appear in the second order pole.

One of them is given by
\bea
G^{+(\frac{5}{2}),0} & = &
\frac{(k^2-4)(k+M+N)(3k+2M+2N)}{k M (k+2M)} \, b_1 \, 
\Bigg[ - 3 \, t^{\al}_{\si \bar{\si}} \, \de_{j \bar{i}} \, J^{\al}
  \, \psi^{(\bar{\si} j)}\, J^{(\si \bar{i})}\nonu \\
  &-& \frac{3}{(k+M)} \, (k+N) \, J^a \, G^{+,a}-
  \frac{3}{k}\, \sqrt{\frac{M+N}{M N}} \, (k+N)
  J^{u(1)}\, G^{+} \nonu \\
  &+& 3 \, (k+N) \, \de_{\rho \bar{\si}}\, \de_{j \bar{i}} \,
  \psi^{(\bar{\si} j)}\, \pa \, J^{(\rho \bar{i})} -
  \frac{(k^2-1)(2k+M+N)}{k(k+M)} \, \pa \, G^{+} \Bigg].
\label{g5half}
\eea
Due to the first, third and fourth terms,
we cannot rewrite (\ref{g5half}) in terms of previous known currents
we have discussed so far \footnote{The overall factor $b_1$ appears in this
expression by using (\ref{bvalues}) for the $W^{(3)}$.}.
The field contents look similar to the one $V^{+(\frac{5}{2}),a}$
(\ref{5halfone}).
There is no $f$ or $d$ symbols in (\ref{g5half}) because
this current should be a singlet field.
We can easily check that
the first term comes from the sixth term of (\ref{W}).

The other is given by
\bea
G^{-(\frac{5}{2}),0} &  = &
\frac{(k^2-4)(k+M+N)(3k+2M+2N)}{k M (k+2M)} \, b_1 \, 
\Bigg[  3 \, t^{\al}_{\si \bar{\si}} \, \de_{j \bar{i}} \, J^{\al}
  \, \psi^{(\si \bar{i})}\, J^{(\bar{\si} \bar{j})}\nonu \\
  &+& \frac{3}{(k+M)} \, (k+N) \, J^a \, G^{-,a}+
  \frac{3}{k}\, \sqrt{\frac{M+N}{M N}} \, (k+N)
  J^{u(1)}\, G^{-} \nonu \\
  &+& 3 \, (k+N) \, \de_{\rho \bar{\si}}\, \de_{j \bar{i}} \,
  \psi^{(\rho \bar{i})}\, \pa \, J^{(\bar{\si} j)} -
  \frac{(k^2-1)(2k+M+N)}{k(k+M)} \, \pa \, G^{-} \Bigg].
\label{g5half1}
\eea
The field contents of (\ref{g5half1})
look similar to the one
 $V^{-(\frac{5}{2}),a}$
(\ref{5halftwo}).
Again the first term comes from the sixth term of (\ref{W}).
Of course, after we determine the lowest component of this
multiplet later, we should check whether the above two primary
spin-$\frac{5}{2}$
currents are really the elements of this multiplet under the action of
the generators of ${\cal N}=2$ superconformal algebra
based on Appendix $B$. In general,
we expect
that the spin-$\frac{5}{2}$ currents obtained by
$W^{(3)}$ from the bosonic coset model
will be different from the ones determined by 
the corresponding currents in ${\cal N}=2$ multiplet.
However, it turns out that they are equivalent to each other.
In next section, we will observe some examples
where this is not the case.

From Appendix $B$,
the second order pole in the OPE of
either $G^{-}(z) \, G^{+(\frac{5}{2}),0}(w)$ or
$G^{+}(z) \, G^{-(\frac{5}{2}),0}(w)$
leads to the following
primary singlet
spin-$2$ current, which is the lowest component of
${\cal N}=2$ multiplet
in this section \footnote{If we act $G^{\pm}(z)$ on $T_{boson}(w)$
  appearing in the footnote \ref{bosonT},
  then we obtain the spin-$\frac{3}{2}$ current at the second
  order pole  which is proportional to $G^{\pm}(w)$. After
  acting on $G^{\mp}(z)$ further, then we obtain the second order pole
  which is proportional to $K(w)$: the lowest component of (\ref{kggt}).
  This is one way to observe that the multiplet of (\ref{kggt})
  is an ${\cal N}=2$
  extension of the stress energy tensor in the bosonic coset model.}
\bea
W^{-(2),0} & = &
\frac{(k^2-4)   (k+M+N) (3 k+2 M+2 N)}{k M (k+2M)}\, b_1 \,  \Bigg[
  \nonu \\
  & - & \frac{2 
  (k^3+2 k^2 M+2 k^2 N+3 k M N+2 k+M+N)}{k  (k+M)}\,
\de_{\rho \bar{\si}} \de_{j \bar{i}} \,
J^{(\rho \bar{i})}  J^{(\bar{\si} j)}
\nonu \\
&+ &
3 \, M \,
J^{\al}\, J^{\al}
-\frac{2  (k^2-1)    (2 k+M+N) }{k  (k+M) } \
J^{\al} \, J_f^{\al}
\nonu \\
&-&
\frac{2  (k^2-1)    (2 k+M+N) }{
  k  (k+M) } \, \sqrt{\frac{M+N}{M N}} 
\, J^{u(1)} J^{u(1)}_f
\nonu \\
& - & \frac{2  (k^2-1)    (2 k+M+N) }{k  (k+M)}\,
J^a \, J^a_f
\nonu \\
&+& \,
\frac{ (k^2-1)    (2 k+M+N) }{  (k+M) }\,
\de_{\rho \bar{\si}} \, \de_{j \bar{i}} \,
\pa \, \psi^{(\rho \bar{i})} \, \psi^{(\bar{\si} j)}
+ \frac{3  N  (k+N) }{  (k+M) }\,
J^a \, J^a 
\nonu \\
&-&
\frac{ (k^2-1)    (2 k+M+N) }{  (k+M) }
\, \de_{\rho \bar{\si}} \, \de_{j \bar{i}} \,
\psi^{(\rho \bar{i})} \, \pa\, \psi^{(\bar{\si} j)}
\nonu \\
&+&
\frac{3  (k+N) (M+N) }{k  }
\, J^{u(1)} \, J^{u(1)}
\nonu \\
&+&
\frac{ N M
  (k^3+2 k^2 M+2 k^2 N+3 k M N+2 k+M+N)}{k (k+M)}\,
 \sqrt{\frac{M+N}{M N}}
\, \pa \, J^{u(1)} \, \Bigg].
\label{singletspin2}
\eea
Note that the second, fifth and ninth terms in (\ref{singletspin2})
do not arise in the stress energy tensor $T$ (\ref{T})
and we cannot express this in terms of the known currents
obtained so far.
The fifth term is characteristic of this spin-$2$ current
(or of modified stress energy tensor in Appendix $A$)
in the sense that we do not see this term from the bosonic
stress energy tensor also. 
In doing this computation, we need to obtain the OPE
 between $G^{-,a}(z)$ and $G^{+,b}(w)$ when we consider the OPE
between the $G^-(z)$ and the second term of (\ref{g5half})
at the coordinate $w$. See also Appendix $D$
where the corresponding OPE
is written in terms of the coset fields and section $6$
where the right hand side of this OPE is expressed in terms of the
known currents explicitly.

In section $7$, we will calculate the OPE between this
singlet spin-$2$ current and itself.

\subsection{Construction of second and third components}

From the Thielemans package \cite{Thielemans} with \cite{mathematica},
we can check that
the OPE between $G^+(z)$ and
$W^{-(2),0}(w)$ does not contain the singular terms
except the first order pole for fixed $N=5, M=4$ values.
It turns out that the first order pole of this OPE
is proportional to $G^{+(\frac{5}{2}),0}(w)$.
Then we should determine the coefficient appearing
in front of this spin-$\frac{5}{2}$ current.
As described before, we can focus on the particular
nonderivative term of  $G^{+(\frac{5}{2}),0}(w)$.
That is, the first term of (\ref{g5half}).
We can observe that the contribution from this term
comes from the first three terms in $W^{-(2),0}(w)$ of
(\ref{singletspin2}) and by collecting all the contributions
we can obtain the final coefficient appearing in
the spin-$\frac{5}{2}$ current $G^{+(\frac{5}{2}),0}(w)$
in the above OPE. This implies that
the spin-$\frac{5}{2}$ current $G^{+(\frac{5}{2}),0}$
belongs to the second component of this ${\cal N}=2$ multiplet.
See also Appendix $B$.

Similarly, we should also check whether
the spin-$\frac{5}{2}$ current $G^{-(\frac{5}{2}),0}$
belongs to the third component of this ${\cal N}=2$ multiplet
or not. In this case, we also
observe that the OPE between $G^-(z)$ and
$W^{-(2),0}(w)$ does not contain the singular terms
except the first order pole for fixed $N=5, M=4$ values.
By considering the first three terms in $W^{-(2),0}(w)$ of
(\ref{singletspin2}) and  collecting all the contributions,
the final coefficient appearing in
the spin-$\frac{5}{2}$ current $G^{-(\frac{5}{2}),0}(w)$
in the above OPE can be fixed completely.
Therefore, the spin-$\frac{5}{2}$ current $G^{-(\frac{5}{2}),0}$
belongs to the third component of this ${\cal N}=2$ multiplet.

Then we
have the following currents obtained so far in this multiplet 
\bea
(W^{-(2),0}, G^{+(\frac{5}{2}),0}, G^{-(\frac{5}{2}),0}, ?),
\label{nonsingmul}
\eea
with the coset field contents given in
(\ref{singletspin2}), (\ref{g5half}) and (\ref{g5half1}).
The last component of (\ref{nonsingmul})
will be described in next subsection and we expect that
the corresponding generalization of (\ref{W}) will appear
and further checks will be given later.

\subsection{Construction of last component}

In this subsection, we would like to construct the
unknown last component in (\ref{nonsingmul}).
First of all, in the previous description of
(\ref{singletspin2}), we have obtained the following result
\bea
G^{-}(z) \,
G^{+(\frac{5}{2}),0}(w)\Bigg|_{\frac{1}{(z-w)^2}} =  W^{-(2),0}(w).
\label{spin2expr}
\eea
According to the explanation of Appendix $B$,
the last component of the ${\cal N}=2$ multiplet
can be obtained from the next first order pole
in the OPE between $G^{-}(z)$ and
$G^{+(\frac{5}{2}),0}(w)$.
After subtracting the descendant term associated with
(\ref{spin2expr}) in the first order pole, we arrive at
the following relation
\bea
W^{+(3),0}(w) & \equiv &
G^{-}(z) \, G^{+(\frac{5}{2}),0}(w)\Bigg|_{\frac{1}{(z-w)}}-
\frac{1}{4}\, \pa \,  W^{-(2),0}(w).
\label{spin3singletexpr}
\eea

Then from the first order pole
in the OPE between $G^{-}(z)$ and
$G^{+(\frac{5}{2}),0}(w)$ with derivative terms (\ref{spin3singletexpr}),
we eventually
obtain the following singlet spin-$3$ current as follows:
\bea
W^{+(3),0}
&=& \frac{(k^2-4)(k+M+N)(3k+2M+2N)}{k M (k+2M)} \, b_1 \, 
\Bigg\{ - 3 \Bigg[ t^{\alpha}_{\si \bar{\si}}\,\de_{j \bar{i}}  \,
  J^{(\bar{\si} j)} \, J^{\alpha} \,
  J^{(\si \bar{i})}\nonu \\
  &-& t^{\al}_{\rho \bar{\nu}} \,
  \de_{m \bar{m}}\, t^{\al}_{\si \bar{\si}}\, \de_{j \bar{i}}\,
  \psi^{(\bar{\si} j)}\, (( \psi^{(\rho \bar{m})}\, J^{(\bar{\nu} m)})\, J^{(\si
    \bar{i})})+ \sqrt{\frac{M+N}{M N}}\,
  t^{\al}_{\si \bar{\si}}\, \de_{j \bar{i}}\, \psi^{(\bar{\si} j)}\,
  J^{\al}\, \psi^{(\si \bar{i})}\, J^{u(1)}
  \nonu \\
  &-& (t^{\beta} \, t^{\al})_{\rho \bar{\si}}\, \de_{j \bar{i}}\,
  J^{\al}\, J^{\beta}\, \psi^{(\rho \bar{i})}\, \psi^{(\bar{\si} j)}
  + t^a_{i \bar{k}}\, t^{\al}_{\si \bar{\si}} \, J^{\al}\, J^a \,
  \psi^{(\si \bar{k})}\, \psi^{(\bar{\si} i)}
  \nonu \\
  &-& k \,  t^{\al}_{\si \bar{\si}} \,\de_{j \bar{i}}\,
  \psi^{(\bar{\si} j)}\, J^{\al}\, \pa \, \psi^{(\si \bar{i})}
  \Bigg] \nonu \\
&-& \frac{3}{(k+M)} \, (k+N) \, \Bigg[
  \frac{1}{2}\, (-N \, J^a \, \pa \, J^a +
  k \, J^a \, \pa \, J_f^a)
  + G^{-,a}\, G^{+,a}+J^a \, W^{+(2),a}\nonu \\
  & + & i \, f^{a b c}\, J^a \, J^b J_f^c
  \Bigg]\nonu \\
& - &
  \frac{3}{k}\, \sqrt{\frac{M+N}{M N}} \, (k+N)
  \Bigg[\sqrt{\frac{M+N}{M N}}\, G^- \, G^+ -\frac{1}{2}\,
    (k+M+N)\, J^{u(1)}\, \pa \, K \nonu \\
    & + & (k+M+N)\, J^{u(1)}\, T
    -\frac{1}{2}\, J^{u(1)}\,  (J^a+J_f^a)^2 \Bigg] \nonu \\
  &+& 3 \, (k+N) \, \Bigg[
    \de_{\rho \bar{\si}}\, \de_{j \bar{i}}\, J^{(\bar{\si} j)}\, \pa \,
    J^{(\rho \bar{i})} + \sqrt{\frac{M+N}{M N}}\,
    \de_{\rho \bar{\si}}\, \de_{j \bar{i}}\,
    \psi^{(\bar{\si} j)}\, \pa \,(
    \psi^{(\rho \bar{i})} \, J^{u(1)})
    \nonu \\
    &+& t^{\al}_{\rho \bar{\si}}\, \de_{j \bar{i}}\,
    \psi^{(\bar{\si} j)} \,
    \pa \, (\psi^{(\rho \bar{i})} \, J^{\al})
    -t^a_{i \bar{k}}\, \de^{i \bar{i}}\,
    \de_{\rho \bar{\si}}\, \de_{j \bar{i}}\,
     \psi^{(\bar{\si} j)} \,
    \pa \, (\psi^{(\rho \bar{k})} \, J^{a})
    \nonu \\
    &-& k \,\de_{\rho \bar{\si}}\, \de_{j \bar{i}}\,
 \psi^{(\bar{\si} j)} \, \pa^2 \, \psi^{(\rho \bar{i})}
 \Bigg]  \nonu \\
  & - & 
  \frac{(k^2-1)(2k+M+N)}{k(k+M)} \, \Bigg[
    -\frac{1}{2}\, (k+M+N)\, \pa^2 \, K +(k+M+N)\,
    \pa \, T
    \nonu \\
    & - & \frac{1}{2}\, \pa \, (J^a +J_f^a)^2
    \Bigg] \Bigg\}
  -
  \frac{1}{4}\, \pa \,  W^{-(2),0}.
  \label{spin3singlet}
\eea
We do not simplify this expression further
because there are not too many common terms.
The second term can be further simplified by using the
rearrangement lemma in \cite{BS}.
Except the last term of (\ref{spin3singlet}),
we can read off the five contributions denoted by each two
brackets inside of curly bracket from (\ref{g5half}).
Note that the first term of (\ref{spin3singlet}) originates
from the $b_6$ term of (\ref{W}) \footnote{
\label{someopeexpression}
  For convenience,
  we present, as in the footnote \ref{iden}, some relevant OPEs
  $J^a(z) \, G^{\pm}(w) =\pm \, \frac{1}{(z-w)}\, G^{\pm,a}(w) +
  \cdots $,
  $J^{\al}(z) G^+(w) =\frac{1}{(z-w)}\, t^{\al}_{\rho \bar{\si}}\,
  \de_{j \bar{k}}\, \psi^{(\bar{\si} j)}\, J^{(\rho \bar{k} )}(w) +
  \cdots $ and
$J^{\al}(z) G^-(w) =-\frac{1}{(z-w)}\, t^{\al}_{\rho \bar{\si}}\,
  \de_{j \bar{k}}\, \psi^{(\rho \bar{k})}\, J^{(\bar{\si} j )}(w)
  + \cdots$. These OPEs will be used at various places of this paper.}.
We can understand this feature from the OPE between
$G^{-}(z)$ and the first term of $G^{+(\frac{5}{2}),0}(w)$
where the OPE between the former and the spin-$\frac{1}{2}$
operator provides the spin-$1$ current transforming as
$(\overline{\bf N},{\bf M})$.
In this sense, the singlet spin-$3$ current
(\ref{spin3singlet}) is a generalization of the spin-$3$ current
(\ref{W}) in the ${\cal N}=2$ supersymmetric coset model.
As mentioned before, some OPEs in Appendix $D$ (or in section
$6$) can be used.

Then
 we
 have the following currents of this multiplet
 we are considering in this section
\bea
(W^{-(2),0}, G^{+(\frac{5}{2}),0}, G^{-(\frac{5}{2}),0}, W^{+(3),0}).
\label{finalmul}
\eea
They also satisfy the regular conditions as in (\ref{REG1}). 
It would be interesting to construct
the lowest singlet spin-$2$ current from the beginning
directly
without using the information of the spin-$3$ current in the
bosonic coset model and by collecting all the possible
composite singlet spin-$2$ operators with arbitrary coefficients.
The nontrivial part in this direction is how to fix these coefficients
which will depend on the three parameters $k, N$ and $M$
explicitly.

In next subsection, in order to check that
this multiplet (\ref{finalmul}) is right ${\cal N}=2$ multiplet,
other OPEs with the generators of ${\cal N}=2$ superconformal
algebra are described \footnote{Compared with the bosonic
  coset model  description, the last component of (\ref{finalmul})
is an ${\cal N}=2 $ supersymmetric version of previous spin-$3$
current (\ref{W}) and the remaining three components are its
superpartners. In (\ref{kggt}), the stress energy tensor
(\ref{T}) is the ${\cal N}=2$ supersymmetric version of $T_{boson}$
of the footnote \ref{bosonT}. There are also three
other components in (\ref{kggt}). It seems that there
is no bosonic analog for
the first ${\cal N}=2$ multiplet of (\ref{simplest}).}.

\subsection{The OPEs with the currents of ${\cal N}=2$
superconformal algebra}

The OPEs between the spin-$1$ current of (\ref{kggt})
and the currents of (\ref{finalmul}), by realizing the
corresponding OPEs for fixed $N$ and $M$, can be summarized by
\bea
K(z) \, W^{-(2),0}(w) &= &
\frac{1}{(z-w)^2}\, \Bigg[ \frac{2 (k^2-1) (k^2-4)
}{k^2 M (k+M) (k+2 M) } \nonu \\
& \times &  (k+M+N)^2 (2 k+M+N) (3 k+2 M+2 N)\,
b_1 \Bigg] \, K(w)  + \cdots, 
\nonu \\
K(z) \, G^{\pm(\frac{5}{2}),0}(w) &= &
\mp \frac{1}{(z-w)^2} \,
\Bigg[ \frac{ (k^2-1)  (k^2-4)  }{
  k^2 M (k+M) (k+2 M) } \nonu \\
&\times & (k+M+N) (2 k+M+N) (3 k+2 M+2 N) \, b_1 \Bigg] \,
G^{\pm}(w) \nonu \\
& \pm & \frac{1}{(z-w)} \,
G^{\pm (\frac{5}{2}),0}(w) + \cdots,
\nonu \\
K(z) \, W^{+(3),0}(w) &= & \frac{1}{(z-w)^2}\, \Bigg[
  -\, W^{-(2),0} \nonu \\
  & - & \frac{  (k^2-1) (k^2-4) (k+M+N)^2 (2 k+M+N) (3 k+2 M+2 N)}{
    k^2 M (k+M) (k+2 M)} \, b_1
  \, \nonu \\
  & \times & \Big( T-\frac{1}{2(k+M+N)}\, (J^a +J_f^a)^2 \Big)
  \Bigg](w)  + \cdots.
\label{kwithsinglet}
\eea
Note that in the last OPE of (\ref{kwithsinglet}),
the combination of stress energy tensor with
$(J^a+J_f^a)^2$ term is exactly the same as the one in
the OPE between the supersymmetry generators in Appendix $A$.
Compared with the ones in Appendix $B$, there are additional terms
in the right hand sides of the OPEs.
Although we can check all the relevant terms in the right hand sides,
the structure constants appearing in the composite operators of
the right hand sides in (\ref{kwithsinglet}) can be
determined by focusing on the particular operators as before.

For example, in the last OPE of (\ref{kwithsinglet}), we can focus on 
the $J^{\al}\, J^{\al}$ term in the second order pole,
which determines the coefficient of
$W^{-(2),0}$ because we do not see this particular term in the
remaining two terms. After that by considering 
the $J^a \, J_f^a$ term, which will appear in
the first and the last terms of the second order pole,
we can fix the coefficient of 
$(J^a+J_f^a)^2$ term because the coefficient of the first term
is already known from the previous analysis.
Finally, from the $J^a \, J^a$ term
in the second order pole, which will appear in
the second and the last terms, the coefficient appearing
in the stress energy tensor (the second term)
can be obtained because the coefficient
of the last terms is known.

Next we summarize the following OPEs between the spin-$\frac{3}{2}$
currents of the ${\cal N}=2$ superconformal algebra and the
currents of ${\cal N}=2$ multiplet in this section
\bea
&& G^{\pm}(z) \, W^{-(2),0}(w)  = 
\frac{1}{(z-w)}\, 2(k+M+N) \, G^{\pm (\frac{5}{2}),0}(w) + \cdots,
\nonu \\
&& G^{\pm}(z) \, G^{\pm (\frac{5}{2}),0}(w)  = 
0 + \cdots,
\nonu \\
&& G^{\pm }(z) \, G^{\mp (\frac{5}{2}),0}(w)  = 
\nonu \\
&& \pm \frac{1}{(z-w)^3} \,
\frac{  (k^2-1) (k^2-4) (k+M+N)^2 (2 k+M+N)
  (3 k+2 M+2 N)}{k^2 M (k+M) (k+2 M)} \, b_1\, 
K(w)\nonu \\
&& +
\frac{1}{(z-w)^2}\,  W^{-(2),0}(w)
+
\frac{1}{(z-w)}\, \Bigg[ \frac{1}{4}\, \pa \, W^{-(2),0} \mp
   W^{+(3),0}
  \Bigg](w) + \cdots,
\nonu \\
&& G^{\pm}(z) \, W^{+(3),0}(w) =
\nonu \\
&& \mp \frac{1}{(z-w)^3}
\frac{
  (k^2-1) (k^2-4)  (k+M+N)^2 (2 k+M+N) (3 k+2 M+2 N)}{
  k^2 M (k+M) (k+2 M)} \, b_1
\,
G^{\pm }(w)\nonu \\
&& \pm \frac{1}{(z-w)^2}  \,
\frac{5}{2} (k+M+N)  \,
\, G^{\pm (\frac{5}{2}),0}(w)
\nonu \\
&& \pm
\frac{1}{(z-w)} \, \frac{1}{5}\,
\frac{5}{2} (k+M+N)  \, \pa\, G^{\pm (\frac{5}{2}),0}(w)
+ \cdots.
\label{gwithsinglet}
\eea
As before, there are additional higher order terms in the right hand
sides of these OPEs
when we compare with the standard ${\cal N}=2$ primary conditions
described in Appendix $B$.
In the third OPE of (\ref{gwithsinglet}), we can also obtain the
singlet spin-$3$ current by using the upper sign.
The lower sign was used in previous analysis. 
In the last OPE of (\ref{gwithsinglet}), we can fix
the structure constant of the second order pole (where there are no descendant
terms)
by focusing on the first terms in (\ref{g5half}) or
(\ref{g5half1}). In other words,
in the computation of the left hand side of these OPEs,
after we  select the possible terms from $W^{+(3),0}$
(\ref{spin3singlet}) for fixed $(N,M)$ in the Thielemans package
and
add all the contributions from those terms
for generic $(N,M)$, we  compute them
manually and obtain $k,N$ and $M$ dependence explicitly.
It turns out that the first order pole is given by
the descendant terms only.

Finally, we can compute the OPEs with stress energy tensor.
It turns out that 
\bea
&& T(z) \, W^{-(2),0}(w) = 
\frac{1}{(z-w)^2}\,  2\, W^{-(2),0}(w) + \frac{1}{(z-w)}\,  \pa \,
W^{-(2),0}(w) + \cdots, 
\nonu \\
&& T(z) \, G^{\pm(\frac{5}{2}),0}(w) = 
\frac{1}{(z-w)^2}\,  \frac{5}{2} \,
G^{\pm(\frac{5}{2}),0}(w) + \frac{1}{(z-w)}\,  \pa \,
G^{\pm(\frac{5}{2}),0}(w) + \cdots,
\nonu \\
&& T(z) \, W^{+(3),0}(w) = \nonu \\
&& - \frac{1}{(z-w)^4} \,
\frac{3  (k^2-1)  (k^2-4) (3 k+2 M+2 N) (
  k+M+N)^2 (2 k+M+N)}{
  2 k^2 M (k+M) (k+2 M)} \, b_1 \, K(w)\nonu \\
&& +
\frac{1}{(z-w)^2}\,  3 \,
W^{+(3),0}(w) + \frac{1}{(z-w)}\,  \pa \,
W^{+(3),0}(w)
+\cdots.
\label{twithsinglet}
\eea
The singlet spin-$3$ current $W^{+(3),0}$ in (\ref{twithsinglet})
is a quasi primary current
under the stress energy tensor.
If we add the extra term to the $W^{+(3),0}$, 
\bea
W^{+(3),0}+ \frac{3}{4}\, K\, W^{-(2),0},
\label{newspin3}
\eea
then we can remove the fourth order pole in the last OPE
of (\ref{twithsinglet}) with
stress energy tensor because the OPE between $K(z)$ and $W^{-(2),0}(w)$
has only a second order pole from (\ref{kwithsinglet}).
Then (\ref{newspin3}) becomes a primary field.
We can easily obtain the OPEs between the
currents of ${\cal N}=2$ superconformal algebra and
the composite operator $ K\, W^{-(2),0}$ and then we will
obtain the corresponding OPEs containing (\ref{newspin3})
\footnote{In the OPE between $K(z)$ and
  $ K\, W^{-(2),0}(w)$, the second order pole has
  $W^{-(2),0}$ and $K \, K$ terms. Similarly, the
  OPE between
  $G^{+}(z)$ and  $ K\, W^{-(2),0}(w)$ provides
  the second order pole with $G^{+(\frac{5}{2}),0}$
  and the first order pole together with
  $G^{+} \, W^{-(2),0}$ and $K \, G^{+(\frac{5}{2}),0}$.
  Moreover,
 the
  OPE between
  $G^{-}(z)$ and  $ K\, W^{-(2),0}(w)$ leads to
  the second order pole with $G^{-(\frac{5}{2}),0}$
  and the first order pole together with
  $G^{-} \, W^{-(2),0}$ and $K \, G^{-(\frac{5}{2}),0}$.
We observe that there are several nonlinear terms.}.

Therefore, the singlet currents are given by (\ref{singletspin2}),
(\ref{g5half}), (\ref{g5half1}) and (\ref{spin3singlet})
in the coset realization and they satisfy
(\ref{kwithsinglet}), (\ref{gwithsinglet}), and (\ref{twithsinglet})
under the action of the generators of
${\cal N}=2$ superconformal algebra.
The right hand sides of these OPEs
contain the currents of ${\cal N}=2$ multiplet, the composite
spin-$2$ operator $(J^a+J_f^a)^2$ nonlinearly
as well as
the currents of ${\cal N}=2$ superconformal algebra.
We will describe some features on the OPEs between
this ${\cal N}=2$ multiplet and itself  in section $7$.

\section{ The nonsinglet
   multiplet of spins $(2,\frac{5}{2},\frac{5}{2},3)$}

We continue to construct the ${\cal N}=2$ multiplet which
has $SU(M)$ index $a$.
From the experience of \cite{Ahn2011}, we realize that
the OPE between the spin-$2$ current
$K^a(z)$ and itself $K^b(w)$ produces
the spin-$3$ current $P^c(w)$. We can start with either
this spin-$2$ current
plus other terms as a candidate for the lowest component 
or this spin-$3$ current plus other terms as a candidate
for the last component of the ${\cal N}=2$ multiplet.
Let us take the latter.

\subsection{Construction of lowest component}

Let us consider the spin-$3$ current found in
bosonic coset model of \cite{CH1812}
 \bea
 P^a & = & a_1 \,
 t^{\al}_{\rho \bar{\si}} \, t^a_{j\bar{i}} \, J^{\al} 
 J^{(\rho \bar{i})}  \, J^{(\bar{\si} j)}
 + a_2 \, J^{\al}  \, J^{\al}  \, J^a+a_3 \, J^b  \, J^b \, J^a
 +a_4 \, J^a  \, J^{u(1)}  \, J^{u(1)}
 \nonu \\
 &+& a_5 \, d^{abc} \, \de_{\rho \bar{\rho}} \,
t^b_{j\bar{i}} \, J^c   (J^{(\rho \bar{i})}   J^{(\bar{\rho} j)} +
J^{(\bar{\rho} j)}   J^{(\rho \bar{i})})  
+ a_7 \, \de_{\rho \bar{\si}} \,
t^a_{j\bar{i}}\, J^{u(1)}   (J^{(\rho \bar{i})}  J^{(\bar{\si} j)} +
J^{(\bar{\si} j)}   J^{(\rho \bar{i})})  
\nonu \\
&+& a_8 \, \de_{\rho \bar{\si}} \, \de_{j\bar{i}}\,
J^a   \, (J^{(\rho \bar{i})} \, J^{(\bar{\si} j)} +
J^{(\bar{\si} j)}  \,  J^{(\rho \bar{i})})  
+ a_9 \,  d^{abc} \, J^b  \, J^c  \,  J^{u(1)}
+ a_{11} \, i \, f^{abc} \, \pa \,   J^b  \,  J^c
\nonu \\
&+& a_{12} \, \de_{\rho \bar{\si}} \,
t^a_{j\bar{i}} \, \pa \, J^{(\rho \bar{i})} \, J^{(\bar{\si} j)}
+ a_{13} \, \de_{\rho \bar{\si}} \,
t^a_{j\bar{i}}  \, \pa  \, J^{(\bar{\si} j)} \,  J^{(\rho \bar{i})}(z)
+a_{16} \, \pa^2  \, J^a
\nonu \\
&+& a_{17} \, 6 \, \mbox{Tr} \, (t^{a} \,
t^{\left(b \right.} \, t^c \, t^{\left. d \right)}) \, 
J^b  \, J^c  \, J^d.
\label{spin3exp}
 \eea
 Here the relative coefficients depend on
 $k, N$ and $M$ as follows:
 \bea
 a_2  & = &
 \frac{a_1}{k}, \qquad a_3= \frac{ N (k+2 N)}{k (k+M) (3 k+2 M)} \, a_1,
 \qquad
 a_4 = \frac{ (k+2 N) (M+N)}{k^2 M} \, a_1,
 \nonu \\
 a_5  & = & -\frac{ (k+2 N)}{4 (k+M)} \, a_1,
 \qquad
 a_7 =   \frac{(k+2 N)}{2 k} \, \sqrt{\frac{M+N}{M N}} \, a_1,
 \qquad
 a_8 = -\frac{ (k+2 N)}{2 k M} \, a_1,
 \nonu \\
 a_9 &=& -  \frac{(k+2 N)N}{2 k (k+M)} \,
 \sqrt{\frac{ (M+N)}{M N}} \, a_1,
\qquad
a_{11} =\frac{ (k^2-8) N (k+2 N)}{4 k (k+M) (3 k+2 M)} \, a_1,
\nonu \\
a_{12} & = & -\frac{1}{2} \, (k+2 N) \, a_1,
\qquad
a_{13}  =\frac{1}{2}  \, (k+2 N) \, a_1,
\label{avalues}
 \\
a_{16} & = & -\frac{ N (6 k^3+9 k^2 M+4 k M^2+12 M)
  (k+2 N)}{12 k (k+M) (3 k+2 M)} \, a_1,
\qquad
a_{17}  =  \frac{ N (k+2 N)}{6 (k+M) (3 k+2 M)} a_1.
\nonu
 \eea 

 First of all this nonsinglet spin-$3$ current satisfies 
\bea
P^{a}(z) \,  ( J^{u(1)} + \sqrt{\frac{M+N}{M N}}\, J^{u(1)}_f )(w) =
0 + \cdots,
\nonu \\
P^{a}(z) \, (J^{\alpha} + J_f^{\alpha})(w) =0 +\cdots.
\label{Reg}
\eea
That is, the regular conditions in the bosonic coset model
are valid in the ${\cal N}=2$ supersymmetric coset model
because the additional terms in (\ref{Reg})
come from the complex fermions which commute with the spin-$3$
currents made by bosonic operators purely.

Let us describe the OPEs between the supersymmetry
generators of the ${\cal N}=2$ superconformal algebra
and the nonsinglet spin-$3$ current of the bosonic coset model.
We use the OPEs between the spin-$1$ currents and
the spin-$3$ current described in  Appendix $C$.
It turns out that with the help of (\ref{spin3exp}) and
(\ref{avalues})
\bea
&& G^{\pm}(z) \, P^a(w)  = \nonu\\
&&
\mp \frac{1}{(z-w)^3} \,
\frac{ (k^2-1) (k^2-4)  (2 k+M+N) (3 k+2 M+2 N)}{
  2k^2  (k+M) (3k+2 M)} \, b_1 \, G^{\pm,a}(w)\nonu \\
&& \pm
\frac{1}{(z-w)^2} \,
\hat{V}^{\pm (\frac{5}{2}),a}(w) + {\cal O} (\frac{1}{(z-w)}).
\label{gp}
\eea
The point here is that the second order pole of (\ref{gp})
possesses exactly the first, second and fifth terms
of $V^{+(\frac{5}{2}),a}$ (\ref{5halfone}) or
 $V^{-(\frac{5}{2}),a}$
(\ref{5halftwo})
with same relative coefficients. Then this implies that
we can express the above second order pole of (\ref{gp})
in terms of previous known operators.
We can absorb those three unwanted terms by using
$V^{\pm (\frac{5}{2}),a}$ and recombine the remaining terms
with the known operators appearing in the second order pole.

Then we obtain the following spin-$\frac{5}{2}$ currents
which are not new because they are previously known expressions
\bea
\hat{V}^{\pm(\frac{5}{2}),a} & = &
\frac{(k^2-4)(3k+2M+2N)}{2k(k+M)} \, a_1 \,
\Bigg[
  \frac{1}{2}\, V^{\pm (\frac{5}{2}),a} 
  \nonu \\
  &\pm & \frac{(2k+M+N)}{k (3k+2M)} \, J^a \, G^{\pm} +
  \frac{(k^2-1) M (2k +M+N)}{3k(2k+M)(3k+2M)} \, \pa \, G^{\pm,a}
  \nonu \\
  &+& \frac{(2k+M+N)}{2(3k+2M)} \, i \, f^{a b c} \, J^b\, G^{\pm,c}
  \mp \frac{M(2k+M+N)}{2(2k+M)(3k+2M)} \, d^{a b c} \, J^b\, G^{\pm,c}
  \Bigg],
\label{5HALF}
\eea
where $V^{\pm (\frac{5}{2}),a} $ are given by (\ref{5halfone}) and
(\ref{5halftwo}).
The last four terms in $\hat{V}^{\pm,(\frac{5}{2}),a}$ (\ref{5HALF})
are  primary
operators under the stress energy tensor
$T$ (\ref{T}) and they can be seen from
(\ref{5halfone}) and
(\ref{5halftwo}).

As done in previous section, the second order pole in the OPEs of
$G^{\mp}(z) \, \hat{V}^{\pm,(\frac{5}{2}),a}(w)$
leads to the following
primary nonsinglet spin-$2$ current
\bea
G^{\mp}(z) \, \hat{V}^{\pm,(\frac{5}{2}),a}(w)\Bigg|_{\frac{1}{(z-w)^2}}
& = & \pm
W^{-(2),a }(w),
\label{opegvhat}
\eea
where the lowest nonsinglet spin-$2$ current can be written as 
\bea
W^{-(2),a } & = & \frac{(k^2-4)(3 k+2 M+2 N)  }{ k (k+M) (3 k+2 M)}\,
a_1 \, 
\Bigg[ 
\frac{1}{4} \, (2 k+M) (3 k+M+3 N)
  \, K^a
\nonu \\
& + & \frac{1}{k} \, (k^2-1)   (2 k+M+N) 
  \, \Bigg( W^{+(2),a} +  \frac{1}{
  2  } \, i \, f^{a b c} \, J^b \, J_f^c \Bigg) \Bigg].
\label{Spin2non}
\eea
Here we have the relations $W^{+(2),a}$ (\ref{nonspin2}) and
$K^a$ (\ref{spin2Ka})
which are given by the coset fields.
Note that this is a new quantity which cannot be written in terms of
the previous known currents obtained so far.
The spin-$2$ current $K^a$ obtained in the bosonic coset model
is not an element of any ${\cal N}=2$ multiplet.
When we compute the 
OPEs between the supersymmetry generators
and the fourth or fifth terms of (\ref{5HALF})
in the computation of (\ref{opegvhat}),
we should use some property appearing in Appendix $D$ or section $6$
together with the footnote \ref{someopeexpression}.
Note that this (\ref{Spin2non}) is
a new current in the sense that
although
it contains both the known current $ W^{+(2),a}$ (\ref{nonspin2})
and known
operator  $i \, f^{a b c} \, J^b \, J_f^c$ which is primary,
it also contains
the spin-$2$ current $K^a$ (\ref{spin2Ka}) \footnote{
That is 
$K^a$ is one of the currents in the
bosonic coset model but this is not an element of
any ${\cal N}=2$ multiplet in the
${\cal N}=2$
supersymmetric coset model.}.
As soon as the spin-$2$ current $K^a$ appears in any OPEs
we are considering, then we should replace it with
other two currents $W^{\pm(2),a}$ and other terms
by using the relation (\ref{Spin2non}).

Therefore, we have obtained the nonsinglet spin-$2$ current
in (\ref{Spin2non}) by acting the supersymmetry generators
of the ${\cal N}=2$ superconformal algebra
on the spin-$3$ current successively. In next subsections,
we would like to construct its superpartners explicitly.

\subsection{Construction of second and third components}

Because we have found the lowest component of the ${\cal N}=2$
multiplet we consider in this section, it is straightforward to
determine the other three components.
For given the lowest component in (\ref{Spin2non}), we calculate the
following OPEs
\bea
G^{\pm}(z) \, W^{-(2),a }(w) & = &
\frac{1}{(z-w)}\, G^{\pm (\frac{5}{2}),a}(w) +\cdots.
\label{opegw}
\eea
It is a good sign for the existence of the pole one term
by realizing Appendix $B$.
When we compute the above
OPE for the first term of (\ref{Spin2non}), we can use
the previous relations in (\ref{gkplus}) and (\ref{gkminus}).
For the second term of (\ref{Spin2non}), we can use
the relations (\ref{class2}) and (\ref{class3}).
For the third term of (\ref{Spin2non}), we obtain 
the composite operators.

After simplifying further,
the first order pole provides the following primary
nonsinglet spin-$\frac{5}{2}$
currents
\bea
G^{\pm (\frac{5}{2}),a} & = &
\frac{  (k^2-4)(3 k+2M+2N)}{ k (k+M) (3 k+2M)} \, a_1 \Bigg[
  \frac{1}{4} \, (2k+M)(3k+M+3N) \, V^{\pm (\frac{5}{2}),a}
  \nonu \\
  &-&
   \frac{1}{2k}\, (k^2-1)(2k+M+N) \, i\, f^{a b c } \, G^{\pm,b}\, J_f^c
+  \frac{1}{2k}\, (k^2-1)(2k+M+N) \, i\, f^{a b c } \, J^{b}\, G^{\pm,c}
\nonu \\
&-& \frac{1}{6 k} \, (k^2-1) \, M \, (2k+M+N) \, \pa G^{\pm,a} \Bigg].
\label{real5half}
\eea
Due to the presence of the third term of (\ref{Spin2non}),
we have the second term of (\ref{real5half}) which does not appear
in (\ref{5HALF}). We can check that the three terms of (\ref{real5half})
except the first one are given by the known operators and are
primary under the stress energy tensor.
Because the first term in (\ref{real5half}) cannot be written in terms of
the known currents, we have the new primary spin-$\frac{5}{2}$ currents. 
Note that
in the OPE of (\ref{opegw}), there are no higher order pole terms
which can be checked for fixed $N$ and $M$ also.

\subsection{Construction of last component}

As done in previous section, we construct the last component
of this multiplet.
We can compute the following OPE
\bea
G^{-}(z) \,
G^{+(\frac{5}{2}),a}(w)\Bigg|_{\frac{1}{(z-w)^2}} =  (2k+M+2N)\,
W^{-(2),a}(w).
\label{ggope}
\eea
The structure constant of (\ref{ggope})
can be determined by keeping track of the particular term in
the nonsinglet spin-$2$ current $W^{-(2),a}(w)$.

Let us introduce the final last component
which is given by
\bea
W^{+(3),a} & \equiv &
G^{-}(z) \, G^{+(\frac{5}{2}),a}(w)\Bigg|_{\frac{1}{(z-w)}}-
\frac{1}{4}\, \pa \, \Bigg(G^{-}(z) \,
G^{+(\frac{5}{2}),a}(w)\Bigg|_{\frac{1}{(z-w)^2}}\Bigg)
\nonu \\
&= &
\frac{  (k^2-4)(3 k+2M+2N)}{ k (k+M) (3 k+2M)} \, a_1 \Bigg\{
  \frac{1}{4} \, (2k+M)(3k+M+3N) \, \Bigg(
  \nonu \\
  & - &
  2(k+N) \Bigg[ \de_{\rho \si} \, t^a_{j \bar{i}} \, J^{(\bar{\si} j)}\,
  \pa \, J^{(\rho \bar{i})}- k \,
  \de_{\rho \bar{\si}} \, t^a_{j \bar{i}} \, \psi^{(\bar{\si} j)}\, \pa^2 \psi^{(\rho
    \bar{i})} \nonu \\
  & + & \sqrt{\frac{M+N}{M N}}\,
  \de_{\rho \si} \, t^a_{j \bar{i}} \,  \psi^{(\bar{\si} j)}\,\pa \,
  (\psi^{(\rho \bar{i})}\, J^{u(1)})+
  \de_{\rho \bar{\si}} \, t^a_{j \bar{i}} \,
  t^{\alpha}_{\si \bar{\rho}} \,
  \de^{\rho \bar{\rho}}\,
  \psi^{(\bar{\si} j)} \, \pa \, (\psi^{(\si \bar{i})} \, J^{\alpha})\nonu \\
  &-&  \de_{\rho \bar{\si}} \, (t^a \, t^b)_{j \bar{k}} \, \psi^{(\bar{\si} j)}
  \, \pa \, (\psi^{(\rho \bar{k})}\, J^b) 
  \Bigg]\nonu \\
&-& \frac{2}{k}\, (k+N)\,  \sqrt{\frac{M+N}{M N}}\,
\Bigg[ \sqrt{\frac{M+N}{M N}}\, G^- \, G^{+,a}-\frac{N}{2}\,
  J^{u(1)}\, \pa \, J^a +\frac{k}{2} \, J^{u(1)}\, \pa \, J_f^a\nonu \\
  &+& J^{u(1)}\, W^{+(2),a} + i \, f^{a b c}\, J^{u(1)}\, J^b \, J_f^c
  \Bigg]
\nonu \\
&-& \frac{2}{k M}\, (k+M+N) \, \Bigg[ G^{-,a} \, G^+ -\frac{1}{2}
  \, (k+M+N) \, J^a \, \pa \, K + (k+M+N)\, J^a \, T \nonu \\
  &-& \frac{1}{2}\, J^a \, (J^b+J_f^b)^2 \Bigg]
\nonu \\
&-& \Bigg[ i \, f^{a b c} \, G^{-,b}\, G^{+,c} +
  \frac{1}{2} (-N \, J^b \, \pa \, J^c + k \, J^b \, \pa \,
  J_f^c) \, i\, f^{a b c} + i \, f^{a b c}\, J^b \, W^{+(2),c}
  \nonu \\
  &-& f^{a b c}\, f^{c d e} \, J^b \, J^d \,J^e_f\Bigg]
\nonu \\
&+& \frac{(2k+M+2N)}{(2k+M)} \Bigg[
  d^{a b c}\, G^{-,b} \, G^{+,c} -\frac{N}{2} \,
  d^{a b c} \, J^b \, \pa \, J^c + \frac{k}{2}\,
  d^{a b c} \, J^b \, \pa \, J_f^c + d^{a b c}\, J^b\,
  W^{+(2),c} \nonu \\
  &+& d^{a b c} \, i\, f^{c d e} \, J^b \, J^d \, J^e_f
  \Bigg] \nonu \\
&+& 2 \Bigg[ t^{\alpha}_{\rho \bar{\si}} \, t^a_{j \bar{i}}\,
  J^{(\bar{\si} j)} \, J^{\alpha}\, J^{(\rho \bar{i})}
  - t^{\alpha}_{\rho \bar{\si}} \, t^a_{j\bar{i}}\,
  t^{\alpha}_{\rho_1 \bar{\nu}} \, \de_{m \bar{m}} \,
  \psi^{(\bar{\si} j)} ((\psi^{(\rho_1 \bar{m})} \, J^{(\bar{\nu} m)})
  J^{(\rho \bar{i})})\nonu \\
  & - & k \, t^{\alpha}_{\rho \bar{\si}} \,t^a_{j \bar{i}}\,
  \psi^{(\bar{\si} j)} \, J^{\alpha}\, \pa \, \psi^{(\rho \bar{i})}
  + \sqrt{\frac{M+N}{M N}}\,  t^{\alpha}_{\rho \bar{\si}} \,t^a_{j \bar{i}}\,
  \psi^{(\bar{\si} j)} \, J^{\alpha}\,\psi^{(\rho \bar{i})}\, J^{u(1)}\
  \nonu \\
  & + & (t^{\beta} \, t^{\alpha})_{\si \bar{\si}}\, t^a_{j \bar{i}}\,
  \psi^{(\bar{\si} j)} \, J^{\alpha}\,\psi^{(\si \bar{i})}\, J^{\beta}
  - t^{\alpha}_{\rho \bar{\si}}\, (t^a \, t^b)_{j \bar{k}}\,
  \psi^{(\bar{\si} j)} \, J^{\alpha}\, \psi^{(\rho \bar{k})}\, J^b\Bigg]
\nonu \\
&-& \frac{2}{3}\, \frac{2(k^2-1)(2k+M+N)}{k(2k+M)} \, \Bigg[
  -\frac{N}{2}\, \pa^2 \, J^a+\frac{k}{2}\, \pa^2 \, J_f^a +
  \pa \, W^{+(2),a} \nonu \\
  & + & i \, f^{a b c} \, \pa \, (J^b \, J_f^c)
  \Bigg]\Bigg) \nonu \\
&-&
\frac{1}{2k}\, (k^2-1)(2k+M+N) \, i \, f^{a b c}\,
\Bigg( -\frac{N}{2} \, \pa \, J^b \, J_f^c +\frac{k}{2}\,
  \pa \, J^b_f \, J_f^c + W^{+(2),b}\, J_f^c \nonu \\
  &+& i \, f^{b c d} (( J^c \, J_f^d) \, J_f^c)+
  G^{+,b} \, G^{-,c}\Bigg)
\nonu \\
& + &  \frac{1}{2k}\, (k^2-1)(2k+M+N) \, \Bigg(
 i \, f^{a b c} \, G^{-,b}\, G^{+,c} +
  \frac{1}{2} (-N \, J^b \, \pa \, J^c + k \, J^b \, \pa \,
  J_f^c) \, i\, f^{a b c} \nonu \\
  & + & i \, f^{a b c}\, J^b \, W^{+(2),c}
  - f^{a b c}\, f^{c d e} \, J^b \, J^d \,J^e_f
\Bigg)
\nonu \\
&-& \frac{1}{6 k} \, (k^2-1) \, M \, (2k+M+N) \,
\Bigg(
 -\frac{N}{2}\, \pa^2 \, J^a+\frac{k}{2}\, \pa^2 \, J_f^a +
 \pa \, W^{+(2),a} \nonu \\
 & + &  i \, f^{a b c} \, \pa \, (J^b \, J_f^c)
\Bigg) 
\Bigg\} \nonu \\
& - & \frac{1}{4}\,(2k+M+2N)\, \pa \,
W^{-(2),a}.
\label{spinthreenon}
\eea
The first seventeen lines of (\ref{spinthreenon})
come from the spin-$\frac{5}{2}$
current $V^{+(\frac{5}{2}),a}$ consisting of seven independent terms
in (\ref{5halfone}) and the remaining six lines
come from the remaining three terms of (\ref{real5half}).
Then we can read off the corresponding OPEs between
the spin-$\frac{3}{2}$ current and the spin-$\frac{5}{2}$
current by looking at the seven pairs of brackets inside the
curly bracket.
We observe that the
$ t^{\alpha}_{\rho \bar{\si}} \, t^a_{j \bar{i}}\,
J^{(\bar{\si} j)} \, J^{\alpha}\, J^{(\rho \bar{i})}$
term appearing inside of the sixth pair of bracket
originates from
the $a_1$ term of (\ref{spin3exp}) and therefore
the above nonsinglet
spin-$3$ current is a generalization of (\ref{spin3exp})
in the ${\cal N}=2$
supersymmetric coset model we are describing in this paper.
As before,
in the OPE between
$G^{-}(z)$ and the sixth term of $V^{+(\frac{5}{2}),a}(w)$,
the OPE between $G^-$ and the spin-$\frac{1}{2}$
operator leads to the spin-$1$ current transforming as
$(\overline{\bf N},{\bf M})$.
By combining this with the two other spin-$2$ currents, 
we obtain the first term inside the sixth pair of brackets.

Then
we
are left with the following currents 
\bea
(W^{-(2),a}, G^{+(\frac{5}{2}),a}, G^{-(\frac{5}{2}),a}, W^{+(3),a}).
\label{fourcurrents}
\eea
As described in the beginning of this section,
it is an open problem to consider the nonsinglet spin-$2$ current
$K^a$ first and obtain the nonsinglet spin-$3$ current
after acting the spin-$\frac{3}{2}$ currents
of the ${\cal N}=2$ superconformal algebra  on $K^a$ successively. 
Once we have obtained  the nonsinglet spin-$3$ current,
then it is straightforward to determine its superpartners step by
step. We can check whether this procedure reproduces the above
${\cal N}=2$ multiplet in (\ref{fourcurrents}) or not.

In next subsections, further checks on the currents under the
action of the  currents of ${\cal N}=2$ superconformal
algebra are given.

\subsection{The OPEs with the currents of ${\cal N}=2$ superconformal
  algebra}

Let us present the OPEs with the spin-$1$ current of the
${\cal N}=2$ superconformal algebra
\bea
&& K(z) \, W^{-(2),a}(w)  =
\nonu \\
&& \frac{1}{(z-w)^2}
\,
\frac{ (k^2-1)  (k^2-4) (3 k+2 M+2 N) (2 k+M+N)}{
  k^2 (k+M) (3 k+2 M)}\, a_1 \, \Bigg[
N \, J^a -k \, J_f^a
  \Bigg](w) + \cdots,
\nonu \\
&& K(z) \, G^{\pm (\frac{5}{2}),a}(w)  =  \nonu \\
&& \mp \frac{1}{(z-w)^2}\,
\frac{  (k^2-1) (k^2-4) (k+N) (2 k+M+N) (3 k+2 M+2 N)}{
  k^2 (k+M) (3 k+2 M)} \, a_1 \, G^{\pm,a}(w)\nonu \\
&& \pm  \frac{1}{(z-w)}\,
 G^{\pm (\frac{5}{2}),a}(w) +\cdots,
\nonu \\
&& K(z) \, W^{+(3),a}(w)  =  \frac{1}{(z-w)^3} \Bigg[
  \nonu \\
  && \frac{ (k^2-1) (k^2-4) M  (2 k+M+N) (3 k+2 M+2 N)}{
    2 k^2 (k+M) (3 k+2 M)} \, a_1\, ( N \, J^a - k \, J_f^a )
  \Bigg](w)\nonu \\
&& +
\frac{1}{(z-w)^2} \Bigg[
  -\frac{1}{2}\,
\pa \, (\mbox{pole-3})-(2k+M+2N)
\,  W^{-(2),a}
\nonu \\
&&
-\frac{  (k^2-1) (k^2-4) (k+N) (2 k+M+N) (3 k+2 M+2 N)}{
  k^2 (k+M) (3 k+2 M)} \, a_1
  \,  i \, f^{a b c } \, J^b \, J_f^c 
\nonu \\
&&
-\frac{ (k^2-1) (k^2-4) (k+N) (2 k+M+N) (3 k+2 M+2 N)}{
  k^2 (k+M) (3 k+2 M)}\, a_1\,
W^{+2(a)}\Bigg](w)
+ \cdots.
\label{kwithnon}
\eea
As before, the easy way to determine the various structure
constants appearing in the right hand sides of the above OPEs
is to figure out how the various composite operators arise from which
parts of each current. For the last OPE of (\ref{kwithnon}),
it is rather nontrivial to
keep track of the contributions from
$ t^{\alpha}_{\rho \bar{\si}} \, t^a_{j\bar{i}}\,
  t^{\alpha}_{\rho_1 \bar{\nu}} \, \de_{m \bar{m}} \,
  \psi^{(\bar{\si} j)} ((\psi^{(\rho_1 \bar{m})} \, J^{(\bar{\nu} m)})
  J^{(\rho \bar{i})})$ appearing in the sixth pair of brackets of the
  nonsinglet spin-$3$ current in (\ref{spinthreenon})
  because of nontrivial normal ordering of the currents.
  Moreover, the presence of $W^{+2(a)}$ (which belongs to other
  ${\cal N}=2$ multiplet) in the second order pole
  can be understood from
  the fact that the above spin-$3$ current contains
  this current in various places of (\ref{spinthreenon}).
  In doing this, we also need the OPEs between $J_f^a(z)$ and
  $G^{\pm,b}(w)$ which will be described in detail later.

Let us describe other OPEs as follows:  
\bea
&& G^{+}(z) \, W^{-(2),a}(w)  =  \frac{1}{(z-w)}\, G^{+(\frac{5}{2}),a}(w)
+ \cdots,
\nonu \\
&& G^{+}(z) \, G^{+(\frac{5}{2}),a}(w)  =  0 + \cdots,
\nonu \\
&& G^{+}(z) \, G^{-(\frac{5}{2}),a}(w)  =
\nonu \\
&&
\frac{1}{(z-w)^3} \,
\frac{ (k^2-1) (k^2-4)  (k+M+N) (2 k+M+N) (3 k+2 M+2 N)}{
  k^2 (k+M) (3 k+2 M)}\,
\Bigg[ N \, J^a - k \, J_f^a \Bigg](w) \nonu \\
&& + 
\frac{1}{(z-w)^2} \, (2k+M+2N)\,
W^{-(2),a}(w) +\frac{1}{(z-w)} \, \Bigg[\frac{1}{4} \,
\pa \, \mbox{(pole-2)}
  \nonu \\
  && +  i\, f^{a b c} \, J^b \, W^{-(2),c}  + i \,
  f^{a b c} \, J_f^b \, W^{-(2),c}  +
  \frac{1}{4}  \, 
  (2 k+3 M+2 N)\,
  \pa \, W^{-(2),a}
  -  W^{+(3),a}
\Bigg](w) \nonu \\
&& + \cdots,
\nonu \\
&& G^{+}(z) \, W^{+(3),a}(w)  = \nonu \\
&& -\frac{1}{(z-w)^3}\,
\frac{(k^2-1) (k^2-4)  (k+N) (k+M+N) (2 k+M+N)
   (3 k+2 M+2 N)}{k^2 (k+M) (3 k+2 M)}
 \, a_1\,
 \nonu \\
 && \times \, G^{+,a}(w) \nonu \\
 && + \frac{1}{(z-w)^2} \,\frac{1}{4} (10 k+7 M+10 N)\, G^{+(\frac{5}{2}),a}
\nonu \\
&& +
\frac{1}{(z-w)}\, \Bigg[
  \frac{1}{5}\, \pa \, \mbox{(pole-2)}
  +
  i \, f^{a b c}\, J^b \, G^{+(\frac{5}{2}),c} +
  i \, f^{a b c}\, J_f^b \, G^{+(\frac{5}{2}),c}
   +  \frac{2 M}{5}\, \pa \,
  G^{+(\frac{5}{2}),a} 
  \Bigg](w)   \nonu \\
&& +  \cdots. 
\label{GWs}
\eea
In the third OPE of (\ref{GWs}),
it is rather nontrivial to observe the first order pole
explicitly. We should write down all the possible terms
of spin-$3$ with free index $a$. It turns out that
the  free index appears in the $f$ symbols and the currents
themselves.
This first order pole is another way to
determine the nonsinglet spin-$3$ current.
For the last OPE of (\ref{GWs}), we can fix the coefficient of the
last term appearing in the first order pole by using the property
of quasi primary condition. That is, after calculating the third
order pole in the OPEs between the stress energy tensor and
the second and third terms of the first order pole and obtaining
$-2M \,  G^{+(\frac{5}{2}),a}(w)$, we can fix
the structure constant of $\pa \,  G^{+(\frac{5}{2}),a}(w)$
as $\frac{2M}{5}$ because we can check that the third order pole
in the OPE  between the stress energy tensor and this derivative term
leads to $5 \,  G^{+(\frac{5}{2}),a}(w)$. Then the last three terms
appearing in the first order pole of the last OPE
satisfy the quasi primary condition
as we expected \footnote{In doing this,
  first we check the field contents for fixed $N$ and $M$ and then
  we need to figure out which parts of the currents of
  ${\cal N}=2$ multiplet contribute to the right hand
  sides of the OPEs.
Next we calculate those contributions manually for generic $N$ and $M$.}.

We continue to describe the next OPEs 
\bea
&& G^{-}(z) \, W^{-(2),a}(w)  = 
\frac{1}{(z-w)}\, G^{-(\frac{5}{2}),a}(w)
 + \cdots,
\nonu \\
&& G^{-}(z) \,  G^{+(\frac{5}{2}),a}(w)  = \nonu \\
&&
-\frac{1}{(z-w)^3} \,
\frac{ (k^2-1) (k^2-4)  (k+M+N) (2 k+M+N) (3 k+2 M+2 N)}{
  k^2 (k+M) (3 k+2 M)}\,
\nonu \\
&& \times \, \Bigg[ N \, J^a - k \, J_f^a \Bigg](w) \nonu \\
&& + 
\frac{1}{(z-w)^2} \, (2k+M+2N)\,
W^{-(2),a}(w) +\frac{1}{(z-w)} \, \Bigg[\frac{1}{4} \,
\pa \, \mbox{(pole-2)}  +  W^{+(3),a}
\Bigg](w) \nonu \\
&& + \cdots,
 \nonu \\
&& G^{-}(z) \, G^{-(\frac{5}{2}),a}(w)  =  0 + \cdots,
 \nonu \\
 && G^{-}(z) \, W^{+(3),a}(w)  = \nonu \\
 && \frac{1}{(z-w)^3} \,
 \frac{(k^2-1) (k^2-4)  (k+N) (k+M+N) (2 k+M+N)
   (3 k+2 M+2 N)}{k^2 (k+M) (3 k+2 M)}
 \, a_1\, 
 \nonu \\
 && \times \, G^{-,a}(w) \nonu \\
 && - \frac{1}{(z-w)^2} \,  \frac{5}{4} (2k+M+2N) \,
 G^{-(\frac{5}{2}),a}(w)
  -\frac{1}{(z-w)} \,\frac{1}{5} \, \pa \, \mbox{(pole-2)}
 +\cdots. 
\label{Gopeother}
 \eea
 The second OPE of (\ref{Gopeother}) provides the way we obtain the
 nonsinglet spin-$3$ current described before.
 In the last OPE of (\ref{Gopeother})
 which will be rather complicated compared to other OPEs above,
 we should also calculate
 all the contributions from the nonsinglet spin-$3$ current.
 Note that
 when we calculate the OPE between $G^-(z)$ and $i \, f^{a b c}\,
 G^{+,b} \, G^{-,c}(w)$ which is one of the terms of $W^{+(3),a}(w)$
 and appears in the sixth line from the below,
 we should obtain the first order pole of the OPE between
 $W^{+(2),b}(z)$ and $G^{-,c}(w)$.
 See also the second OPE of (\ref{class3}).
 Then for example, the contribution of
 the fifth nonderivative term in $V^{-(\frac{5}{2}),a}$
 (\ref{5halftwo}) with
 (\ref{real5half}) arises in this particular
 pole. The point here is, in general,
  the first order pole of the OPE between
  $W^{+(2),b}(z)$ and $G^{-,c}(w)$ produces a new quasi primary operator
  but with the contraction of indices $b$ and $c$
  appearing in the $f$ symbol
  we can write down this first order pole in terms of the known
  currents.
  Therefore, there will be no new (quasi)primary field
  in the second order pole of the last OPE in (\ref{Gopeother})
  \footnote{This feature also appears in the previous last OPE
  of (\ref{GWs}).}.
  We will return to this issue around
(\ref{gwOPE1})
  in next section.

  After collecting  all the contributions on
  the fifth nontrivial term in $V^{-(\frac{5}{2}),a}$
  (\ref{5halftwo}) in this
  calculation of the OPE between $G^-(z)$ and $W^{+(3),a}(w)$,
  we are left with the structure constant in the second order pole
  as above. Moreover, the first order pole has the descendant terms
  only which can be checked by fixed $N$ and $M$.

  Finally, the OPEs with the stress energy tensor
  are described by
\bea
&& T(z) \, W^{-(2),a}(w) = 
\frac{1}{(z-w)^2}\,  2\, W^{-(2),a}(w) + \frac{1}{(z-w)}\,  \pa \,
W^{-(2),a}(w) + \cdots, 
\nonu \\
&& T(z) \, G^{\pm (\frac{5}{2}),a}(w) = 
\frac{1}{(z-w)^2}\,  \frac{5}{2} \,
G^{\pm (\frac{5}{2}),a}(w) + \frac{1}{(z-w)}\,  \pa \,
G^{\pm (\frac{5}{2}),a}(w) + \cdots,
\nonu \\
&& T(z) \, W^{+(3),a}(w) = \nonu \\
&&
\frac{1}{(z-w)^4}\, \Bigg[
  \nonu \\
  &&
  -\frac{3(k^2-1)(k^2-4)(k+M+N)(2k+M+N)(3k+2M+2N)}{2k^2(k+M)(3k+2M)}\,
a_1\,  ( N \, J^a - k \, J_f^a)
  \Bigg](w)
\nonu \\
&& +\frac{1}{(z-w)^2}\,  3 \,
W^{+(3),a}(w) + \frac{1}{(z-w)}\,  \pa \,
W^{+(3),a}(w)
+ \cdots.
\label{Twithother}
\eea
For the nonsinglet spin-$3$ current, there is a fourth order pole in
(\ref{Twithother}). This implies that
 the nonsinglet spin-$3$ current is quasi primary field.
By introducing the composite operator
$ K \, W^{-(2),a}$, we can make the following
nonsinglet spin-$3$ current
\bea
W^{+(3),a} + \frac{3}{2}\, (k+M+N) \, K \, W^{-(2),a}
\label{PRI}
\eea
be primary.
In doing this, the first OPE of (\ref{kwithnon}) is used in (\ref{PRI})
\footnote{In the OPE between $K(z)$ and
  $ K\, W^{-(2),a}(w)$, the second order pole contains
  $W^{-(2),a}$ and $K \, (N\, J^a-k \, J_f^a)$ terms. The
  OPE between
  $G^{+}(z)$ and  $ K\, W^{-(2),a}(w)$ leads to
  the second order pole with $G^{+(\frac{5}{2}),a}$
  and the first order pole with
  $G^{+} \, W^{-(2),a}$ and $K \, G^{+(\frac{5}{2}),a}$.
  Furthermore,
 the
  OPE between
  $G^{-}(z)$ and  $ K\, W^{-(2),a}(w)$ becomes
  the second order pole with $G^{-(\frac{5}{2}),a}$
  and the first order pole together with
  $G^{-} \, W^{-(2),a}$ and $K \, G^{-(\frac{5}{2}),a}$.
There exist several nonlinear terms.}.

Therefore, the nonsinglet currents are given by (\ref{Spin2non}),
(\ref{real5half}) and (\ref{spinthreenon})
in the coset realization and they satisfy
(\ref{kwithnon}), (\ref{GWs}), (\ref{Gopeother}) and (\ref{Twithother})
under the action of the currents of
${\cal N}=2$ superconformal algebra.
Also note that the right hand side of these OPEs contain
the currents of ${\cal N}=2$ multiplet in (\ref{simplest})
as well as the currents of  ${\cal N}=2$ multiplet in
(\ref{fourcurrents}).

\section{The OPEs between the nonsinglet
   multiplet of spins $(1,\frac{3}{2},\frac{3}{2},2)$
and itself}

The simplest nontrivial OPEs between the nonsinglet ${\cal N}=2$
multiplets
can be described in this section.

\subsection{ The OPEs with lowest component }

From the explicit coset realizations in (\ref{threeJf}), (\ref{gpma}) and
(\ref{nonspin2}), we determine the following OPEs
(the OPEs between the lowest component and the four components in
(\ref{simplest}))
\bea
J_f^a(z) \, J_f^b(w) & = & \frac{1}{(z-w)^2}\, N \, \de^{a b} +
\frac{1}{(z-w)} \, i \, f^{a b c} \, J_f^c(w) + \cdots,
\nonu \\
J_f^a(z) \, G^{+,b}(w) & = & 
\frac{1}{(z-w)} \, \Bigg[
  -\frac{1}{M} \, \de^{a b} \, G^{+}+ \frac{1}{2} \, ( i \,
  f + d)^{a b c} \, G^{+,c} \Bigg](w) + \cdots,
\nonu \\
J_f^a(z) \, G^{-,b}(w) & = & 
\frac{1}{(z-w)} \,  \Bigg[
  \frac{1}{M} \, \de^{a b} \, G^{-}+ \frac{1}{2} \, ( i \,
  f - d)^{a b c} \, G^{-,c} \Bigg](w) + \cdots,
\nonu \\
J_f^a(z) \, W^{+(2),b}(w) & = & \frac{1}{(z-w)^2}\,
\Bigg[ -\frac{1}{M}\,(k+M+N)\,
  \de^{a b}\, K -\frac{N}{2}
  \, i\, f^{a b c}\, J^c \nonu \\
  & + & \frac{1}{2} \,
  d^{a b c} (
  N\, J^c- k \, J_f^c )\Bigg](w) \nonu \\
& + &
\frac{1}{(z-w)}
\, \Bigg[ \frac{1}{2}\, (i \, f +d)^{c b d}\, i \, f^{a d e} \,
  J^c \, J_f^e \nonu \\
  & + & i \, f^{a b c} \Bigg(
  W^{+(2),c} -\frac{1}{k M} \, (k+M+N)\, J^c K -
  \frac{1}{2} \, (i \, f+ d)^{d c e} \, J^d \, J_f^e +
  \frac{N}{2} \, \pa \, J^c 
   \nonu \\
  &+&
\frac{1}{2}\,
  (K^c +\frac{N}{(2k+M)} \, d^{c d e} \, J^d \, J^e - N \,
  \pa \, J^c)
  \Bigg) \Bigg](w) + \cdots.
\label{ope1}
\eea
In the last OPE of (\ref{ope1}),
we should replace the nonsinglet spin-$2$ current $K^c$ with
the other spin-$2$ currents by using the nonsinglet spin-$2$
current (\ref{Spin2non})
because the $K^c$ is not an element of ${\cal N}=2$ multiplet
we are considering in the ${\cal N}=2$ supersymmetric coset model.
It is obvious to see 
$
J_f^a(z) \, K^{b}(w)  =  0 +\cdots $ from Appendix $A$.
The presence of the spin-$2$ current $K^c$
can be understood from the fact that in doing this, the unwanted terms,
consisting of the second, third, fourth, and sixth terms
of $W^{+(2),c}$ (\ref{nonspin2}) with $f$ symbol after simplifying,
are replaced by the nonsinglet
spin-$2$ current $W^{+(2),c}$ and other terms
which contain the nonsinglet spin-$2$ current $K^c$.
In the last line of (\ref{ope1}), we replace
the first term of $K^c$ (\ref{spin2Ka}) with other terms
where $J^c \, J^{u(1)}$ term is combined with $J^c \, K$
term and to appear in the second line from the below.

From the last OPE of (\ref{ope1}),
we observe that the right hand side of the OPE
contains also
the lowest component of the ${\cal N}=2$
multiplet in (\ref{fourcurrents}).

\subsection{The OPEs with the second component }

Now let us consider the following OPEs (the OPEs between the second component
and the three components in (\ref{simplest}))
\bea
&& G^{+,a}(z) \, G^{+,b}(w) = 0 + \cdots,
\nonu \\
&& G^{+,a}(z) \, G^{-,b}(w) = \frac{1}{(z-w)^3} \, k \, N \, \de^{a b}
\nonu \\
&& + 
\frac{1}{(z-w)^2} \, \Bigg[ \frac{(k+M+N)}{M} \,
  \de^{a b} \, K +
  \frac{N}{2} \, (i f - d)^{a b c} \, J^c +\frac{k}{2} \,
  ( i f + d)^{a b c} \, J_f^c \Bigg](w) \nonu \\
&& +  \frac{1}{(z-w)} \, \Bigg[ \de^{a b} \Bigg( \frac{1}{M} \,
  (k+M+N) \, T -\frac{1}{2M} \, J^a \, J^a -\frac{1}{2M}\, J_f^a\,
  J_f^a \nonu \\
  & & +  \frac{1}{2M} \, (k+M+N) \, \pa \, K \Bigg)
  + i \, f^{a b c} \Bigg( -\frac{1}{2} \, W^{+(2),c} -
  \frac{N}{4} \, \pa \, J^c \nonu \\
  && + \frac{k}{4} \, \pa \, J_f^c -\frac{1}{2}\,
  (K^c +\frac{N}{(2k+M)} \, d^{c d e} \, J^d \, J^e - N \,
  \pa \, J^c)\nonu \\
  && + \frac{1}{4} \, d^{c d e} \, J^d \, J_f^e -
  \frac{1}{4} \, i \, f^{c d e} \, J^d \, J_f^e  + \frac{(k+M+N)}{k M}
  \, J^c \, K \Bigg) \nonu \\
  && +  d^{a b c} \Bigg(-\frac{1}{2} \, W^{+(2),c} -
  \frac{N}{4} \, \pa \, J^c + \frac{k}{4} \, \pa \, J_f^c
  + \frac{1}{4} \, d^{c d e} \, J^d \, J_f^e -
  \frac{1}{4} \, i \, f^{c d e} \, J^d \, J_f^e \Bigg)
  \nonu \\
  && - \frac{1}{M}\, J^b \, J_f^a -\frac{1}{4} \, (i f +d)^{e b c}
  \, (i f +d)^{a c d} \, J^e\, J_f^d
  \Bigg](w) + \cdots,
\nonu \\
&& G^{+,a}(z) \, W^{+(2),b}(w) = \frac{1}{(z-w)^2}\,
\Bigg[ \frac{1}{2M}\, (3k+2M+3N)\, \de^{a b}\, G^+ +
  \frac{1}{4} \, (N-k)\, i \, f^{a b c}\, G^{+,c}
  \nonu \\
  && - \frac{3}{4}\, (k+N)\, d^{a b c}\, G^{+,c}
  \Bigg](w)
+\frac{1}{(z-w)}\, \Bigg[ \frac{1}{3}\, \pa \, \mbox{(pole-2)} +
  R^{+(\frac{5}{2}),a b} \Bigg](w) + \cdots.
\label{ope2}
\eea
In Appendix $D$, we present the coset realization for the 
second OPE of (\ref{ope2}).

The question is how we obtain the above result from Appendix $D$?
For the second order pole, we can check them without any difficulty.
For the first order pole, we can compute the
$\de^{a b}$, $i \, f^{a b c}$ and $d^{a b c}$ with the first order pole
appearing in Appendix $D$. In other words, we have a free index
$c$ for the last two cases while we do not have any free index
for the first case.
From the expression of 
$\de^{a b} \,  G^{+,a}(z) \, G^{-,b}(w)\Big|_{\frac{1}{(z-w)}}$,
we can use the expression of stress energy tensor $T$ (\ref{T}). Then
it turns out that we have four independent terms appearing in
the first two lines of the first order pole having
$\de^{a b}$ in (\ref{ope2}).

From the expression of 
$i \, f^{a b c } \,  G^{+,a}(z) \, G^{-,b}(w)\Big|_{\frac{1}{(z-w)}}$,
we can use the expressions $W^{+(2),c}$ (\ref{nonspin2}) and
$K^c$ (\ref{spin2Ka}) in order to remove
the unwanted terms. Then we have nine terms appearing in the second, third
and fourth lines of the first order pole having $i \, f^{a b c}$
as well as $J^c \, J_f^{u(1)}$ term.
From the expression of 
$ d^{a b c } \,  G^{+,a}(z) \, G^{-,b}(w)\Big|_{\frac{1}{(z-w)}}$,
the previous relation (\ref{nonspin2}) can be used. Then
we have the five terms appearing in the fifth line of
the first order pole having $ d^{a b c}$
as well as $J^c \, J_f^{u(1)}$ term in (\ref{ope2}).

Furthermore, the last term in Appendix $D$ having
the factor $t^a \, t^c \, t^b$ can be simplified
and will appear in the last line of the first order pole as well as
$J^c \, J_f^{u(1)}$ term. However, this  $J^c \, J_f^{u(1)}$-term
can be cancelled by the above two contributions. Therefore,
we are left with the above final result
where there is no $J^c \, J_f^{u(1)}$ term
\footnote{For convenience,
  we present the OPEs $J^a(z) \, G^{\pm,b}(w) =\frac{1}{(z-w)}\,
  \Bigg[ \pm \frac{1}{M}\, \de^{a b}\, G^{\pm} \mp \frac{1}{2}
    ( \mp i \, f +d)^{a b c}\, G^{\pm,c}\Bigg](w) + \cdots$.}.

In the last OPE of (\ref{ope2}), there exist the following
terms, after subtracting the descendant terms,
which is a primary 
\bea
R^{+(\frac{5}{2}),a b} & = &
\frac{(3 k+N)}{2 M}\, \de^{a b}\, \pa \, G^+
+ \frac{1}{4} (- k-N) \, i \, f^{a b c}\, \pa \, G^{+,c} +
\frac{1}{4} (-3 k-N)\,  d^{a b c}\, \pa \, G^{+,c}
\nonu \\
&- & \sqrt{\frac{M+N}{M N}}\,  i \, f^{a b c}\,J^{u(1)}\, G^{+,c}
-\frac{(k M N+M+2 N)}{M N}\, \de_{\rho \bar{\si}}\, \de^{a b}
\, \de_{k \bar{i}}\, \psi^{(\bar{\si} k)}\, \pa \,
J^{(\rho \bar{i})}
\nonu \\
&- & \frac{(k N+1)}{2 N}\, i \, f^{a b c}
\de_{\rho \bar{\si}}\, 
\, t^c_{k \bar{i}}\, \psi^{(\bar{\si} k)}\, \pa \,
J^{(\rho \bar{i})}
-\frac{(k M N+M+2 N)}{2 M N}\,
d^{a b c}
\de_{\rho \bar{\si}}\, 
\, t^c_{k \bar{i}}\, \psi^{(\bar{\si} k)}\, \pa \,
J^{(\rho \bar{i})}
\nonu \\
&+& i \, f^{a b c}\,  t^c_{k \bar{i}}\,
t^{\al}_{\rho \bar{\mu}}\, J^{(\rho \bar{i})}\,
\psi^{(\bar{\mu} k)}\, J^{\al}+
\frac{1}{M}\, J^a \, G^{+,b}-\frac{k}{M}\,
\de^{a b}\, \de_{\tau \bar{\mu}} \, \de_{j \bar{k}}\,
J^{(\tau \bar{k})}\, \pa \, \psi^{(\bar{\mu} j)}
\nonu \\
&+& \frac{k}{2}\,
i \, f^{a b c}\,
 \de_{\tau \bar{\mu}} \, t^c_{j \bar{k}}\,
 J^{(\tau \bar{k})}\, \pa \, \psi^{(\bar{\mu} j)}
 -\frac{k}{2} \, d^{a b c}\,
 \de_{\tau \bar{\mu}} \, t^c_{j \bar{k}}\,
 J^{(\tau \bar{k})}\, \pa \, \psi^{(\bar{\mu} j)}
 \nonu \\
 &- & \frac{M+N}{M N}\,
 t^a_{l \bar{k}}\, \de_{\tau \bar{\mu}} ((\psi^{(\bar{\mu} l)}\,
 J^{(\tau \bar{k})}) \, J_f^b)-
 t^{\al}_{\rho \bar{\si}}\, t^b_{j \bar{i}}\,
 t^a_{l \bar{k}}\, t^{\al}_{\mu \bar{\nu}}\,
 \psi^{(\bar{\si} j)} \, \psi^{(\rho \bar{i})}
 \, \psi^{(\bar{\nu} l)} \, J^{(\mu \bar{k})}
 \nonu \\
 &+& \frac{1}{M^2}\, \de^{a b}\, G^{+}\, J_f^{u(1)}
 -\frac{1}{2M}\, i \, f^{a b c}\, G^{+,c}\, J_f^{u(1)}
 -\frac{1}{2M}\,  d^{a b c}\, G^{+,c}\, J_f^{u(1)}
-\frac{1}{M}\, J^b \, G^{+,a}
\nonu \\
&-& \frac{1}{2M}\, (i \, f+ d)^{a b c}\, G^{+}\, J_f^c
+ \frac{1}{4}\, (i \, f+ d)^{a c e}(i \, f +d)^{c b d}\, G^{+,e}\, J_f^d
\nonu \\
&+& \frac{1}{2M} (i \, f +d )^{c b a}\, J^c \, G^+
-\frac{1}{4}\,  (i \, f+ d)^{d a e}(i \, f +d)^{c b d}\, J^c \, G^{+,e}
\nonu \\
&-& \frac{1}{2} \,  (i \, f+ d)^{a  c d} \, (t^d \, t^b)_{k \bar{i}}\,
\de_{\rho \bar{\si}}\, J^{(\rho \bar{i})}\, \psi^{(\bar{\si} k)} \, J^{c}
-\frac{1}{3}\, \pa \, \Bigg(G^{+,a}(z) \, W^{+(2),b}(w)
\Bigg|_{\frac{1}{(z-w)^2}}\Bigg).
\label{newRplus}
\eea
The question is
whether this can be written in terms of the known
currents or not.
Let us
look at $
t^{\al}_{\rho \bar{\si}}\, t^b_{j \bar{i}}\,
 t^a_{l \bar{k}}\, t^{\al}_{\mu \bar{\nu}}\,
 \psi^{(\bar{\si} j)} \, \psi^{(\rho \bar{i})}
 \, \psi^{(\bar{\nu} l)} \, J^{(\mu \bar{k})} $ term
 appearing in the sixth line of
 (\ref{newRplus}).
 Although the index $\al$ is summed but
 the product of the generators of $SU(M)$ has the
 free indices $a$ and $b$ with four different
 lower indices contracted with the coset fields.
 As far as I know, there is no identity in
 the product of the two generators of $SU(M)$
 with two different free indices
 \footnote{There is a relation
   $t^c_{j \bar{i}}\, t^{d}_{l\bar{m}}=
   \de^{cd}\, \de_{j \bar{m}} \, \de_{l \bar{i}}+
   \frac{M}{2}\, (i \, f+d)^{c d e}\, t^e_{j \bar{m}}\, \de_{l \bar{i}}-
   \frac{1}{M} \, \de^{c d}\, \de_{l \bar{m}}\, \de_{j \bar{i}}-
   \frac{1}{2} \, (i \, f+d)^{c d e}\, t^e_{l \bar{m}}\, \de_{j\bar{i}}-
   \frac{1}{2} \, (i \, f+d)^{c d e}\, t^e_{j \bar{i}}\, \de_{l \bar{m}}-
   \frac{M}{4} \, (i \, f +d)^{c a b} (i \, f+d)^{a d e}\,
   t^{b}_{j \bar{i}}\, t^e_{l \bar{m}}$. From this we obtain
   $i \, f^{a b c} \, t^b_{j \bar{i}}\, t^a_{l \bar{k}}=
   t^c_{j \bar{k}}\, \de_{l \bar{i}}-t^c_{l \bar{i}}\, \de_{j \bar{k}}$
   and $d^{a b c}\, t^b_{j \bar{i}}\, t^a_{l \bar{k}} =
   t^c_{j \bar{k}}\, \de_{l \bar{i}}-\frac{2}{M}\,
   t^c_{j \bar{i}}\, \de_{l \bar{k}}-\frac{2}{M}\,
   t^c_{l \bar{k}}\, \de_{j \bar{i}} + t^c_{l \bar{i}}\, \de_{j \bar{k}}$.
   In other words, when we contract with $f$ or $d$ symbols, then
   the product of two generators leads to several single generators
   with appropriate indices. We have seen these features in (\ref{GWs})
   and (\ref{Gopeother}) also.}.

 Therefore, we cannot express
 (\ref{newRplus}) in terms of the previous known currents
 obtained so far (although we have tried to rewrite it by using
 the various invariant tensors appearing in \cite{Ahn2011}).
 According to (\ref{Ws}),
 each  ${\cal N}=2$ current has a single $SU(M)$
 index. It would be interesting to study how we can obtain
 the superpartners of (\ref{newRplus}) explicitly
 if they exist.
 It is natural to consider (\ref{newRplus}) as the second component
 of the ${\cal N}=2$ multiplet and then
 the lowest, the third and last
 components are not known so far. 

\subsubsection{The OPE with spin-$2$ current}

Although the nonsinglet
spin-$2$ current $K^b$ in the bosonic coset model
does not belong to the component of the ${\cal N}=2$ multiplet,
it is very useful to calculate the OPE between
the nonsinglet spin-$\frac{3}{2}$ current
$G^{+,a}$ of (\ref{simplest})
and this spin-$2$ current. It turns out, from Appendix $D$ where
the coset field realizations are given, that
we have
\bea
&& G^{+,a}(z) \, K^{b}(w) = \frac{1}{(z-w)^2} \,
\frac{2(k^2-1)(2k+M+N)}{k(2k+M)} \,
\Bigg[
 \frac{1}{M} \, \de^{a b} \, G^{+}- \frac{1}{2} \, ( i \,
  f + d)^{a b c} \, G^{+,c}
  \Bigg](w)\nonu \\
&& + 
\frac{1}{(z-w)}\, \Bigg[ -\frac{1}{3}\,
  \frac{2(k^2-1)(2k+M+N)}{k(2k+M)} \nonu \\
  & & \times  
    \Bigg(
 \frac{1}{M} \, \de^{a b} \, \pa \, G^{+}- \frac{1}{2} \, ( i \,
  f + d)^{a b c} \, \pa \, G^{+,c}
  \Bigg)
  \nonu \\
  && -  \frac{1}{2} \, ( i f +d)^{a b c} \,
  \Bigg( V^{+(\frac{5}{2}),c}
  +\frac{2}{k M} \, (k+M+N)  \, J^c\, G^+ \nonu \\
  & & +
  (i f -\frac{(2k+M+2N)}{2k+M} \, d)^{ c d e}\,
  J^d \, G^{+,e} + \frac{2}{3}\,
  \frac{2(k^2-1)(2k+M+N)}{k(2k+M)} \, \pa \, G^{+,c} \Bigg)
  \nonu \\
  && -    \frac{2}{k M}\, (k+M+N)\,
  J^b \, G^{+,a}  \nonu \\
  && +  \frac{2k(k+2M)}{3(k^2-4)(k+M+N)(3k+2M+2N) b_1}\,
  \de^{a b} \, G^{+(\frac{5}{2}), 0} \nonu \\
  && - (i f -\frac{(2k+M+2N)}{2k+M} \, d)^{b c d}\,
  \Bigg( \frac{1}{M} \, \de^{a d} \, J^c\, G^+ -
  \frac{1}{2} \, (i f +
  d)^{a d e} \, J^c \, G^{+,e} \Bigg) \nonu \\
&& -  \frac{2}{3}\,
  \frac{2(k^2-1)(2k+M+N)}{k(2k+M)} \, 
    \Bigg(
 \frac{1}{M} \, \de^{a b} \, \pa \, G^{+}- \frac{1}{2} \, ( i \,
  f + d)^{a b c} \, \pa \, G^{+,c}
  \Bigg)
  \Bigg](w)
+ \cdots.
\label{gkOPE}
\eea
The first two lines of the first order pole in (\ref{gkOPE})
are the descendant terms associated with the
currents in the second order pole.
Recall that the first, second and fifth
terms in $V^{+(\frac{5}{2}),c}$
(\ref{5halfone}) appear as unwanted terms
in the sense that they cannot be written in terms of the
known currents.
We can check that the second, fourth and last terms of the
first order pole in the corresponding OPE in Appendix $D$
can be written in terms of $V^{+(\frac{5}{2}),c}$
with $f$ and $d$ symbols  plus other terms. 
Moreover,
the spin-$\frac{5}{2}$ current $G^{+(\frac{5}{2}),0}$ (\ref{g5half})
contains similar first, third and fourth terms and they can be written
in terms of $G^{+(\frac{5}{2}),0}$ and other known operators.
Finally, we arrives at the above results of the first order pole
in (\ref{gkOPE}) by matching those unwanted terms with
$V^{+(\frac{5}{2}),c}$,  $G^{+(\frac{5}{2}),0}$ and other known composite
operators.
Of course, we can write down $V^{+(\frac{5}{2}),c}$
in terms of $G^{+(\frac{5}{2}),c}$ with other terms by using
(\ref{real5half}).

In summary, 
as before,
from the second OPE in (\ref{ope2}),
we observe that the right hand side of the OPE
contains the lowest component of the ${\cal N}=2$
multiplet in (\ref{fourcurrents}) by noting the presence of
the nonsinglet spin-$2$ current. Moreover,
from the last OPE,
we observe that $
 i \, f^{a b c}\,  t^c_{k \bar{i}}\,
t^{\al}_{\rho \bar{\mu}}\, J^{(\rho \bar{i})}\,
\psi^{(\bar{\mu} k)}\, J^{\al}$ term of (\ref{newRplus})
can be interpreted as $ i \, f^{a b c}\, V^{+(\frac{5}{2}),c}$
(or $i \, f^{a b c}\, G^{+(\frac{5}{2}),c}$)
plus other terms. This implies that
 the right hand side of the OPE
contains the second component of the ${\cal N}=2$
multiplet in (\ref{fourcurrents}) as we expect. 

\subsection{ The OPEs with the third  component }

The remaining two OPEs 
are given by
\bea
&& G^{-,a}(z) \, G^{-,b}(w) = 0 + \cdots,
\nonu \\
&& G^{-,a}(z) \, W^{+(2),b}(w) = \frac{1}{(z-w)^2}\,
\Bigg[ \frac{1}{2M}\, (3k+3N)\, \de^{a b}\, G^- -
  \frac{1}{4} \, (N-k)\, i \, f^{a b c}\, G^{-,c}
  \nonu \\
  && - \frac{3}{4}\, (k+N)\, d^{a b c}\, G^{-,c}
  \Bigg](w) +\frac{1}{(z-w)}\,
 \Bigg[ \frac{1}{3}\, \pa \, \mbox{(pole-2)}+
R^{-(\frac{5}{2}),a b}\Bigg](w) + \cdots.
\label{gwOPE1}
\eea
The first order pole in the last OPE
of (\ref{gwOPE1}), after subtracting
the descendant terms, contains
the following primary spin-$\frac{5}{2}$ current
with free indices $a$ and $b$
\bea
R^{-(\frac{5}{2}),a b} &= &
\frac{(k+N)}{2 M}\, \de^{a b}\, \pa \, G^{-}
-\sqrt{\frac{M+N}{M N}}\, i \, f^{a b c}\,
J^{u(1)}\, G^{-,c}
\nonu \\
&- & \frac{(M+2 N)}{M^2 N}\, \de^{a b}\, \de_{\rho \bar{\si}}\,
\de_{i \bar{k}}\, \psi^{(\si \bar{k})}\, \pa \, J^{(\bar{\rho} i)}
\nonu \\
&+& \frac{(2 k N+1)}{2 N}\, i \,
f^{a b c} \, t^c_{k \bar{i}}\, \de_{\rho \bar{\si}}\,
\psi^{(\rho \bar{i})}\, \pa \, J^{(\bar{\si} k)}
-\frac{(M+2 N)}{2 M N}\, d^{a b c} \, t^c_{k \bar{i}}\, \de_{\rho \bar{\si}}\,
\psi^{(\rho \bar{i})}\, \pa \, J^{(\bar{\si} k)}
\nonu \\
&+& i \, f^{a b c}\, t^{c}_{i \bar{k}}\, t^{\al}_{\mu \bar{\rho}}\,
J^{(\bar{\rho} i)}\, \psi^{(\mu \bar{k})}\, J^{\al}-
\frac{1}{M}\, J^a \, G^{-,b}
+ \frac{(M+N)}{M N}\, t^a_{k \bar{l}}\, \de_{\mu \bar{\tau}}\, ((\psi^{(\mu
  \bar{l})} \, J^{(\bar{\tau} k)})\, J_f^b)
\nonu \\
&-& t^{\al}_{\si \bar{\rho}}\, t^b_{i \bar{j}}\, t^a_{k \bar{l}}\, t^{\al}_{
  \nu \bar{\mu}} \, \psi^{(\si \bar{j})}\,\psi^{(\bar{\rho} i)}\,
\psi^{(\nu \bar{l})}\, J^{(\bar{\mu} k)} -\frac{1}{M^2}\,
\de^{a b}\, G^- \, J_f^{u(1)}-\frac{1}{2M}\, i \, f^{a b c}\,
G^{-,c}\, J_f^{u(1)}
\nonu \\
& + &
\frac{1}{2M}\,d^{a b c}\,
G^{-,c}\, J_f^{u(1)}+\frac{1}{M}\, J^b \, G^{-,a}
+ \frac{1}{2M} \, (i \, f +d)^{a b d}\, G^- \, J_f^d
\nonu \\
&-& \frac{1}{4}\, (i \, f+d)^{c a e} (i\, f +d)^{c b d}\, G^{-,e}\,
J_f^d -\frac{1}{2M}\, (i \, f+d)^{c b a} \, J^c \, G^-
\nonu \\
& - & \frac{1}{4}\, (i \, f-d)^{d a e} (i\, f +d)^{c b d}\,
J^c \, G^{-,e}+
\frac{1}{4} (3 k+N)\, i \, f^{a b c}\, \pa \, G^{-,c}
\nonu \\
&+&\frac{1}{4} (-k-N)\,
d^{a b c}\, \pa \, G^{-,c}
+ \frac{1}{2}\, (i \, f+ d)^{c a d}\, t^b_{j\bar{i}}\,
\de_{\rho \bar{\si}}\, ( t^b \, t^d)_{j \bar{l}}\,
J^{(\bar{\si} j)}\, \psi^{(\rho \bar{l})}\, J^c
\nonu \\
& - &\frac{1}{3}\, \pa \, \Bigg(G^{-,a}(z) \, W^{+(2),b}(w)
\Bigg|_{\frac{1}{(z-w)^2}}\Bigg).
\label{newRminus}
\eea
Again, due to the existence of
$t^{\al}_{\si \bar{\rho}}\, t^b_{i \bar{j}}\, t^a_{k \bar{l}}\, t^{\al}_{
  \nu \bar{\mu}} \, \psi^{(\si \bar{j})}\,\psi^{(\bar{\rho} i)}\,
\psi^{(\nu \bar{l})}\, J^{(\bar{\mu} k)}$ in (\ref{newRminus}),
we cannot express this in terms of the known currents. 
It is an open problem whether
the above spin-$\frac{5}{2}$ current (\ref{newRminus})
is a third component of
any ${\cal N}=2$ multiplet with free two indices.
It is natural to consider the OPE between the spin-$\frac{3}{2}$
current $G^-$ of the ${\cal N}=2$ superconformal algebra and the previous
spin-$\frac{5}{2}$ current with two indices
$R^{+(\frac{5}{2}), a b}$ (\ref{newRplus})
and obtain the possible lowest component
of ${\cal N}=2$ multiplet.
After that we need to check whether the OPE
between  the spin-$\frac{3}{2}$
current $G^-$ and this lowest component (which
will contain the quartic complex fermions with generators of
$SU(N)$ and $SU(M)$)
will give us the above
spin-$\frac{5}{2}$ current (\ref{newRminus}) plus other terms or not.
See also the footnote \ref{fourpsi} for the four product of
complex fermions.

\subsubsection{The OPE with the spin-$2$ current}

As before, the corresponding OPE can be described as 
\bea
&& G^{-,a}(z) \, K^{b}(w) =
 -\frac{1}{(z-w)^2} \,
\frac{2(k^2-1)(2k+M+N)}{k(2k+M)} \,
\Bigg[
 \frac{1}{M} \, \de^{ b a} \, G^{-}- \frac{1}{2} \, ( i \,
  f + d)^{b a c} \, G^{-,c}
  \Bigg](w)\nonu \\
&& + 
\frac{1}{(z-w)}\, \Bigg[ -\frac{1}{3}\,
  \frac{2(k^2-1)(2k+M+N)}{k(2k+M)} \nonu \\
  & & \times  
    \Bigg(
 \frac{1}{M} \, \de^{ b a} \, \pa \, G^{-}- \frac{1}{2} \, ( i \,
  f + d)^{ b a c} \, \pa \, G^{-,c}
  \Bigg)
  \nonu \\
&  & -  \frac{1}{2} \, ( i f +d)^{ b a c} \,
  \Bigg( V^{-(\frac{5}{2}),c}
  -\frac{2}{k M} \, (k+M+N)  \, J^c\, G^- \nonu \\
 & &+
  (i f +\frac{(2k+M+2N)}{2k+M} \, d)^{ c d e}\,
  J^d \, G^{-,e} + \frac{2}{3}\,
  \frac{2(k^2-1)(2k+M+N)}{k(2k+M)} \, \pa \, G^{-,c} \Bigg)
  \nonu \\
  && +    \frac{2}{k M}\, (k+M+N)\,
  J^b \, G^{-,a}  \nonu \\
  && -  \frac{2k(k+2M)}{3(k^2-4)(k+M+N)(3k+2M+2N) b_1}\,
  \de^{a b} \, G^{-(\frac{5}{2}), 0} \nonu \\
  & & - (i f +\frac{(2k+M+2N)}{2k+M} \, d)^{b c d}\,
  \Bigg( \frac{1}{M} \, \de^{ d a } \, J^c\, G^- -
  \frac{1}{2} \, (i f +
  d)^{ d a e} \, J^c \, G^{-,e} \Bigg) \nonu \\
&& -  \frac{2}{3}\,
  \frac{2(k^2-1)(2k+M+N)}{k(2k+M)} \, 
    \Bigg(
 \frac{1}{M} \, \de^{ b a } \, \pa \, G^{-}- \frac{1}{2} \, ( i \,
  f + d)^{ b a c} \, \pa \, G^{-,c}
  \Bigg)
  \Bigg](w)
+ \cdots.
\label{GKope}
\eea
The second, fourth and last terms of the
first order pole in the corresponding OPE in Appendix $D$
can be written in terms of $V^{-(\frac{5}{2}),c}$
with $f$ and $d$ symbols  plus other terms. 
By using the first, second and fifth
terms in $V^{-(\frac{5}{2}),c}$
(\ref{5halftwo}) appearing as unwanted terms,
we can reexpress the corresponding terms in the first order pole
in terms of $V^{-(\frac{5}{2}),c}$ and other known operators.
From  the first, third and fourth terms
in the spin-$\frac{5}{2}$ current  $G^{-(\frac{5}{2}),0}$
(\ref{g5half1}),
the  corresponding terms in the first order pole
can be written
in terms of $G^{-(\frac{5}{2}),0}$ and other known operators
by focusing on the $\de^{a b}$ factor in the product of
two $SU(M)$ generators with indices $a$ and $b$.
Finally, we obtain the above results (\ref{GKope})
where we can write down $V^{-(\frac{5}{2}),c}$
in terms of $G^{-(\frac{5}{2}),c}$ with other terms by using
(\ref{real5half}).

\subsection{ The final OPE}

Now we describe the final OPE between the spin-$2$ current and itself
with Appendix $E$
\bea
&& W^{+(2),a}(z) \, W^{+(2),b}(w)  =  \frac{1}{(z-w)^4}\,
\frac{3}{2}\, k \, N \, (k + N)\, \de^{a b} \nonu \\
&& + 
\frac{1}{(z-w)^3} \, \Bigg[
   \frac{1}{2} N (2 k+N)\, i \, f^{a b c}\, J^c +
   \frac{1}{2} k (k+2 N)\,  i \, f^{a b c}\,J_f^c \Bigg](w) \nonu \\
&& + 
\frac{1}{(z-w)^2} \, \Bigg[ \de^{a b} \,
  2(k+N) \Bigg( \frac{1}{M} \, (k+M+N) \, T -\frac{1}{2M}\,
  J^c \, J^c
 -
 \frac{1}{2M} \, J_f^c \, J_f^c \Bigg)
 \nonu \\
 && + d^{a b c}\, \Bigg( -(k+N)\, W^{+(2),c}
 +\frac{1}{2}\, (i \, f +d)^{d c e}\, J^d \, J_f^e
 +\frac{k}{2}\, (1-k-N)\, \pa \, J_f^c \nonu \\
 && +
\frac{1}{2} N (k+N-1)\, \pa \, J^c
\Bigg) + i\, f^{a b c}\, \Bigg( \frac{k}{4}\, (k+2N)\,
\pa \, J_f^c \nonu \\
&&+ \frac{N
  (2 k^3 M+k^2 M^2+k^2 M N+2 k M+4 k N+M^2+M N)}{
  2 k M (2 k+M)}\, \pa \, J^c\Bigg) 
\nonu \\
&& -\frac{(k+2 M+N)}{M}\,J^b \, J^a_f
-\frac{(k+N)}{M}\, J^a \, J^b_f
-\frac{k}{4}\,
(i \, f + d)^{d a e}\, (i \, f+d)^{c b d} \, J^c \,J_f^e
\nonu \\
&& -\frac{k}{4}\,
(i \, f + d)^{d b e}\, (i \, f+d)^{c a d} \, J^c \,J_f^e
-\frac{1}{4}\, f^{d h f} \, f^{e c g}\,
(i \, f + d)^{h a c}\, (i \, f+d)^{d b e} \, J^f \,
J_f^g\nonu \\
&& -\frac{k}{4}\,
i \, f^{b c g}\, (i \, f+d)^{d a c} \, J^d \,J_f^g
-\frac{N}{4}\,
d^{e b c}\, (i \, f+d)^{e a d} \, J^c \,J_f^d
\nonu \\
&& -\frac{k}{4}\,
i \, f^{a c g}\, (i \, f+d)^{d b c} \, J^d \,J_f^g
-\frac{N}{4}\,
 d^{e a c}\, (i \, f+d)^{e b d} \, J^c \,J_f^d
\Bigg](w)
\nonu \\
&& +
\frac{1}{(z-w)} \, \Bigg[ \frac{1}{2}\, \pa \, \mbox{(pole-2)}
  + R^{+(3),a b}
  \Bigg](w) + \cdots.
\label{WW}
\eea
In the second order pole of (\ref{WW}),
we observe that there exists
$d^{a b c}\, W^{+(2),c}$ term.
Moreover, there are no quartic terms in the complex fermions
by collecting all the contributions 
\footnote{
\label{fourpsi}
  There is a relation 
  $k \, \frac{M+N}{M N}\, J_f^a \, J_f^b + k \, t^{\al}_{\rho \bar{\si}}
  \, t^{\al}_{\mu \bar{\nu}}\, t^a_{j \bar{i}}\, t^b_{k \bar{l}}\,
  (\psi^{(\rho \bar{i})}\, \psi^{(\bar{\si} j)})(\psi^{(\mu \bar{l})}\,
  \psi^{(\bar{\nu} k)})+\frac{k}{M^2}\,
  \de^{a b}\, J_f^{u(1)}\, J_f^{u(1)}-
  \frac{k}{M}\, d^{a b c}\, J_f^c \,J_f^{u(1)}
  + \frac{k}{4}\, (i \, f+ d)^{d a c}\,
  (i \, f+ d)^{d b e}\, J_f^c \, J_f^e=0$. That is,
  the exact coefficients appearing in these five terms
  with the appropriate complex fermion terms
  lead to zero. Note that the particular combination
  of this quartic fermions (with two indices $a$ and $b$)
  $ t^{\al}_{\rho \bar{\si}}
  \, t^{\al}_{\mu \bar{\nu}}\, t^a_{j \bar{i}}\, t^b_{k \bar{l}}\,
  \psi^{(\rho \bar{i})}\, \psi^{(\bar{\si} j)}\,\psi^{(\mu \bar{l})}\,
  \psi^{(\bar{\nu} k)}$
  is a candidate term of the lowest component of ${\cal N}=2$
multiplet with two free indices.}.
In the first order pole of (\ref{WW}),
after subtracting the descendant terms, we are left with
$R^{+(3),a b}$ terms.
In particular, we observe that by checking the purely bosonic terms
in $R^{+(3), a b}$, there is a term
\bea
\frac{2k(k+M)(3k+2M)}{(k^2-4)(2k+M)(3k+2M+2N)(3k+M+3N)}\,
\frac{1}{a_1}\, i \, f^{a b c}\, W^{+(3),c}.
\label{Wplusthree}
\eea
This can be seen from the OPE between $K^a$ and $K^b$
and the $a_1$ term of $P^c$ appears in the
inside of the sixth pair of bracket in
(\ref{spinthreenon}).
The nonsinglet spin-$3$ current $ W^{+(3),c}$ in (\ref{Wplusthree})
is the last component of the ${\cal N}=2$ multiplet in
(\ref{fourcurrents}).
This is reasonable because
the OPE between $K^a(z)$ and $K^b(w)$
leads to $i \, f^{a b c}\, P^c$ term at the first order pole
in the bosonic coset model \cite{Ahn2011}.

One of the reasons why we cannot write down
this  $R^{+(3), a b}$ in terms of the known currents is
that there is a term from the OPE between the fourth term of
$W^{+(2),a}$
(\ref{nonspin2})and itself  (which is presented in Appendix $E$)
\bea
i \, f^{\al \beta \ga} \, t^{\al}_{\rho \bar{\si}}\,
t^{\beta}_{\si \bar{\nu}} \, t^b_{j \bar{i}}\, t^a_{l \bar{k}} \,
J^{\ga}\, \psi^{(\rho \bar{i})} \,\psi^{(\bar{\si} j)} \,\psi^{(\si \bar{k})}
\,\psi^{(\bar{\nu} l)},
\label{threespin}
\eea
which can be obtained by acting $G^-(z)$ on
$t^{\al}_{\rho \bar{\si}}\, t^b_{j \bar{i}}\,
 t^a_{l \bar{k}}\, t^{\al}_{\mu \bar{\nu}}\,
 \psi^{(\bar{\si} j)} \, \psi^{(\rho \bar{i})}
 \, \psi^{(\bar{\nu} l)} \, J^{(\mu \bar{k})}(w)$.
 This spin-$\frac{5}{2}$ operator
 is the characteristic term for the $R^{+(\frac{5}{2}), a b}$
 in (\ref{newRplus}). Then the OPE
 between $G^-$ and $R^{+(\frac{5}{2}), a b}$ contains the above
 term (\ref{threespin}).
 Note from the footnote \ref{someopeexpression}
 that the OPE between $G^{-}$ and $J^{\ga}$
 leads to the composite field of spin-$\frac{1}{2}$
 and spin-$1$ operators transforming as
 $({\bf N}, \overline{\bf M})$ and $(\overline{\bf N},
 {\bf M})$ respectively.
 Therefore we obtain (\ref{threespin}) after contracting the indices
 properly. In this way, there is a connection between
 $R^{+(\frac{5}{2}), a b}$ and $R^{+(3), a b}$ via a supersymmetry generator.
 It is an open problem to see whether there exists
 any ${\cal N}=2$ multiplet of $(R^{-(2),a b}, R^{+(\frac{5}{2}), a b},
 R^{-(\frac{5}{2}),a b }, R^{+(3), a b})$ or not.
 Probably, if we compute the OPEs between the currents of high spins,
 then the lowest current $R^{-(2), a b }$ can be seen
 \footnote{Because we present all the OPEs
   between the nonsinglet spin-$2$ operators in Appendix $E$,
   by reversing the orders of some OPEs
   we can write all the contributions on the first order poles
   in (\ref{WW}) and read off
   the nonsinglet spin-$3$ current $R^{+(3), a  b}$ explicitly
   which covers a several pages.}.
 
 In summary, we have
 the complete OPEs in (\ref{ope1}), (\ref{ope2}), (\ref{gwOPE1})
 and (\ref{WW}).
 The nontrivial part of these OPEs is that in the right hand side of
 these OPEs, the components of the third ${\cal N}=2$
 multiplet (\ref{fourcurrents})
 as well as the components of the first
${\cal N}=2$ multiplet
 (\ref{simplest}) and the currents
 of ${\cal N}=2$ superconformal algebra 
 arise.
 We have seen the new primary currents having two free indices of
 $SU(M)$. The various quasi primary operators in the first order pole
 of (\ref{WW})
 will appear as in the bosonic case.
 
\section{Towards the OPEs between the singlet
  multiplet of spins $(2,\frac{5}{2},\frac{5}{2},3)$
and itself}

The simplest OPE between the singlet currents
is given by the OPE between the singlet spin-$2$ current and itself.
We obtain the following OPE with the help of Appendix $F$
\bea
&& W^{-(2),0}(z) \, W^{-(2),0}(w)  = \nonu \\
&&
\frac{1}{(z-w)^4}\,\Bigg[ \frac{ 6   (k^2-1) (k^2-4)^2 N (k+M+N)^3 (
  2 k+M+N) (3 k+2 M+2 N)^2}{k^3 M (k+M)^2 (k+2 M)^2}
\nonu \\
&& \times (k^3+2 k^2 M+2 k^2 N+3 k M N+2 k+M+N) \, b_1^2
\Bigg] \nonu \\
&& +
\frac{1}{(z-w)^2}\, \Bigg[\frac{
 8   (k^2-1)  (k^2-4)^2 (k+M+N)^4 (2 k+M+N)
  (3 k+2 M+2 N)^2
  }{
    k^4 M^2 (k+M)^2 (k+2 M)^2} \nonu \\
  && \times (k^3+2 k^2 M+2 k^2 N+3 k M N+2 k+M+N) \, b_1^2 \, \Bigg(T-
  \frac{1}{2(k+M+N)} \, (J^a+J_f^a)^2\Bigg)
  \nonu \\
  &&+
  \frac{ 4  (k^2-4)  (k+M+N)^2 (3 k+2 M+2 N)}{k^2 M (k+M) (k+2 M)}\nonu \\
  && \times (k^3-k^2 M-k^2 N-3 k M N-4 k-2 M-2 N) \, b_1\,
  W^{-(2),0} \Bigg](w)\nonu \\
  && + \frac{1}{(z-w)} \, \frac{1}{2} \, \pa \, (\mbox{pole-2})(w)
  +\cdots.
\label{spin2spin2ope}
\eea
From the fourth order pole of (\ref{spin2spin2ope}),
we can fix the normalization of the singlet spin-$2$ current
by taking the unknown coefficient $b_1^2$ properly.
For example, one way to fix is such that the fourth order pole
is given by $\frac{c}{2}$ where $c$ is the central charge in (\ref{central}).
The self coupling constant of the singlet spin-$2$ current
which depends on the $k, N$ and $M$ explicitly
appears in the second order pole of (\ref{spin2spin2ope}).
As we expected, the right sides of this OPE
consist of 1) the central term, 2) stress energy tensor,
3) $(J^a+J_f^a)^2$
term, 4) the singlet spin-$2$ current and 5) their descendant terms
\footnote{It is useful to use the following
  relation $\de_{\rho \bar{\si}}\, \de_{j \bar{i}}\, J^{(\rho \bar{i})}\,
  J^{(\bar{\si} j)} = (k+M+N)\, T_{boson} + \frac{M}{2(k+N)}\, J^{\al }\,
  J^{\al} +\frac{(M+N)}{2k}\, J^{u(1)} \, J^{u(1)}-\frac{1}{2}\, J^a \,
  J^a + \frac{1}{2}\, M N \, \sqrt{\frac{M+N}{M N}}\, \pa \,
  J^{u(1)}$ together with the footnote \ref{bosonT}
  in order to calculate the OPE between the first term of
  $W^{-(2),0}$ and itself. Then we know the OPE between
  $T_{boson}$ and itself and the four
  remaining composite spin-$2$ operators of above
  belong to  $W^{-(2),0}$ and we can also
  compute the OPEs between them.}.

We can check that
the singlet spin-$2, \frac{5}{2}, 3$ currents,  $W^{-(2),0}$,
  $G^{\pm (\frac{5}{2}),0}$ and
  $W^{+(3),0}$
have the regular terms in the OPEs
with $(J^a+J_f^a)$. Moreover,
the OPE between the combination of
$T-
  \frac{1}{2(k+M+N)} \, (J^a+J_f^a)^2$
  and  $(J^a+J_f^a)$ does not have any singular terms.
  This implies that the singlet currents,
  \bea
  T-
  \frac{1}{2(k+M+N)} \, (J^a+J_f^a)^2, \qquad  W^{-(2),0},\qquad
  G^{\pm (\frac{5}{2}),0}, \qquad 
  W^{+(3),0},
  \label{Singletcurrent}
  \eea
  are decoupled from the spin-$1$ current
  $(J^a+J_f^a)$.
  It is easy to observe that the previous (quasi)primary conditions
  for these  singlet spin-$2, \frac{5}{2}, 3$ currents still hold
  under the modified stress energy tensor because
  the  $(J^a+J_f^a)^2$ term does not produce any singular terms
  \footnote{We can see that the OPEs between $(J^a+J_f^a)$ and
  $K, G^{\pm}$ do not produce any singular terms.}. 

  After the OPEs
  between the currents of $(W^{-(s),0}, G^{+(s+\frac{1}{2}),0},
  G^{-(s+\frac{1}{2}),0},W^{+(s+1),0})$ in (\ref{Ws}) where
  $s=2,3,4, \cdots $ are obtained,
  we expect that
  the right hand sides of these OPEs
  will consist of the composite operators in terms of these
  singlet currents as well as the modified stress energy tensor
  (\ref{Singletcurrent}) (and maybe $K$ and $G^{\pm}$ also).
  The new singlet (quasi)primary current will have the regular OPE
  with  the spin-$1$ current
  $(J^a+J_f^a)$.

  Then the algebra
  from $(W^{-(s),0}, G^{+(s+\frac{1}{2}),0},
  G^{-(s+\frac{1}{2}),0},W^{+(s+1),0})$ will close
  by themselves up to the presence of
  the currents of ${\cal N}=2$ superconformal algebra
  with modified stress energy tensor. In other words,
  the right hand sides of these OPEs do not contain
  the $SU(M)$ nonsinglet fields appearing in (\ref{Ws}).
  On the other hands, the OPEs between the nonsinglet currents
  do contain the $SU(M)$ singlet fields.
  
\section{ The
  extension of the large ${\cal N}=4$ nonlinear superconformal algebra
for $M=2$}

We describe in this section how we can realize the extension
of  the large ${\cal N}=4$ nonlinear superconformal algebra for $M=2$.

\subsection{Four spin-$\frac{3}{2}$ currents}

The four supersymmetry generators of
the large ${\cal N}=4$ nonlinear superconformal algebra
can be obtained by combining $G^{+,a}$ with $G^{-,a}$
and also combining $G^+$ and $G^-$ properly
\bea
\hat{G}_{11} & = & \sqrt{\frac{1}{k+N+2}} \, \Bigg[ G^{+,1}-i \, G^{+,2}
  + G^{-,1}-i \, G^{-,2} \Bigg],
\nonu \\
\hat{G}_{12} & = & -\frac{1}{\sqrt{2}} \,
\sqrt{\frac{1}{k+N+2}} \, \Bigg[ G^{+}+\sqrt{2} \, G^{+,3}
  - G^{-}+\sqrt{2} \, G^{-,3} \Bigg],
\nonu \\
\hat{G}_{21} & = & \frac{1}{\sqrt{2}} \,
\sqrt{\frac{1}{k+N+2}} \, \Bigg[ G^{+}-\sqrt{2} \, G^{+,3}
  - G^{-}-\sqrt{2} \, G^{-,3} \Bigg],
\nonu \\
\hat{G}_{22} & = & \sqrt{\frac{1}{k+N+2}} \,
\Bigg[ G^{+,1}+i \, G^{+,2}
  + G^{-,1}+i \, G^{-,2} \Bigg].
\label{fourg}
\eea
These expressions (\ref{fourg}) satisfy the fundamental relations
Appendix $(D.1)$ of \cite{AK1411} or Appendix
(\ref{ggopenonlinear}) in this paper
\footnote{
\label{spin1rel}
  We have
  the following six spin-$1$ currents 
  $\hat{A}_{\pm}= \frac{1}{\sqrt{2}}\, (-i\, J^1 \mp J^2)$,
  $\hat{A}_{3}= -\frac{i}{\sqrt{2}}\, J^3$ and
 $\hat{B}_{\pm}= \frac{1}{\sqrt{2}}\, (i\, J_f^1 \mp J_f^2)$,
  $\hat{B}_{3}= \frac{i}{\sqrt{2}}\, J_f^3$. 
  The defining relations for these $SU(2)$
  currents are given in $(2.13)$ and $(2.16)$
  of \cite{AK1411} where we can fix the normalizations
  of the spin-$1$ currents. The defining
  OPEs between these spin-$1$ current
  and four spin-$\frac{3}{2}$ currents are given in Appendix $C$ of
  \cite{AK1411}. Furthermore, the spin-$1, \frac{3}{2}$ currents
  are primary under the stress energy tensor. The central charge
  (\ref{central}) gives us $c=\frac{3(k+N+2 k N)}{(k+2+N)}$ for
$M=2$.}.
From the explicit expressions in $G^{\pm}$ (\ref{gpm}) and
$G^{\pm,a}$ (\ref{gpma}),
the summation over the fundamental and antifundamental
$SU(N)$ indices appears
in the Kronecker delta with the coset fields. On the other hands,
the summation over the fundamental and antifundamental $SU(M=2)$ indices
appears in the matrix elements of the $SU(M=2)$ generators 
with the coset fields.
For example, in the spin-$\frac{3}{2}$ current $\hat{G}_{11}$,
the index $j=1$ and the index $\bar{i}=2$ survives in
the first combination of $ (G^{+,1}-i \, G^{+,2})$ and the
index $j=2$ and the index $\bar{i}=1$ survives
in the second combination of $ (G^{-,1}-i \, G^{-,2})$.
This corresponds to the last $4N \times 4N$ matrix in
Appendix $(B.2)$ of \cite{AK1411} where the nonzero elements appear
in the two $N \times N $ identity matrices inside of this matrix.
We can analyze the other three currents similarly from
$G^{\pm}$
(\ref{gpm}) and
$G^{\pm,a}$ (\ref{gpma}).

\subsection{Higher spin-$\frac{3}{2}$ currents}

Then the next spin-$\frac{3}{2}$ currents
can be determined by using the relations Appendix $(G.1)$ of
\cite{AK1411} (or Appendix (\ref{spin3halfgenerating}) in this paper)
and it turns out that 
\bea
T_{+}^{(\frac{3}{2})} & = &
-\frac{1}{\sqrt{2}} \,
\sqrt{\frac{1}{k+N+2}} \, \Bigg[ G^{+}-\sqrt{2} \, G^{+,3} \Bigg],
\nonu \\
T_{-}^{(\frac{3}{2})} & = &
\frac{1}{\sqrt{2}} \,
\sqrt{\frac{1}{k+N+2}} \, \Bigg[ G^{-}-\sqrt{2} \, G^{-,3} \Bigg],
\nonu \\
U^{(\frac{3}{2})} & = &
-
\sqrt{\frac{1}{k+N+2}} \, \Bigg[ G^{+,1}-i \, G^{+,2} \Bigg],
\nonu \\
V^{(\frac{3}{2})} & = &
\sqrt{\frac{1}{k+N+2}} \, \Bigg[ G^{-,1}+ i \, G^{-,2} \Bigg].
\label{spin3halfm2}
\eea
Again, from
(\ref{gpm}) and (\ref{gpma}),
we observe that 
in the spin-$\frac{3}{2}$ current $T_{+}^{(\frac{3}{2})}$,
the index $j=1$ and the index $\bar{i}=1$ survives
and 
this corresponds to  the nonzero elements
in the $N \times N $ identity matrix inside of the $4N \times 4N$ matrix
in the last expression of Appendix $B$ of \cite{AK1411}.
Here, each of spin-$\frac{3}{2}$ currents in (\ref{fourg})
is reduced to further here. Note that the eight rank two tensors
which are $4N \times 4N$ matrices and
the nonzero elements appear in four $N \times N$ identity matrices
in \cite{AK1411}. In (\ref{spin3halfm2}),
we are left with these $4N \times 4N$ matrices
where only $N \times N$ identity matrix arises. 

Then the eight spin-$\frac{3}{2}$ currents of the left hand sides
in (\ref{fourg}) and (\ref{spin3halfm2}) for generic $k$ and $N$
can be realized by two $G^{\pm}$ and six
$G^{\pm, a}$ (where $a=1,2,3$) in the right hand sides.

\subsection{Higher spin-$2$ currents}

By using Appendix $(G.2)$ of \cite{AK1411} together with
(\ref{fourg}) and (\ref{spin3halfm2}) (or Appendix
(\ref{spin2generating1})),
we can determine the following spin-$2$ currents
 for generic $k$ and $N$
as follows:
\bea
U_{-}^{(2)} &=& -\frac{1}{2(k+N+2)} \Bigg[ -
  i \, f^{3 1 c} \, K^c + \frac{(k+N+2)}{k}\,  i \, f^{3 1 c} \,
  J^c \, K + 2 \, J^1 \, J_f^3 \nonu \\
   &- &  f^{3 2 c} \, K^c + \frac{(k+N+2)}{k} \, f^{3 2 c} \,
  J^c \, K - 2 i \, J^2 \, J_f^3 \Bigg],
\nonu \\
V_{+}^{(2)} &=& \frac{1}{2(k+N+2)} \Bigg[ 
  i \, f^{3 1 c} \, K^c - \frac{(k+N+2)}{k}\,  i \, f^{3 1 c} \,
  J^c \, K + 2 \, J^1 \, J_f^3 \nonu \\
   &- &  f^{3 2 c} \, K^c + \frac{(k+N+2)}{k}\, f^{3 2 c} \,
  J^c \, K + 2 i \, J^2 \, J_f^3 \Bigg],
\nonu \\
U_{+}^{(2)} &=& -\frac{1}{2(k+N+2)} \Bigg[
2 \, i\, f^{1 3 c} \, W^{+(2),c}
 + 
  i \, f^{1 3 c} \, K^c - \frac{1}{k}\,  i \, f^{1 3 c} \,
  J^c \, K + 4 \, J^3 \, J_f^1\nonu \\
  & + &
    f^{a 3 c} \, f^{1 c d} \,
  J^a \, J_f^d
+2  \, f^{2 3 c} \, W^{+(2),c}
  \nonu \\
   &+ &  f^{2 3 c} \, K^c - \frac{(k+N+2)}{k} \, f^{2 3 c} \,
  J^c \, K - 4 i \, J^3 \, J_f^2
- i\,  f^{a 3 c} \, f^{2 c d} \,
  J^a \, J_f^d
  \Bigg],
\nonu \\
V_{-}^{(2)} &=& \frac{1}{2(k+N+2)} \Bigg[
2\, i\, f^{3 1 c} \, W^{+(2),c}
 + 
  i \, f^{3 1 c} \, K^c - \frac{(k+N+2)}{k}\,  i \, f^{3 1 c} \,
  J^c \, K + 2 \, J^1 \, J_f^3 \nonu \\
 &
- & 2 \, f^{3 2 c} \, W^{+(2),c}
  -   f^{3 2 c} \, K^c + \frac{(k+N+2)}{k} \, f^{3 2 c} \,
  J^c \, K + 2 i \, J^2 \, J_f^3
  \Bigg],
\nonu \\
T^{(2)} & = &-\frac{1}{2(k+N+2)} \Bigg[
  -\frac{2 (k+N)(k+N+2)}{(k+N+2 k N)}\, \de^{3 3} \, T
  +2 \, J^3 \, J_f^3
  + \sqrt{2} \, i \, f^{3 a b} \, J^a \, J_f^b
  \nonu \\
  & + &
   J^a \, J^a +  J_f^a \, J_f^a
  + 2 \sqrt{2}\, W^{+(2),3}
  \Bigg],
\nonu \\
W^{(2)} & = &\frac{1}{2(k+N+2)} \Bigg[
  -2 \, J^1 \, J_f^1
 - 2 \, f^{2 1 c} \, W^{+(2),c}
   - 2 \, f^{2 1 c} \, K^{c}
   \nonu \\
   & +& \frac{2 (k+N+2)}{k}  \, f^{2 1 c} \, J^c \, K
  \nonu \\
  & + &
  2 i \, J^1 \, J_f^2 +
 i \,  f^{a 1 c} \, f^{2 c d} \,
 J^a \, J_f^d
+ 2 (k+N+2)\, \de^{2 2} \, T
- \de^{2 2}\, J^a \, J^a
\nonu \\
& - &
   \de^{2 2}\,J_f^a \, J_f^a
  +   f^{a 2 c} \, f^{2 c d} \,
 J^a \, J_f^d
  \Bigg].
\label{sixspin2}
\eea
The way we obtain (\ref{sixspin2}) is
that for fixed $(N,M)=(5,2)$,
we can determine the field contents explicitly.
After that, we read off the generic $(N,M)$ dependence manually.
If we compute the relevant OPEs manually  from the beginning, then
we will obtain different expressions. However, we can check
that eventually those become the above results (\ref{sixspin2})
by using  some identities in the structure constants.
In (\ref{sixspin2}), although there exist
the spin-$2$ current $K^a$ dependent terms,
we can replace them by using (\ref{Spin2non})
with $W^{\pm (2),a}$ term and others.

Therefore, the six spin-$2$ currents
 for generic $k$ and $N$
can be realized by the six spin-$2$ currents
$W^{\pm (2),a}$ where $a=1,2,3$ and other composite operators \footnote{
  \label{spin1andspin2}
  By considering the second order poles of the last four OPEs
  in Appendix $(G.2)$ of
  \cite{AK1411}, the spin-$1$ current of the lowest ${\cal N}=4$
  higher spin multiplet can be realized by $K$ in the
  ${\cal N}=2$ superconformal algebra in (\ref{kggt}). Moreover, the
  stress energy tensor of the ${\cal N}=4$ large nonlinear
  superconformal algebra is identified with the stress energy tensor
  in (\ref{kggt}).
  The stress energy tensor consists of
  purely bosonic part, purely fermionic part and boson fermionic part.
  The purely fermionic part can be summarized by the first two terms in the
  second line of $T$ (\ref{T}) and $J_f^a \, J_f^a$ term
  after using the identity appearing in the last
  equation of Appendix $A$.
  The  boson fermionic part is given by both $J^{\al}\, J^{\al}_f$
  term and $J^{u(1)}\, J^{u(1)}_f$ term.
  The purely bosonic part consists of the second, third and fourth terms
  of $T$ (\ref{T}). Then we can make the correspondences between
  the stress energy tensor in \cite{AK1411} and the one in (\ref{T})
  by focusing on these three parts.
}.

\subsection{Higher spin-$\frac{5}{2},3$ currents}

Now we can compute Appendix $(G.4)$ of \cite{AK1411} by using
(\ref{fourg}) and (\ref{sixspin2}) and taking the first order poles
(or Appendix (\ref{spin5halfgenerating})),
we will obtain the four spin-$\frac{5}{2}$ currents.
Then we need to calculate the OPEs between $G^{\pm,a}(z)$
and $W^{\pm (2),b}(w)$ as done in section $6$.
In other words, we should compute the OPEs
between the second ${\cal N}=2$ multiplet and itself
(and the OPEs between the first ${\cal N}=2$ multiplet
and the second ${\cal N}=2$ multiplet).
For fixed $N$ with $M=2$, we have checked that
we can write down the spin-$\frac{5}{2}$ currents explicitly.
See Appendix (\ref{5halfm2case}).
It turns out that they can be written in terms of
$V^{\pm (\frac{5}{2}),a}$ (or $G^{\pm (\frac{5}{2}),a}$) plus other terms
\footnote{In the next lowest ${\cal N}=4$ multiplet,
there are also other four spin-$\frac{5}{2}$ currents.}.

Eventually, the spin-$3$ current can be realized by
the equation $(3.50)$ of \cite{AK1411}.
That is, after determining the spin-$\frac{5}{2}$
current $ W_{-}^{(\frac{5}{2})}$ for generic $N$ and $M$,
we use 
\bea
W^{(3)}  &=& \hat{G}_{21}(z) \, W_{-}^{(\frac{5}{2})}(w) \Bigg|_{\frac{1}{(z-w)}}
-\Bigg[ \frac{1}{4} \, \pa  \,   \Bigg(
  \hat{G}_{21}(z) \, W_{-}^{(\frac{5}{2})}(w)
  \Bigg|_{\frac{1}{(z-w)^2}}\Bigg)
  \nonu \\
  & + &
  \frac{8 i N(3k+1)}{(N+k+2) ( 5N+4+6k N+5k )} \,
  (-\frac{i}{\sqrt{2}})
  \, \Big( T \, J^3-
\frac{1}{2} \pa^2 \, J^3 \Big)  
\nonu \\
& + &   
\frac{8 i k (3N+1)}{(N+k+2) ( 5N+4+6 k N+5k )} \,
 (\frac{i}{\sqrt{2}})
 \, \Big( T \, J_f^3-
 \frac{1}{2} \pa^2 \, J_f^3 \Big)\nonu \\
 & + & 
 \frac{8(k-N)}{ (5N+4+6 k N+5k) }
 \, \Big( T \, K-
\frac{1}{2} \pa^2 K \Big)  \Bigg].
\label{w3exp}
\eea
Of course,  from the first line of (\ref{w3exp}),
we should calculate the corresponding OPEs.
In the last three lines, 
the relations appearing in the footnotes \ref{spin1rel}
and \ref{spin1andspin2} are used.

\subsection{For $M=3$}

What happens for $M=3$ case?
In particular, how do the supersymmetry generators appear when we
increase $M$ by $1$?
The field contents for the spin-$\frac{3}{2}$ currents
$G^{\pm, a}$ remain the same when we select
the right choice for the index $a$ as $SU(2)$.
The previous indices $1,2,3$ in (\ref{fourg}) correspond to
$1,4,7$ for $M=3$ case. From the definition of
$G^{\pm,a}$ (\ref{gpma}),
the summations over $j$ and $\bar{i}$ contain the index $3$.
However, any generators $t^a$ for indices $a=1,4,7$
do not have any rows and columns having an index $3$. Then we
are left with  the spin-$\frac{3}{2}$ currents
$G^{\pm, a}$ with same field contents for $M=2$ case.
Now we look at the other spin-$\frac{3}{2}$ currents
in $G^{\pm}$ (\ref{gpm}).
In this case, there exist the summations over
$j$ and $\bar{i}$ having the index $3$.
Therefore, as we increase the $M$ value, the field contents
for these spin-$\frac{3}{2}$ currents are increasing.
In other words, the field contents of  spin-$\frac{3}{2}$ currents
for $M=3$ are the same as the one for $M=2$ and other terms.

This implies that for example, the OPE
between $\hat{G}_{11}(z)$ and $\hat{G}_{12}$, where the numerical factor
$2$ inside of the square root is replaced by $3$ and
the indices $1,2,3$ are replaced by $1,4,7$,
does not lead to the one of the known relation 
in the large ${\cal N}=4$ nonlinear superconformal algebra
due to the contribution from other terms we mentioned above.
Therefore, we do not have ${\cal N}=4$ supersymmetry for $M=3$
and we expect that this holds for $M >2$.

In summary,
the currents of
${\cal N}=4$ nonlinear superconformal algebra 
are given by (\ref{fourg}) together with other currents
appearing in the footnotes \ref{spin1rel} and \ref{spin1andspin2}.
Moreover, the higher spin-$\frac{3}{2}, 2$ currents are
described in (\ref{spin3halfm2}) and (\ref{sixspin2}).
The spin-$\frac{5}{2}$ currents are in Appendix $H$ for fixed
$N=5$. Finally, the spin-$3$ current is given by (\ref{w3exp}) implicitly.

\section{ Conclusions and outlook}

The highest component of the first ${\cal N}=2$ multiplet
is found in (\ref{nonspin2}).
The second ${\cal N}=2$ multiplet is found by
(\ref{singletspin2}), (\ref{g5half}), (\ref{g5half1}) and
(\ref{spin3singlet}).
The third ${\cal N}=2$ multiplet is obtained from
(\ref{Spin2non}), (\ref{real5half}) and (\ref{spinthreenon}).
Their OPEs between the currents of the ${\cal N}=2$ superconformal algebra
and these above currents are determined completely
in the sections $3, 4$ and $5$.
Moreover the OPEs between the first
${\cal N}=2$ multiplet and itself are described with the observation
of three kinds of new primary operators.
Similarly, we describe the OPE between the
lowest singlet spin-$2$ current and itself
in the second ${\cal N}=2$ multiplet.
Finally, the extension of the large ${\cal N}=4$ nonlinear
superconformal algebra for $M=2$ is realized from the coset fields
living in (\ref{coset2}). 

Therefore, we have obtained some currents (and their OPEs)
in the supersymmetric coset model for generic $k, N$ and $M$
which correspond to the generators of the ${\cal N}=2$
``rectangular'' $W$-algebra in the $AdS_3$ bulk theory according to
a holography \cite{CH1906}. We have obtained
the ${\cal N}=2$ version of \cite{Ahn2011} and obtained
the generalization of \cite{AK1411} (which is for $M=2$
case) to a generic $M$. In this paper, we have $f$, $d$ symbols
or Kronecker delta in the structure constants of the OPEs compared with
the ones of \cite{AK1411} with various invariant tensors.
For generic $M > 2$, we do not have to worry about the identities
between these invariant tensors although the supersymmetry
is given by ${\cal N}=2$.

We present the future directions related to the results of this paper
as follows:

$\bullet$ More supersymmetric cases

It is obvious that when we further restrict the level for $SU(N+M)_k$
to be the
dual Coxeter number of $SU(N+M)$,
there arises an enhancement of the ${\cal N}=3$ supersymmetry \cite{CHR1406}.
It would be interesting to see
how we can construct the relevant coset fields
explicitly by using the additional fermions.
Note that in the coset of (\ref{coset2}), there is no
$SU(M)_{k+N}$ factor in the denominator and this will lead to more general case
because we should add this factor into the one of \cite{CHR1406}.

$\bullet$ At the critical level with $M=2$

In this case (by fixing the $M$ value further in previous consideration),
we should check whether the original
${\cal N}=4$  supersymmetry is
enhanced to more supersymmetric cases or
not. The point is
how to obtain the additional supersymmetry generator by using
the additional
spin-$\frac{1}{2}$ fermions without spoiling the
present ${\cal N}=4$ supersymmetry generators.

$\bullet$ More OPEs between the ${\cal N}=2$ multiplets

So far we have considered the simplest OPEs in section $6$.
However, there are other OPEs between the ${\cal N}=2$ multiplets
we should consider.  
It is an open problem to study their OPEs systematically.

$\bullet$ Are there any ${\cal N}=2$ primary basis?

We have seen that the ${\cal N}=2$ superconformal algebra
we describe in this paper has the modified stress energy tensor.
It is an open problem to find whether there are any ${\cal N}=2$
primary basis where the description of
Appendix $B$ is satisfied or not. What happens if we use the modified
stress energy tensor rather than the stress energy tensor
$T$ (\ref{T})?

$\bullet$ How do we construct the higher spin algebra in the
$AdS_3$ gravity side via holography?

According to the result of \cite{CH1812} on the holography,
it is natural to ask whether we can construct the corresponding
${\cal N}=2$
higher spin algebra in the supersymmetric coset model in this paper.  
For $M=2$, there is a construction in \cite{AK2009}. For generic
$M$, it is an interesting problem to obtain them explicitly.
The additional $M$ dependence will play the role newly. 

$\bullet$ The orthogonal case

Once we have found the higher spin currents in the
bosonic orthogonal case, then its  supersymmetric version
can be obtained similarly.
If we use the embedding between the unitary and orthogonal
groups, then
maybe the $f$ and $d$ symbols in the unitary group
can be used.
The previous works \cite{Ahn1106,Ahn1202,AP1301,AP1310} can be used.
See also \cite{CHU}.

$\bullet$ Are there any systematical construction for the currents
having more than two free indices?

We have seen the new primary currents in the section $6$
having the two free indices. Then the question is how
we can determine the spin contents of this kind of primary currents.
As a first step, we should obtain the ${\cal N}=2$ description
for the above three spin-$\frac{5}{2},3$ currents. That is, can we
obtain the possible lowest (and highest) components explicitly? 

$\bullet$ Extension of the large ${\cal N}=4$ linear superconformal
algebra

We can think of the large ${\cal N}=4$ linear superconformal algebra
\cite{STVplb,npb1988,Schoutensnpb,Ivanov1,Ivanov2,Ivanov3,
Ivanov4,ST,Saulina}
by introducing the four spin-$\frac{1}{2}$ operators and the
spin-$1$ current with fixed $M=2$.
For the spin-$1$ current we can think of
some linear combination of the spin-$1$ operator
$J^{u(1)}$ and $J^{u(1)}_f$ appearing in the spin-$1$
current of (\ref{kggt}) as a candidate.
For the spin-$\frac{1}{2}$ operators, it is not clear how to
construct them by some contractions of the indices.
It would be interesting to
study further.

\vspace{.7cm}

\centerline{\bf Acknowledgments}

We
thank
Y. Hikida for the discussions.
This work was supported by
the National Research Foundation of Korea(NRF) grant
funded by the Korea government(MSIT)(No. 2020R1F1A1066893).

\newpage

\appendix

\renewcommand{\theequation}{\Alph{section}\mbox{.}\arabic{equation}}

\section{ The ${\cal N}=2$ superconformal algebra with a modified
stress energy tensor}

In this Appendix, we describe the ${\cal N}=2$ superconformal algebra
discussed in section $2$.

\subsection{The OPEs between the spin-$\frac{1}{2}$ operators and
the spin-$1$ operators}

By using (\ref{psipsi}) and (\ref{threeJf}) we obtain
\bea
\psi^{(\rho \bar{i})}(z) J_f^{\alpha}(w) &= & -\frac{1}{(z-w)} \,
t^{\alpha}_{\si \bar{\tau}} \, \de^{\rho \bar{\tau}}\,  \psi^{(\si \bar{i})}(w)
+ \cdots,
\nonu \\
\psi^{(\bar{\si} j)}(z) J_f^{\alpha}(w) &= & \frac{1}{(z-w)} \,
t^{\alpha}_{\tau \bar{\nu}} \, \de^{\tau \bar{\si}}\,  \psi^{(\bar{\nu} j)}(w)
+ \cdots,
\nonu \\
\psi^{(\rho \bar{i})}(z) J_f^{a}(w) &= & \frac{1}{(z-w)} \,
t^{a}_{ l \bar{k}} \, \de^{l \bar{i}}\,  \psi^{(\rho \bar{k})}(w)
+ \cdots,
\nonu \\
\psi^{(\bar{\si} j)}(z) J_f^{a}(w) &= & -\frac{1}{(z-w)} \,
t^{a}_{ l \bar{k}} \, \de^{j \bar{k}}\,  \psi^{(\bar{\si} l)}(w)
+ \cdots,
\nonu \\
\psi^{(\rho \bar{i})}(z) J_f^{u(1)}(w) &= & - \frac{1}{(z-w)} \,
\psi^{(\rho \bar{i})}(w) + \cdots,
\nonu \\
\psi^{(\bar{\si} j)}(z) J_f^{u(1)}(w) &= &  \frac{1}{(z-w)} \,
\psi^{(\bar{\si} j)}(w) + \cdots.
\nonu
\eea

\subsection{ The OPEs between the spin-$1$ operators}

Similarly, from the relations
(\ref{psipsi}) and (\ref{threeJf}),
the following OPEs satisfy
\bea
J^{\al}_f (z) \, J^{\beta}_f (w) & = & \frac{1}{(z-w)^2} \, M \de^{\al \beta}
+ \frac{1}{(z-w)} \, i \, f^{\al \beta \ga} \, J_f^{\ga}(w) +
\cdots,
\nonu \\
J^{\al}_f(z) \, J^{a}_f(w) & = &  0 + \cdots,
\nonu \\
J_f^{\al}(z) \, J^{u(1)}_f(w) & = & 0 + \cdots,
\nonu \\
J^{a}_f (z) \, J^{b}_f (w) & = & \frac{1}{(z-w)^2} \, N \, \de^{a b} +
\frac{1}{(z-w)}\, i \, f^{a b c} \, J_f^c(w) +
\cdots,
\nonu \\
J^{a}_f(z) \, J^{u(1)}_f(w) & = &  0 + \cdots,
\nonu \\
J_f^{u(1)}(z) \, J_f^{u(1)}(w) & = & \frac{1}{(z-w)^2}\, M N + \cdots.
  \nonu
\eea
We can read off various levels in these OPEs.

  \subsection{The OPEs between the spin-$\frac{1}{2}$ operators and
  the currents of ${\cal N}=2$ superconformal algebra }

  By using (\ref{T}), (\ref{gpm}), (\ref{K}) and (\ref{psipsi})
  we determine the following OPEs
  \bea
  \psi^{(\rho \bar{i})}(z) \, K(w) &= &
  \frac{1}{(z-w)} \, \frac{k}{(k+M+N)}\, \psi^{(\rho \bar{i})}(w)
  +\cdots,
  \nonu \\
   \psi^{(\bar{\si} j)}(z) \, K(w) &= &
  -\frac{1}{(z-w)} \, \frac{k}{(k+M+N)}\, \psi^{(\bar{\si} j)}(w)
  +\cdots,
  \nonu \\
   \psi^{(\rho \bar{i})}(z) \, G^{+}(w) &= &
   \frac{1}{(z-w)} \, J^{(\rho \bar{i})}(w) + \cdots,
   \nonu \\
   \psi^{(\bar{\si} j)}(z) \, G^{+}(w) &= & 0 + \cdots,
   \nonu \\
     \psi^{(\rho \bar{i})}(z) \, G^{-}(w) &= &
   0 + \cdots,
   \nonu \\
   \psi^{(\bar{\si} j)}(z) \, G^{-}(w) &= &   \frac{1}{(z-w)} \,
   J^{(\bar{\si} j)}(w)+ \cdots,
   \nonu \\
 \psi^{(\rho \bar{i})}(z) \, T(w) & = &  \frac{1}{(z-w)^2}
    \, \frac{(k M+M^2-1)}{2 M (k+M+N)} \,
    \psi^{(\rho \bar{i})}(w) \nonu \\
    & + &
    \frac{1}{(z-w)} \, \Bigg[
    \frac{1}{(k+M+N)}  \,
     t^{\al}_{\si \bar{\tau}}\,
      \de^{\rho \bar{\tau}}\,  J^{\al} \, \psi^{(\si \bar{i} )}  \nonu \\
      & + &   \frac{1}{(k+M+N)} 
       \,   t^{\al}_{\si \bar{\tau}}\,
      \de^{\rho \bar{\tau}}\, \psi^{(\bar{\si} i)} \, J_f^{\al}
      +
 \frac{1}{(k+M+N)}\, \sqrt{\frac{M+N}{M N}}\,
      \,  J^{u(1)}\, \psi^{(\bar{\si} j)} \nonu \\
      &+&
      \frac{1}{(k+M+N)}\, \frac{(M+N)}{M N}
  \, \psi^{(\rho \bar{i})} \,  J_f^{u(1)}
   - \frac{(k M+M^2-1)}{2 M (k+M+N)}  \pa \,  \psi^{(\rho \bar{i})}
      \Bigg](w) + \cdots, 
    \nonu \\
 \psi^{(\bar{\si} j)}(z) \, T(w) & = &  \frac{1}{(z-w)^2}
    \, \frac{(k M+M^2-1)}{2 M (k+M+N)} \,
    \psi^{(\bar{\si} j)}(w) \nonu \\
    & + &
    \frac{1}{(z-w)} \, \Bigg[
    -\frac{1}{(k+M+N)}  \,
     t^{\al}_{\tau \bar{\nu}}\,
      \de^{\tau \bar{\si}}\,  J^{\al} \, \psi^{(\bar{\nu} j )}  \nonu \\
      & - &   \frac{1}{(k+M+N)} 
       \,   t^{\al}_{\tau \bar{\nu}}\,
      \de^{\tau \bar{\si}}\, \psi^{(\bar{\nu} j)} \, J_f^{\al}
      -
 \frac{1}{(k+M+N)}\, \sqrt{\frac{M+N}{M N}}\,
      \,  J^{u(1)}\, \psi^{(\bar{\si} j)} \nonu \\
      &-&
      \frac{1}{(k+M+N)}\, \frac{(M+N)}{M N}
  \, \psi^{(\bar{\si} j)} \,  J_f^{u(1)}
   - \frac{(k M+M^2-1)}{2 M (k+M+N)}  \pa \,  \psi^{(\bar{\si} j)}
      \Bigg](w) + \cdots.
    \nonu
  \eea
  
\subsection{The OPEs between the spin-$1$ operators and
  the currents of ${\cal N}=2$ superconformal algebra}

We have some OPEs between
 the spin-$1$ operators and
  the currents of ${\cal N}=2$ superconformal algebra as follows:
\bea
J_f^{u(1)}(z) \, K(w) & = &
- \frac{1}{(z-w)^2}\,
   \frac{k M N}{(k+M+N)} +\cdots,\nonu \\
J_f^{u(1)}(z) \, G^{+}(w) & = &
- \frac{1}{(z-w)}\,
  G^{+} + \cdots,
  \nonu
  \\
  J_f^{u(1)}(z) \, G^{-}(w) & = &
 \frac{1}{(z-w)}\,
  G^{-} + \cdots,
  \nonu \\
   J_f^{u(1)}(z) \, T(w) & = & -\frac{1}{(z-w)^2} \,
   K(w)+ \cdots,
   \nonu  \\
   J^{u(1)}(z) \, K(w) & = & \frac{1}{(z-w)^2} \,
   \frac{k M N}{(k+M+N)} \, \sqrt{\frac{M+N}{M N}} + \cdots,
\nonu \\
J^{u(1)}(z) \, G^{+}(w) & = &
\frac{1}{(z-w)}\, \sqrt{\frac{M+N}{M N} }\,
  G^{+} + \cdots,
  \nonu \\
  J^{u(1)}(z) \, G^{-}(w) & = &
-\frac{1}{(z-w)}\, \sqrt{\frac{M+N}{M N} }\,
  G^{-} + \cdots,
  \nonu \\
 J^{u(1)}(z) \, T(w) & = & \frac{1}{(z-w)^2} \,
  \sqrt{\frac{M+N}{M N} }\, K(w)+ \cdots,
  \nonu \\
  J^{(\rho \bar{i})}(z) \, K(w) & = & -\frac{1}{(z-w)} \,  \frac{(M+N)}{(
    k+M+N)}\, 
  J^{(\rho \bar{i})}(w) +\cdots,
  \nonu \\
  J^{(\rho \bar{i})}(z) \, G^{+}(w) & = & 0 + \cdots,
  \nonu \\
  J^{(\rho \bar{i})}(z) \, G^{-}(w) & = & \frac{1}{(z-w)^2} \,
  k \, \psi^{(\rho \bar{i})}(w) \nonu \\
  & +&  \frac{1}{(z-w)} \,
  \Bigg[ \sqrt{\frac{M+N}{M N}} \, \psi^{(\rho \bar{i})}\, J^{u(1)}+
 t^{\al}_{\si \bar{\rho}}\,
  \de^{\rho \bar{\rho}}\, \psi^{(\si \bar{i})} \,  J^{\al}
    - t^{a}_{i \bar{k}}\,
    \de^{i \bar{i}}\, \psi^{(\rho \bar{k})} \,  J^{a} \Bigg](w) \nonu \\
  & + & \cdots,
  \nonu \\
  J^{(\rho \bar{i})}(z) \, T(w) & = &  \frac{1}{(z-w)^2}
    \, \Bigg[ \frac{2 k M+2 M^2+M N-1}{2 M (k+M+N)} \Bigg] \,
    J^{(\rho \bar{i})}(w) \nonu \\
    & + &
    \frac{1}{(z-w)} \, \Bigg[
      \frac{1}{(k+M+N)}\, \sqrt{\frac{M+N}{M N}}\,
      J^{(\rho \bar{i})} \, J^{u(1)} \nonu \\
      & + & 
      \frac{(M+N)}{M N (k+M+N)} \,   J^{(\rho \bar{i})} \, J_f^{u(1)}
      + \frac{1}{(k+M+N)} \, t^{\al}_{\si \bar{\rho}}\,
      \de^{\rho \bar{\rho}}\, J^{\al} \, J^{(\si \bar{i})} \nonu \\
      &+&
  \frac{1}{(k+M+N)}\,  t^{\al}_{\si \bar{\rho}}\,
  \de^{\rho \bar{\rho}}\, J^{(\si \bar{i})} \,  J_f^{\al}
  -\frac{(M N^2-2 M-N)}{2 M N (k+M+N)} \, \pa \,  J^{(\rho \bar{i})}
  \Bigg](w) \nonu \\
    & + & \cdots, 
    \nonu \\
    J^{(\bar{\si} j)}(z) \, K(w) & = & \frac{1}{(z-w)} \,
    \frac{(M+N)}{(
    k+M+N)}\, 
  J^{(\bar{\si} j)}(w) +\cdots,
    \nonu \\
    J^{(\bar{\si} j)}(z) \, G^{+}(w) & = &
\frac{1}{(z-w)^2} \,
  k \, \psi^{(\bar{\si} j)}(w) \nonu \\
  & +&  \frac{1}{(z-w)} \,
  \Bigg[ -\sqrt{\frac{M+N}{M N}} \, \psi^{(\bar{\si} j)}\, J^{u(1)}-
 t^{\al}_{\rho \bar{\tau}}\,
  \de^{\rho \bar{\si}}\, \psi^{(\bar{\tau} j)} \,  J^{\al}
    + t^{a}_{k \bar{k}}\,
    \de^{j \bar{k}}\, \psi^{(\bar{\si} k)} \,  J^{a} \Bigg](w) \nonu \\
  & + & \cdots,
  \nonu \\
  J^{(\bar{\si} j)}(z) \, G^{-}(w) & = &  0 +\cdots,
  \nonu \\
    J^{(\bar{\si} j)}(z) \, T(w) & = &  \frac{1}{(z-w)^2}
    \, \Bigg[ \frac{2 k M+2 M^2+M N-1}{2 M (k+M+N)} \Bigg] \,
    J^{(\bar{\si} j)}(w) \nonu \\
    & + &
    \frac{1}{(z-w)} \, \Bigg[
      -\frac{1}{(k+M+N)}\, \sqrt{\frac{M+N}{M N}}\,
      J^{(\bar{\si} j)} \, J^{u(1)} \nonu \\
      & - & 
      \frac{(M+N)}{M N (k+M+N)} \,   J^{(\bar{\si} j)} \, J_f^{u(1)}
      - \frac{1}{(k+M+N)} \, t^{\al}_{\si \bar{\tau}}\,
      \de^{\si    \bar{\si}}\, J^{\al} \, J^{(\bar{\tau} j)} \nonu \\
      &-&
  \frac{1}{(k+M+N)}\,  t^{\al}_{\si \bar{\tau}}\,
  \de^{\si \bar{\si}}\, J^{(\bar{\tau} j)} \,  J_f^{\al}
  -\frac{(M N^2-2 M-N)}{2 M N (k+M+N)} \, \pa \,  J^{(\bar{\si} j)}
      \Bigg](w) + \cdots, 
    \nonu \\
    J^a_f(z) \, K(w) & = & 0 +\cdots,
    \nonu \\
    J^a_f(z) \, G^{+}(w) & = & -\frac{1}{(z-w)} \, G^{+,a}(w) +\cdots,
    \nonu \\
     J^a_f(z) \, G^{-}(w) & = & \frac{1}{(z-w)} \, G^{-,a}(w) +\cdots,
     \nonu \\
     J^{a}_f(z) \, T(w) & = & \frac{1}{(z-w)^2}\, J^a_f(w)   +\cdots,
     \nonu \\
      J^{\al}_f(z) \, K(w) & = & 0 +\cdots,
    \nonu \\
    J^{\al}_f(z) \, G^{+}(w) & = & -\frac{1}{(z-w)} \, t^{\al}_{\rho \bar{\nu}}
    \, \de_{j \bar{i}} \, J^{(\rho \bar{i})}\, \psi^{(\bar{\nu} j)}(w)+\cdots,
    \nonu \\
    J^{\al}_f(z) \, G^{-}(w) & = &
    \frac{1}{(z-w)} \, t^{\al}_{\si \bar{\si}}
    \, \de_{j \bar{i}} \,  \psi^{(\si \bar{i})}\, J^{(\bar{\si} j)}\,(w)
    +\cdots,
     \nonu \\
     J^{\al}_f(z) \, T(w) & = & \frac{1}{(z-w)^2} \, 
     \frac{1}{(k+M+N)} \, \Bigg[  k \, J_f^{\al} - M \, J^{\al}
       \Bigg](w)\nonu \\
     & - & 
     \frac{1}{(z-w)} \,
\frac{1}{(k+M+N)} \, i \,
     f^{\al \beta \ga} \, J^{\beta} \, J_f^{\ga}(w)+
         \cdots.
    \nonu 
\eea

We can calculate the following OPEs 
\bea
 J_f^{u(1)}(z) \, 
\de_{\rho \bar{\si}} \, \de_{j \bar{i}} \,
\psi^{(\rho \bar{i})} \, \pa\, \psi^{(\bar{\si} j)}(w) &=&
\frac{1}{(z-w)^3} \, M N - \frac{1}{(z-w)^2} \, J_f^{u(1)}(w) +
\cdots,
\nonu \\
 J_f^{u(1)}(z) \, 
\de_{\rho \bar{\si}} \, \de_{j \bar{i}} \,
\pa \, \psi^{(\rho \bar{i})} \,  \psi^{(\bar{\si} j)}(w) &=&
\frac{1}{(z-w)^3} \, M N + \frac{1}{(z-w)^2} \, J_f^{u(1)}(w) +
\cdots,
\nonu \\
J_f^{u(1)}(z) \, J_f^{u(1)} \, J_f^{u(1)}(w) & = & \frac{1}{(z-w)^2}\,
2 M N \, J_f^{u(1)}(w)+ \cdots.
\nonu
\eea

\subsection{The ${\cal N}=2$ superconformal algebra}

As before, we determine the following OPEs from
(\ref{T}), (\ref{gpm}) and (\ref{K})
\bea
K(z) \, K(w) & = & \frac{1}{(z-w)^2} \, \frac{k M N}{(k+M+N)} + \cdots,
\nonu\\
K(z) \, G^{+}(w) & = &
\frac{1}{(z-w)} \, G^{+}(w) + \cdots,
\nonu \\
K(z) \, G^{-}(w) & = &
-\frac{1}{(z-w)} \, G^{-}(w) + \cdots,
\nonu \\
K(z) \, T(w) &= & \frac{1}{(z-w)^2} \, K(w) + \cdots,
\nonu \\
G^{+}(z) \, G^{+}(w) &= & 0 + \cdots,
\nonu \\
G^{+}(z) \, G^{-}(w) & = & \frac{1}{(z-w)^3}\, k M N+
\frac{1}{(z-w)^2}\, (k+M+N) \, K(w) \nonu \\
& + &
\frac{1}{(z-w)}\, \Bigg[ \frac{1}{2}\,  (k+M+N) \, \pa\,
  K \nonu \\
  & + & (k +M+N) \, \Bigg( T -\frac{1}{2(k+M+N)} (J^a+J^a_f)(J^a+J^a_f)
  \Bigg) \Bigg](w) + \cdots,
\nonu \\
G^{+}(z) \, T(w) & = & \frac{1}{(z-w)^2} \, \frac{3}{2}\, G^{+}(w)
+\frac{1}{(z-w)} \, \frac{1}{2}\, \pa \, G^{+}(w)
+ \cdots,
\nonu \\
G^{-}(z) \, T(w) & = & \frac{1}{(z-w)^2} \, \frac{3}{2}\, G^{-}(w)
+\frac{1}{(z-w)} \, \frac{1}{2}\, \pa \, G^{-}(w)
+ \cdots,
\nonu \\
T(z) \, T(w) & = &
\frac{1}{(z-w)^4} \, \frac{1}{2} \,
\frac{(k M^2+3 k M N-k+M^2 N-N)}{(k+M+N)}
\nonu \\
& + & \frac{1}{(z-w)^2}\, 2 T(w) +
\frac{1}{(z-w)}\, \pa T(w) + \cdots.
\nonu 
\eea  
The central term in the OPE of $K(z) \, K(w)$ is the same as
the standard expression $\frac{c}{3}$ for only $M=1$ value.
We can multiply $\frac{1}{(k+M+N)}$ in the OPE of
$G^{+}(z)\, G^{-}(w)$ in order to have the same central term
as the one of the OPE of $K(z) \, K(w)$.

In obtaining  the above result, we observe that
there exist the relations
\bea
    J^{\al}_f \, J^{\al}_f & = & t^{\al}_{\rho \bar{\si}} \, t^{\al}_{\tau \bar{\nu}}\,
    \de_{j \bar{i}} \, \de_{l \bar{k}}
    \psi^{(\rho \bar{i})} \, \psi^{(\bar{\si} j)} \, \psi^{(\tau \bar{k})}
    \, \psi^{(\bar{\nu} l)} + \frac{(N^2-1)}{N} \, 
\de_{\rho \bar{\si}} \, \de_{j \bar{i}} \,
\pa \, \psi^{(\rho \bar{i})} \,  \psi^{(\bar{\si} j)}
\nonu \\
&-&\frac{(N^2-1)}{N} \, 
\de_{\rho \bar{\si}} \, \de_{j \bar{i}} \,
 \, \psi^{(\rho \bar{i})} \,  \pa \, \psi^{(\bar{\si} j)},
 \nonu \\
 J_f^{u(1)} \, J_f^{u(1)} & = &
 \de_{\rho \bar{\si}} \, \de_{\tau \bar{\nu}}\,
    \de_{j \bar{i}} \, \de_{l \bar{k}}
    \psi^{(\rho \bar{i})} \, \psi^{(\bar{\si} j)} \, \psi^{(\tau \bar{k})}
    \, \psi^{(\bar{\nu} l)} +  
\de_{\rho \bar{\si}} \, \de_{j \bar{i}} \,
\pa \, \psi^{(\rho \bar{i})} \,  \psi^{(\bar{\si} j)}
\nonu \\
&-&
\de_{\rho \bar{\si}} \, \de_{j \bar{i}} \,
 \, \psi^{(\rho \bar{i})} \,  \pa \, \psi^{(\bar{\si} j)},
 \nonu \\
  J^{a}_f \, J^{a}_f & = & 
  t^a_{j \bar{i}} \, t^a_{l \bar{k}} \,
  \de_{\rho \bar{\si}} \, \de_{\tau \bar{\nu}}\,
    \psi^{(\rho \bar{i})} \, \psi^{(\bar{\si} j)} \, \psi^{(\tau \bar{k})}
    \, \psi^{(\bar{\nu} l)} + \frac{(M^2-1)}{M} \, 
\de_{\rho \bar{\si}} \, \de_{j \bar{i}} \,
\pa \, \psi^{(\rho \bar{i})} \,  \psi^{(\bar{\si} j)}
\nonu \\
&-&\frac{(M^2-1)}{M} \, 
\de_{\rho \bar{\si}} \, \de_{j \bar{i}} \,
 \, \psi^{(\rho \bar{i})} \,  \pa \, \psi^{(\bar{\si} j)}.
 \label{threejf}
  \eea

  We can check the following identity
  between the spin-$1$ currents in (\ref{threeJf})
  by using (\ref{threejf})
  \bea
&&  J^{\al}_f \, J^{\al}_f + (\frac{1}{M} + \frac{1}{N})\,
  J_f^{u(1)} \, J_f^{u(1)} +  J^{a}_f \, J^{a}_f =
  \nonu \\
  && (M+N) \,
  (\de_{\rho \bar{\si}} \, \de_{j \bar{i}} \,
\pa \, \psi^{(\rho \bar{i})} \,  \psi^{(\bar{\si} j)}
-
\de_{\rho \bar{\si}} \, \de_{j \bar{i}} \,
 \, \psi^{(\rho \bar{i})} \,  \pa \, \psi^{(\bar{\si} j)}).
  \nonu
  \eea 

\section{The ${\cal N}=2$ primary conditions}

The ${\cal N}=2$ primary conditions for the
currents $(\Phi^{-(h)}, \Phi^{+(h+\frac{1}{2})}, \Phi^{-(h+\frac{1}{2})},
\Phi^{+(h+1)})$ are summarized by
\bea
K(z) \, \Phi^{-(h)}(w) & = & 0 + \cdots,
\nonu \\
K(z) \, \Phi^{+(h+\frac{1}{2})}(w) & = & \frac{1}{(z-w)}\,
\Phi^{+(h+\frac{1}{2})}(w)+ \cdots,
\nonu \\
K(z) \, \Phi^{-(h+\frac{1}{2})}(w) & = & -\frac{1}{(z-w)}\,
\Phi^{-(h+\frac{1}{2})}(w)+ \cdots,
\nonu \\
K(z) \, \Phi^{+(h+1)}(w) & = & \frac{1}{(z-w)^2}\, h\, 
\Phi^{-(h)}(w)+ \cdots,
\nonu \\
G^+(z) \, \Phi^{-(h)}(w) & = & -\frac{1}{(z-w)}\,
\Phi^{+(h+\frac{1}{2})}(w) + \cdots,
\nonu \\
G^+(z) \, \Phi^{+(h+\frac{1}{2})}(w) & = & 0 + \cdots,
\nonu \\
G^+(z) \, \Phi^{-(h+\frac{1}{2})}(w) & = & \frac{1}{(z-w)^2}\,
h\, \Phi^{-(h)}(w)+\frac{1}{(z-w)}\,
\Bigg[\Phi^{+(h+1)}+ \frac{1}{2} \, \pa \, \Phi^{-(h)} \Bigg](w)+ \cdots,
\nonu \\
G^+(z) \, \Phi^{+(h+1)}(w) & = & \frac{1}{(z-w)^2}\, (h+\frac{1}{2})\, 
\Phi^{+(h+\frac{1}{2})}(w) + \frac{1}{(z-w)}\, \frac{1}{2}\, 
\pa \, \Phi^{+(h+\frac{1}{2})}(w) + \cdots,
\nonu \\
G^-(z) \, \Phi^{-(h)}(w) & = & \frac{1}{(z-w)}\,
\Phi^{-(h+\frac{1}{2})}(w) + \cdots,
\nonu \\
G^-(z) \, \Phi^{+(h+\frac{1}{2})}(w) & = & -\frac{1}{(z-w)^2}\,
h\, \Phi^{-(h)}(w)+\frac{1}{(z-w)}\,
\Bigg[\Phi^{+(h+1)}- \frac{1}{2} \, \pa \, \Phi^{-(h)} \Bigg](w)+ \cdots,
\nonu \\
G^-(z) \, \Phi^{-(h+\frac{1}{2})}(w) & = & 0 + \cdots,
\nonu \\
G^-(z) \, \Phi^{+(h+1)}(w) & = & \frac{1}{(z-w)^2}\, (h+\frac{1}{2})\, 
\Phi^{-(h+\frac{1}{2})}(w) + \frac{1}{(z-w)}\, \frac{1}{2}\, 
\pa \, \Phi^{-(h+\frac{1}{2})}(w) + \cdots,
\nonu \\
T(z) \, \Phi^{-(h)}(w) & = & \frac{1}{(z-w)^2}\, h\, 
\Phi^{-(h)}(w) +\frac{1}{(z-w)}\, 
\pa \, \Phi^{-(h)}(w) + \cdots,
\nonu \\
T(z) \, \Phi^{+(h+\frac{1}{2})}(w) & = & \frac{1}{(z-w)^2}\,
(h+\frac{1}{2}) \, \Phi^{+(h+\frac{1}{2})}(w)
+\frac{1}{(z-w)}\,
\pa \, \Phi^{+(h+\frac{1}{2})}(w)
+ \cdots,
\nonu \\
T(z) \, \Phi^{-(h+\frac{1}{2})}(w) & = &
\frac{1}{(z-w)^2}\,
(h+\frac{1}{2}) \, \Phi^{-(h+\frac{1}{2})}(w)
+\frac{1}{(z-w)}\,
\pa \, \Phi^{-(h+\frac{1}{2})}(w)+ \cdots,
\nonu \\
T(z) \, \Phi^{+(h+1)}(w) & = & \frac{1}{(z-w)^2}\, (h+1)\, 
\Phi^{+(h+1)}(w)
+ \frac{1}{(z-w)}\, 
\pa \, \Phi^{+(h+1)}(w)
+ \cdots.
\nonu
\eea
In this paper, we observe that
the currents we are considering do not satisfy these
OPEs. That is, some of the higher order terms arise
and the structure constants appear differently.

\section{Some OPEs between the spin-$1$ currents and the
  singlet and nonsinglet spin-$3$ currents in sections $4$ and $5$
}

In order to obtain the OPEs between the supersymmetry
generators and the nonsinglet and singlet spin-$3$ currents
obtained in the bosonic coset model, we need to calculate the
following OPEs  between the spin-$1$ currents and the
  singlet and nonsinglet spin-$3$ currents.  

\subsection{The OPE between $J^{(\rho \bar{i})}(z) \, W^{(3)}(w)$}

The OPE between the spin-$1$ current and the spin-$3$ current
is summarized by
\bea
&& J^{(\rho \bar{i})}(z) \, W^{(3)}(w)  = 
\nonu \\
&& -\frac{1}{(z-w)^3} \,
\frac{ (k^2-1) (k^2-4) (k+M+N) (2 k+M+N) (3 k+2 M+2 N)}{
  k^2 M (k+M) (k+2 M)} \, b_1 \, J^{(\rho \bar{i})}(w)
\nonu \\
&& + \frac{1}{(z-w)^2} \frac{(k^2-4) (k+M+N)  ( 3 k+2 M+2 N)}{
  k M  (k+2 M)} \, b_1 \nonu \\
&& \times \, \Bigg[-\sqrt{\frac{M+N}{M N}}\, \frac{
 \, 3 (k+N)   }{
  k}  \, J^{u(1)} \, J^{(\rho \bar{i})}  
  + \frac{3(k+N)}{(k+M)} \,  t^a_{i \bar{k}} \, \de^{i \bar{i}}\,
  J^a \, J^{(\rho \bar{k})}
  -3 \,  t^{\alpha}_{\si \bar{\rho}} \, \de^{\rho \bar{\rho}}\, J^{\alpha}\,
   J^{(\si \bar{i})}
   \nonu \\
   && + \frac{(k^3+2 k^2 M+2 k^2 N+3 k M N+2 k+M+N)}{k(k+M)}
   \pa \,  J^{(\rho \bar{i})} \Bigg](w)\nonu \\
&&+ \frac{1}{(z-w)} \, \frac{  (k+M+N)}{M}\, b_1\, \Bigg[
  - 3\,
  d^{\alpha \beta \ga} \, t^{\ga}_{\si \bar{\rho}}\, \de^{\rho \bar{\rho}}\,
  J^{\alpha} \, J^{\beta}\, J^{(\si \bar{i})}
  \nonu \\
  && -\frac{3  (k+N) (k+2 N)}{ (k+M) (k+2 M)} \,
  d^{a b c}\,  t^{c}_{i \bar{k}}\, \de^{i \bar{i}}\,
  J^{a} \, J^{b}\, J^{(\rho \bar{k})}
-\frac{6  (k+N) (k+2 N) (M+N) }{k^2 M N}
J^{u(1)}\, J^{u(1)}\, J^{(\rho \bar{i})}
\nonu \\
&& -\frac{6  (k+N)}{k  N}\, J^{\alpha}\, J^{\alpha}\,  J^{(\rho \bar{i})}
-\sqrt{\frac{M+N}{M N}}\, \frac{12  (k+N) }{k } \,
 t^{\alpha}_{\si \bar{\si}}\, \de^{\rho \bar{\si}}\,
 J^{\alpha}\, J^{u(1)} \,  J^{(\si \bar{i})} \nonu \\
 &&
 -\frac{6  (k+N) (k+2 N) }{k M (k+2 M)}\, J^a \, J^a \,
 J^{(\rho \bar{i})}+\sqrt{\frac{M+N}{M N}}\,
 \frac{12  (k+N) (k+2 N)  }{k  (k+2 M)}\,
 t^a_{i \bar{k}}\, \de^{i\bar{i}}\, J^a \, J^{u(1)}\, J^{(\rho \bar{k})}
 \nonu \\
 && +
 \frac{12  (k+N) }{ (k+2 M)}\, \de_{\nu \bar{\tau}} \,
 \de_{k \bar{j}} \, J^{(\nu \bar{i})}\, J^{(\rho \bar{j})}\,J^{(\bar{\tau}
   k      )}+
 \frac{24  (k+N) }{k  (k+2 M)}\,
 \de_{\si \bar{\tau}} \,
 \de_{k \bar{j}} \, J^{(\rho \bar{i})}\, J^{(\si \bar{j})}\,J^{(\bar{\tau}
   k      )} \nonu \\
 && -\sqrt{\frac{M+N}{M N}}\, \frac{12  (k+N)  }{k }
 \,
t^{\al}_{\si \bar{\rho}} \,
 \de^{\rho \bar{\rho}} \, J^{\al}\, J^{u(1)}\,J^{(\si
   \bar{j}      )}
 +\frac{12  (k+N) }{ (k+2 M)}\, t^{a}_{i \bar{j}} \,
 \de^{i \bar{i}} \,
t^{\al}_{\si \bar{\rho}} \,
 \de^{\rho \bar{\rho}} \,
 J^{\al}\, J^a \, J^{(\si
   \bar{j}      )}
 \nonu \\
 &&
 + \frac{3  (k^2+2 k N+4) }{k }\,
 t^{\al}_{\tau \bar{\rho}} \,
 \de^{\rho \bar{\rho}} \,
 J^{\al}\, \pa \, J^{(\tau \bar{i})} 
 -\frac{3 (k^2+2 k M+4) (k+N) (k+2 N) }{k  (k+M) (k+2 M)}
 \,
 t^{a}_{i \bar{l}} \,
 \de^{i \bar{i}} \,
 J^{a}\, \pa \, J^{(\rho \bar{l})} 
 \nonu \\
 && + \sqrt{\frac{M+N}{M N}} \,
\frac{3  (k+N) (k^3+2 k^2 M+2 k^2 N+8 k M N+4 k+8 M+8 N)}{k^2  (k+2 M)}
\, J^{u(1)}\, \pa \,   J^{(\rho \bar{i})}
\nonu \\
&& -3 \, (k+N) \,  t^{\al}_{\tau \bar{\rho}} \,
 \de^{\rho \bar{\rho}} \,
 \pa \, J^{\al} \, J^{(\tau \bar{i})} 
 + \frac{3  (k+N) (k+2 N) }{ (k+2 M)}\,
  t^{a}_{i \bar{j}} \,
 \de^{i \bar{i}} \,
 \pa \, J^{a} \, J^{(\rho \bar{j})} \nonu \\
 &&
 - \sqrt{\frac{M+N}{M N}}\, \frac{3  (k+N) (k+2 N) }{k }
 \,
 \pa \, J^{u(1)}\,  J^{(\rho \bar{i})} \nonu \\
 &&
 -\frac{(k^2-4) }{2 k^2  (k+M) (k+2 M)}\,
 (k^3 M+k^3 N+2 k^2 M^2-2 k^2 M N+2 k^2 N^2-6 k^2-7 k M\nonu \\
 && -7 k N-2 M^2-4 M N-2 N^2) \,
 \pa^2 \,  J^{(\rho \bar{i})}
 \Bigg](w) + \cdots.
\nonu
\eea

\subsection{The OPE between $J^{(\bar{\si} j)}(z) \, W^{(3)}(w)$}

Similarly,
the OPE between the other spin-$1$ current and the spin-$3$ current
is summarized by
\bea
&& J^{(\bar{\si} j)}(z) \, W^{(3)}(w)  = 
\nonu \\
&& \frac{1}{(z-w)^3} \,
\frac{ (k^2-1) (k^2-4) (k+M+N) (2 k+M+N) (3 k+2 M+2 N)}{
  k^2 M (k+M) (k+2 M)} \, b_1 \, J^{(\bar{\si} j)}(w)
\nonu \\
&& + \frac{1}{(z-w)^2} \frac{(k^2-4) (k+M+N)  ( 3 k+2 M+2 N)}{
  k M  (k+2 M)} \, b_1 \nonu \\
&& \times \, \Bigg[-\sqrt{\frac{M+N}{M N}}\, \frac{
 \, 3 (k+N)   }{
  k}  \, J^{u(1)} \, J^{(\bar{\si} j)}  
  + \frac{3(k+N)}{(k+M)} \,  t^a_{l \bar{j}} \, \de^{j \bar{j}}\,
  J^a \, J^{(\bar{\si} l)}
  -3 \,  t^{\al}_{\rho \bar{\tau}} \, \de^{\rho \bar{\rho}}\, J^{\alpha}\,
   J^{(\bar{\tau} j)}
   \nonu \\
   && - \frac{(k^3+2 k^2 M+2 k^2 N+3 k M N+2 k+M+N)}{k(k+M)}
   \pa \,  J^{(\bar{\si} j)} \Bigg](w)\nonu \\
&&+ \frac{1}{(z-w)} \,\frac{ (k+M+N)}{M}\, b_1\,  \Bigg[
  3 \,
  d^{\alpha \beta \ga} \, t^{\ga}_{\rho \bar{\tau}}\, \de^{\rho \bar{\si}}\,
   J^{\alpha} \, J^{\beta}\, J^{(\bar{\tau} j)}
  \nonu \\
  && \, +\frac{3  (k+N) (k+2 N) }{ (k+M) (k+2 M)}\,
  d^{a b c}\,  t^{c}_{k \bar{j}}\, \de^{j \bar{j}}\,
  J^{a} \, J^{b}\, J^{(\bar{\si} k)}
+\frac{6  (k+N) (k+2 N) (M+N) }{k^2 M N}
J^{u(1)}\, J^{u(1)}\, J^{(\bar{\si} j)}
\nonu \\
&&\, + \frac{6  (k+N) }{k  N}\,
J^{\alpha}\, J^{\alpha}\,  J^{(\bar{\si} j)}
+ \sqrt{\frac{M+N}{M N}}\, \frac{12  (k+N)  }{k }\,
 t^{\alpha}_{\si \bar{\tau}}\, \de^{\si \bar{\si}}\,
 J^{\alpha}\, J^{u(1)} \,  J^{(\bar{\tau} j)} \nonu \\
 &&
+\frac{6  (k+N) (k+2 N) }{k M (k+2 M)}\, J^a \, J^a \,
J^{(\bar{\si} j)}
-\sqrt{\frac{M+N}{M N}}\,
\frac{12  (k+N) (k+2 N)  }{k  (k+2 M)}
\,
 t^a_{k \bar{j}}\, \de^{j\bar{j}}\, J^a \, J^{u(1)}\, J^{(\bar{\si} k)}
 \nonu \\
 && -\frac{12  (k+N) }{ (k+2 M)}
 \, \de_{\si \bar{\nu}} \,
 \de_{k \bar{j}} \, J^{(\si \bar{j})}\, J^{(\bar{\nu} j)}\,J^{(\bar{\si}
   k)}-\frac{24  (k+N) }{k  (k+2 M)}
 \,
 \de_{\si \bar{\tau}} \,
 \de_{k \bar{j}} \, J^{(\si \bar{j})}\, J^{(\bar{\si} j)}\,J^{(\bar{\tau}
   k      )} \nonu \\
 && +\sqrt{\frac{M+N}{M N}}\,
 \frac{12  (k+N)  }{k }
 \,
t^{\al}_{\si \bar{\tau}} \,
 \de^{\si \bar{\si}} \, J^{\al}\, J^{u(1)}\,J^{(\bar{\tau}
   j      )}
 -\frac{12  (k+N) }{ (k+2 M)}\, t^{a}_{k \bar{j}} \,
 \de^{j \bar{j}} \,
t^{\al}_{\si \bar{\tau}} \,
 \de^{\si \bar{\si}} \,
 J^{\al}\, J^a \, J^{(\bar{\tau}
   k      )}
 \nonu \\
 &&
 +\frac{3  (k^2+2 k N+4) }{k }\,
 t^{\al}_{\rho \bar{\tau}} \,
 \de^{\rho \bar{\si}} \,
 J^{\al}\, \pa \, J^{(\bar{\tau} j)} 
 -\frac{3 (k^2+2 k M+4) (k+N) (k+2 N) }{k  (k+M) (k+2 M)}
 \,
 t^{a}_{k \bar{j}} \,
 \de^{j \bar{j}} \,
 J^{a}\, \pa \, J^{(\bar{\si} k)} 
 \nonu \\
 && +
\sqrt{\frac{M+N}{M N}}\,
 \frac{3  (k+N)  (k^3+2 k^2 M+2 k^2 N+8 k M N+4 k+8 M+8 N)}{k^2  (k+2 M)}
\, J^{u(1)}\, \pa \,   J^{(\bar{\si} j)}
\nonu \\
&&  -\frac{3 (k^2-2 k M-8) (k+N) }{k  (k+2 M)}\,
t^{\al}_{\rho \bar{\tau}} \,
 \de^{\rho \bar{\si}} \,
 \pa \, J^{\al} \, J^{(\bar{\tau} j)} 
 -\frac{3  (k+N) (-k^2+2 k N+8) }{k  (k+2 M)}\,
  t^{a}_{k \bar{j}} \,
 \de^{j \bar{j}} \,
 \pa \, J^{a} \, J^{(\bar{\si} k)} \nonu \\
 &&
 -\sqrt{\frac{M+N}{M N}}\, \frac{3  (k+N)  (k+M+N)
   (k^2-2 k M-2 k N-4 M N-8)}{k  (k+2 M)}
 \,
 \pa \, J^{u(1)}\,  J^{(\bar{\si} j)} \nonu \\
 &&
 + \frac{1}{2 k^2  (k+M) (k+2 M)} \,
 (k^5 M+k^5 N+2 k^4 M^2+10 k^4 M N+2 k^4 N^2+6 k^4+12 k^3 M^2 N
 \nonu \\
 && +12 k^3 M N^2+25 k^3 M+25 k^3 N+12 k^2 M^2 N^2+14 k^2 M^2+
 64 k^2 M N+
 14 k^2 N^2+24 k^2\nonu \\
 && +24 k M^2 N+24 k M N^2+28 k M+28 k N+8 M^2+16 M N+8 N^2)\,
 \pa^2 \,  J^{(\bar{\si} j)}
  \Bigg](w) + \cdots.
\nonu
\eea

\subsection{The OPE between $J^{(\rho \bar{i})}(z) \, P^{a}(w)$}

The OPE between the spin-$1$ current and the other spin-$3$ current
is summarized by
\bea
&& J^{(\rho \bar{i})}(z) \, P^{a}(w) =\nonu \\
&& \frac{1}{(z-w)^3} \,
\frac{  (k^2-1)  (k^2-4) (2 k+M+N) (3 k+2 M+2 N)}{
  2 k^2 (k+M) (3 k+2 M)} \, a_1 \,
t^a_{i \bar{k}} \, \de^{i \bar{i}} \, J^{(\rho \bar{k})}(w)
\nonu \\
&&+ \frac{1}{(z-w)^2} \, \frac{  (k^2-4)(3 k+2 M+2 N)}{2 k (k+M)}\,a_1\,
\Bigg[ \sqrt{\frac{M+N}{M N}}\,
  \frac{  (k+N)
    }{ k} \, t^a_{i \bar{k}} \, \de^{i \bar{i}} \, J^{u(1)} \, J^{(\rho \bar{k})}
  \nonu \\
  &&
  -\frac{    (3 k^2+3 k M+3 k N+M^2+M N)}{
     k M  (3 k+2 M)}\, J^a \, J^{(\rho \bar{i})}
  +  t^{\al}_{\si \bar{\rho}} \, \de^{\rho \bar{\rho}}\,
  t^a_{i \bar{k}} \, \de^{i \bar{i}}\, J^{\al}\, J^{(\si \bar{k})}
  \nonu \\
  &&
  +\frac{ (-k-M+N) }{2  (3 k+2 M)}\, i\, f^{b a c} \, t^c_{i \bar{k}}\,
  \de^{i \bar{i}}\, J^b \, J^{(\rho \bar{k})}
  -\frac{ (3 k+M+3 N)}{2  (3 k+2 M)}\,
   d^{ a b c} \, t^c_{i \bar{k}}\,
  \de^{i \bar{i}}\, J^b \, J^{(\rho \bar{k})}
  \nonu \\
  &&
  -\frac{ (k^3+k^2 M+2 k^2 N+2 k M N+2 k+M+N)}{ k (3 k+2 M)}\,
  t^a_{i \bar{k}} \, \de^{i \bar{i}}\,  \pa \, J^{(\rho \bar{k})}
  \Bigg](w)\nonu \\
&& +   \frac{1}{(z-w)} \,J^{(\rho \bar{i})}(z) \,
P^{a}(w)\Bigg|_{\frac{1}{(z-w)}}   + \cdots.
\nonu
\eea
The first order pole above is given by
\bea
&& J^{(\rho \bar{i})}(z) \, P^{a}(w)\Bigg|_{\frac{1}{(z-w)}} =
-a_1 \, t^{a}_{k \bar{j}}\, \de_{\nu \bar{\tau}}\,
J^{(\rho \bar{j})}\, J^{(\nu \bar{i})}\, J^{(\bar{\tau} k)}
+\frac{a_1}{N}\,
 t^{a}_{k \bar{j}}\, \de_{\si \bar{\tau}}\,
J^{(\si \bar{j})}\, J^{(\rho \bar{i})}\, J^{(\bar{\tau} k)}
\nonu \\
&& + (\sqrt{\frac{M+N}{M N}} \,a_1+2 \, a_{7})\,
 t^{\al}_{\si \bar{\rho}}\, \de^{\rho \bar{\rho}}\,
 t^a_{i \bar{k}}\, \de^{i \bar{i}} \,
 J^{\al}\, J^{u(1)}\, J^{(\si \bar{k})}
 +(\frac{a_1}{N}+a_2)\, t^{a}_{i \bar{k}}\, \de^{i \bar{i}} \,
 J^{\al}\, J^{\al}\, J^{(\rho \bar{k})}
 \nonu \\
 && + \frac{a_1}{2}\, (i \, f+ d)^{\al \beta \ga}\,
 t^{\ga}_{\si \bar{\rho}}\, \de^{\rho \bar{\rho}}\,
 t^a_{i \bar{k}}\, \de^{i \bar{i}} \,
 J^{\al}\, J^{\beta}\, J^{(\si \bar{k})}
 +(-\frac{a_1}{M}-2 a_2+ 2 a_8)\,
  t^{\al}_{\si \bar{\rho}}\, \de^{\rho \bar{\rho}} \,
 J^{\al}\, J^{a}\, J^{(\si \bar{i})}
 \nonu \\
 &&- \frac{a_1}{2}\,
  (i \, f+ d)^{b a c}\,
 t^{\al}_{\si \bar{\rho}}\, \de^{\rho \bar{\rho}}\,
 t^c_{i \bar{k}}\, \de^{i \bar{i}} \,
 J^{\al}\, J^{b}\, J^{(\si \bar{k})}
 +(a_3 + \frac{6}{M}\, a_{17})\,
  t^{a}_{i \bar{k}}\, \de^{i \bar{i}} \,
 J^{b}\, J^{b}\, J^{(\rho \bar{k})}
 \nonu \\
 && +(2 a_3 - 2 a_8 + \frac{6}{M}\, a_{17})\,
  t^{b}_{i \bar{k}}\, \de^{i \bar{i}} \,
 J^{a}\, J^{b}\, J^{(\rho \bar{k})}
 +(a_4 + 2 \,\sqrt{\frac{M+N}{M N}} \, a_7 )\,
  t^{a}_{i \bar{k}}\, \de^{i \bar{i}} \,
 J^{u(1)}\, J^{u(1)}\, J^{(\rho \bar{k})}
 \nonu \\
 && +(-2 \,\sqrt{\frac{M+N}{M N}} \,a_4-\frac{2}{M}\, a_7+2
 \sqrt{\frac{M+N}{M N}} \, a_8)
 \,J^{a}\, J^{u(1)}\, J^{(\rho \bar{i})}
 \nonu \\
 &&+ a_5 \, d^{a b c}\,
  t^{c}_{i \bar{k}}\, \de^{i \bar{i}}\,
 t^b_{k \bar{j}}\, \de_{\si \bar{\si}} \,
 J^{(\rho \bar{k})}\, J^{(\si \bar{j})}\, J^{( \bar{\si} k)}
 +  a_5 \, d^{a b c}\,
  t^{c}_{i \bar{k}}\, \de^{i \bar{i}}\,
 t^b_{k \bar{j}}\, \de_{\si \bar{\si}} \,
 J^{(\rho \bar{k})}\, J^{( \bar{\si} k)}  \, J^{(\si \bar{j})}
 \nonu \\
 && - \sqrt{\frac{M+N}{M N}} \, a_7 \,
   t^{a}_{k \bar{j}}\, 
  \de_{\si \bar{\tau}} \,
  J^{(\rho \bar{i})}\, J^{(\si \bar{j})}\,  J^{( \bar{\tau} k)}
  -\sqrt{\frac{M+N}{M N}} \, a_7 \,
   t^{a}_{k \bar{j}}\, 
  \de_{\si \bar{\tau}} \,
  J^{(\rho \bar{i})}\,  J^{( \bar{\tau} k)} \, J^{(\si \bar{j})}
  \nonu \\
  &&
  +( 2 \,\sqrt{\frac{M+N}{M N}} \, a_5 -a_7+2 \,
  a_9)\, d^{a b c}\,
  t^{b}_{i \bar{k}}\, \de^{i \bar{i}}\,
  J^{c}\, J^{u(1)}  \, J^{(\rho \bar{k})}
  + 2 \, a_5 \,  d^{a b c}\,
  t^{\al}_{\si \bar{\rho}} \, \de^{\rho \bar{\rho}}\,
  t^{b}_{i \bar{k}}\, \de^{i \bar{i}}\,
  J^{\al}\, J^c  \, J^{(\si \bar{k})}
    \nonu \\
    && +(-\frac{2}{M}\, a_5 -\sqrt{\frac{M+N}{M N}} \, a_9 )
    \,  d^{a b c}\,
    J^{c}\, J^{b}  \, J^{(\rho \bar{i})}
    -a_5\, (i \, f+d)^{e b d}\,
   d^{a b c}\, t^d_{i \bar{k}}\,\de^{i \bar{i}}\,
    J^{c}\, J^{e}  \, J^{(\rho \bar{k})}   
    \nonu \\
    &&+ a_7 \,  i \, f^{a b c}\, t^c_{i \bar{k}}\,
    \de^{i \bar{i}} \, J^{b}\, J^{u(1)}  \, J^{(\rho \bar{k})}   
+ a_8 \,  t^a_{i \bar{l}}\,
\de^{i \bar{i}} \, \de_{\si \bar{\tau}} \, \de_{k \bar{j}}\,
J^{(\rho \bar{l})}\, J^{(\si \bar{j})}  \, J^{(\bar{\tau} k)}  
\nonu \\
&&+a_8 \,  t^a_{i \bar{l}}\,
\de^{i \bar{i}} \, \de_{\si \bar{\tau}} \, \de_{k \bar{j}}\,
J^{(\rho \bar{l})}\, J^{(\bar{\tau} k)} \,  J^{(\si \bar{j})}
+ \frac{3}{2}\, a_{17}\,
(i \, f+d)^{a b e}\,
   d^{e c d}\, t^b_{i \bar{k}}\,
   \de^{i \bar{i}} \,  J^{c}\, J^{d}  \, J^{(\rho \bar{k})}   
   \nonu \\
   &&+
   \frac{3}{2}\, a_{17}\,
(i \, f+d)^{a d e}\,
   d^{e b c}\, t^b_{i \bar{k}}\,
   \de^{i \bar{i}} \,  J^{c}\, J^{d}  \, J^{(\rho \bar{k})}
 +
   \frac{3}{2}\, a_{17}\,
(i \, f+d)^{a c e}\,
   d^{e b d}\, t^b_{i \bar{k}}\,
   \de^{i \bar{i}} \,  J^{c}\, J^{d}  \, J^{(\rho \bar{k})}  
   \nonu \\
   &&+ \frac{6}{M}\, a_{17}\,
    t^b_{i \bar{k}}\,
\de^{i \bar{i}} \, 
J^{b}\, J^{a} \,  J^{(\rho \bar{k})}
+\Big(\frac{1}{2}\, (N -\frac{1}{N})\, a_2+
\frac{1}{2}\, (M -\frac{1}{M})\, a_3+
\frac{1}{2}\, \frac{M+N}{M N}\, a_4
\nonu \\
&& - \sqrt{\frac{M+N}{M N}} \,\frac{(M^2-4)}{2 M}\, a_9
-\frac{M}{2}\, a_{11}-\frac{N}{2}\, a_{12}+ a_{16}
+ \frac{18-6M^2+M^4}{2 M^2} \,a_{17} \Big)\,
\nonu \\
& & \times t^a_{i \bar{k}}\,
\de^{i \bar{i}} \,  \pa^2  \, J^{(\rho \bar{k})}
+(-2 \, a_2 + a_{12})\,
 t^{\al}_{\si \bar{\rho}}\,
\de^{\rho \bar{\rho}} \, t^a_{i \bar{k}} \, \de^{i \bar{i}}\,
J^{\al}\, \pa \,  J^{(\si \bar{k})} +
\Big( (N-\frac{1}{N})\, a_2 + \frac{(M+N)}{M N}\,
a_4 \nonu \\
&& -(M+N)\, a_8 -\frac{1}{M}\, a_{12}+ \frac{6(2M^2-3)}{
  M^2}\, a_{17} + (M+\frac{1}{M})\, a_3 
\Big)\, J^a \, \pa \, J^{(\rho \bar{i})}
\nonu \\
&&+(a_3 -2 \, a_3 -a_{11}+\frac{1}{2}\, a_{12})\,
 i\, f^{a b c} \, t^c_{i \bar{k}}\,
\de^{i \bar{i}} \,  J^b \, \pa  \, J^{(\rho \bar{k})}
\nonu \\
&& + \Big( a_3 - N\, a_5 -2 \, \sqrt{\frac{M+N}{M N}} \,a_9
-\frac{1}{2}\, a_{12}+ \frac{3(M^2-6)}{M} \, a_{17} \Big)\,
 d^{a b c} \, t^c_{i \bar{k}}\,
\de^{i \bar{i}} \,  J^b \, \pa  \, J^{(\rho \bar{k})}
\nonu \\
&& +(-2 \,a_3-a_{11}-\frac{1}{2}\, a_{13})\,
 i\, f^{a b c} \, t^b_{i \bar{k}}\,
\de^{i \bar{i}} \,  \pa \, J^c   \, J^{(\rho \bar{k})}
\nonu \\
&& +(-2\sqrt{\frac{M+N}{M N}} \,a_4 -N \, a_7 + \frac{(M^2-4)}{M}\,
a_9 + \sqrt{\frac{M+N}{M N}} \,a_{12})\,
 t^a_{i \bar{k}}\,
\de^{i \bar{i}}  \, J^{u(1)}\, \pa    \, J^{(\rho \bar{k})}
\nonu \\
&& +\sqrt{\frac{M+N}{M N}} \,a_{13} \,
 t^a_{i \bar{k}}\,
\de^{i \bar{i}}  \, \pa \, J^{u(1)}    \, J^{(\rho \bar{k})}
+a_{13}\, 
 t^{\al}_{\si \bar{\rho}}\,
\de^{\rho \bar{\rho}} \, t^a_{i \bar{k}} \, \de^{i \bar{i}}\,
\pa \, J^{\al} \,  J^{(\si \bar{k})}
-\frac{1}{M}\, a_{13}\, \pa \, J^{a} \,
J^{(\rho \bar{i})}
\nonu \\
&&-\frac{1}{2}\, a_{13}\, 
d^{a b c} \, t^b_{i \bar{k}}\,
\de^{i \bar{i}} \,  \pa \, J^c   \, J^{(\rho \bar{k})},
\nonu
\eea
where we do not substitute the coefficients appearing in
(\ref{avalues}) and then we can observe each contribution by
looking at each coefficient term.

\subsection{The OPE between $J^{(\bar{\si} j)}(z) \, P^{a}(w)$}

The OPE between the other spin-$1$ current and the
other spin-$3$ current
is summarized by
\bea
&& J^{(\bar{\si} j)}(z) \, P^{a}(w) =\nonu \\
&& -\frac{1}{(z-w)^3} \,
\frac{  (k^2-1)  (k^2-4) (2 k+M+N) (3 k+2 M+2 N)}{
  2 k^2 (k+M) (3 k+2 M)} \, a_1 \,
t^a_{k \bar{j}} \, \de^{j \bar{j}} \, J^{(\bar{\si} k)}(w)
\nonu \\
&&+ \frac{1}{(z-w)^2} \, \frac{  (k^2-4)(3 k+2 M+2 N)}{2 k (k+M)}\,a_1\,
\Bigg[ \sqrt{\frac{M+N}{M N}}\,
  \frac{  (k+N)
    }{ k} \, t^a_{k \bar{j}} \, \de^{j \bar{j}} \, J^{u(1)} \, J^{(\bar{\si} k)}
  \nonu \\
  &&
  -\frac{    (3 k^2+3 k M+3 k N+M^2+M N)}{
     k M  (3 k+2 M)}\, J^a \, J^{(\bar{\si} j)}
  +  t^{\al}_{\si \bar{\tau}} \, \de^{\si \bar{\si}}\,
  t^a_{l \bar{k}} \, \de^{j \bar{k}}\, J^{\al}\, J^{(\bar{\tau} l)}
  \nonu \\
  &&
  +\frac{ (-k-M+N) }{2  (3 k+2 M)}\, i\, f^{ a b c} \, t^c_{k \bar{l}}\,
  \de^{j \bar{l}}\, J^b \, J^{(\bar{\si} k)}
  -\frac{ (3 k+M+3 N)}{2  (3 k+2 M)}\,
   d^{ a b c} \, t^c_{k \bar{l}}\,
  \de^{j \bar{l}}\, J^b \, J^{(\bar{\si} k)}
  \nonu \\
  &&
  +\frac{ (k^3+k^2 M+2 k^2 N+2 k M N+2 k+M+N)}{ k (3 k+2 M)}\,
  t^a_{k \bar{l}} \, \de^{j \bar{l}}\,  \pa \, J^{(\bar{\si} k)}
  \Bigg](w)\nonu \\
&& + \frac{1}{(z-w)}\,
 J^{(\bar{\si} j)}(z) \, P^{a}(w)\Bigg|_{\frac{1}{(z-w)}}
 + \cdots.
\nonu
\eea
The first order pole can be summarized by
\bea
&&
J^{(\bar{\si} j)}(z) \, P^{a}(w)\Bigg|_{\frac{1}{(z-w)}}=
(t^a_{l \bar{k}} \, \de_{\si \bar{\nu}} \,
J^{(\bar{\nu} j)}\, J^{(\si \bar{k})}\, J^{(\bar{\rho} l)}
-\frac{1}{N}\, t^a_{l \bar{k}} \, \de_{\si \bar{\tau}} \,
J^{(\bar{\rho} j)}\, J^{(\si \bar{k})}\, J^{(\bar{\tau} l)}\nonu \\
&& -
\sqrt{\frac{M+N}{M N}}\,
t^a_{l \bar{k}} \, t^{\al}_{\si \bar{\tau}} \,
\de^{\si \bar{\rho}} \, \de^{j \bar{k}} \,
J^{\al}\, J^{u(1)}\, J^{(\bar{\tau} l)}
-\frac{1}{N}\, t^a_{l \bar{k}}  \,
\de^{j \bar{k}}\, J^{\al}\, J^{\al}\, J^{(\bar{\rho} l)}
\nonu \\
&& -\frac{1}{2}\, (i \, f +d)^{\beta \al \ga} \,
t^a_{l \bar{k}} \, t^{\ga}_{\nu \bar{\tau}} \,
\de^{\nu \bar{\rho}} \, \de^{j \bar{k}} \,
J^{\al}\, J^{\beta}\, J^{(\bar{\tau} l)}
+ \frac{1}{M} \,  t^{\al}_{\si \bar{\tau}} \,
\de^{\si \bar{\rho}} \,
J^{\al}\, J^{a}\, J^{(\bar{\tau} j)} \nonu \\
&& +
\frac{1}{2}\, (i \, f +d)^{a b c} \,
t^c_{l \bar{i}} \, t^{\al}_{\si \bar{\tau}} \,
\de^{\si \bar{\rho}} \, \de^{j \bar{i}} \,
J^{\al}\, J^{b}\, J^{(\bar{\tau} l)}
)
\, a_1 + (-
 t^{a}_{k \bar{j}}  \, \de^{j \bar{j}} \,
 J^{(\bar{\rho} k)} \, J^{\al}\, J^{\al}
 + t^{\al}_{\rho \bar{\si}}  \, \de^{\rho \bar{\rho}} \,
 J^a \, J^{(\bar{\si} j)} \, J^{\al}\nonu \\
 && +  t^{\al}_{\rho \bar{\si}}  \, \de^{\rho \bar{\rho}} \,
 J^a \, J^{\al} \, J^{(\bar{\si} j)} 
)\, a_2 + (-  t^{a}_{k \bar{j}}  \, \de^{j \bar{j}} \,
 J^{(\bar{\rho} k)} \, J^{b}\, J^{b} -
  t^{b}_{k \bar{j}}  \, \de^{j \bar{j}} \,
  J^a \, J^{(\bar{\rho} k)} \, J^{b}-
 t^{b}_{k \bar{j}}  \, \de^{j \bar{j}} \,
 J^a \, J^b\, J^{(\bar{\rho} k)} \nonu \\
 && + M \,
  t^{a}_{k \bar{j}}  \, \de^{j \bar{j}} \,
\pa^2 \, J^{(\bar{\rho} k)} 
-2 \, i\, f^{b a c}\,
t^{b}_{k \bar{j}}  \, \de^{j \bar{j}} \,
J^{(\bar{\rho} k)}\, \pa \, J^c
-2 \, i\, f^{b a c}\,
t^{c}_{k \bar{j}}  \, \de^{j \bar{j}} \,
 J^b\, \pa \, J^{(\bar{\rho} k)} 
 )\, a_3 \nonu \\
 && (-
 t^{a}_{k \bar{j}}  \, \de^{j \bar{j}} \,
 J^{(\bar{\rho} k)}\, J^{u(1)}\, J^{u(1)}+
 \sqrt{\frac{M+N}{M N}}\,J^a \, J^{(\bar{\rho} j)}\, J^{u(1)}
 +  \sqrt{\frac{M+N}{M N}}\,J^a \, J^{u(1)} \,
 J^{(\bar{\rho} j)}
 )\, a_4 \nonu \\
 && + (
- d^{ a b c}\,
t^{c}_{l \bar{l}}  \, \de^{j \bar{l}} \,
t^b_{k \bar{j}}\, \de_{\si \bar{\si}}\, 
 J^{(\bar{\rho} l)} \, J^{(\si \bar{j})} \, J^{(\bar{\si} k)} 
 -d^{ a b c}\,
t^{c}_{l \bar{l}}  \, \de^{j \bar{l}} \,
t^b_{k \bar{j}}\, \de_{\si \bar{\si}}\, 
J^{(\bar{\rho} l)} \, J^{(\bar{\si} k)}  \, J^{(\si \bar{j})}
\nonu \\
&& -2 \,  \sqrt{\frac{M+N}{M N}}\,
d^{ a b c}\,
t^{b}_{k \bar{l}}  \, \de^{j \bar{l}} \,
J^{c} \, J^{(\bar{\rho} k)}  \, J^{u(1)}
-2 \,
d^{ a b c}\,
t^{\al}_{\rho \bar{\si}}  \, \de^{\rho \bar{\rho}} \,
t^b_{k \bar{l}}\, \de^{j \bar{l}}\, 
J^{c} \, J^{(\bar{\si} k)}  \, J^{\al}
\nonu \\
&& +\frac{2}{M}\,
d^{ a b c}\,
J^{c} \, J^{(\bar{\rho} j)}  \, J^{b}
+ 
 (i \, f +d)^{b e d} \, d^{a b c}\, 
t^d_{k \bar{j}} \, \de^{j \bar{j}} \,
J^{c}\, J^{(\bar{\rho} k)}\, J^e
+ N \,d^{a b c}
t^b_{k \bar{j}} \, \de^{j \bar{j}} \,
J^{c}\, \pa \, J^{(\bar{\rho} k)}
)\, a_5
\nonu \\
&& +(
\sqrt{\frac{M+N}{M N}}\,
t^{a}_{l \bar{k}}  \, 
 \de_{\si \bar{\tau}}\, 
 J^{(\bar{\rho} j)} \, J^{(\si \bar{k})} \, J^{(\bar{\tau} l)}
 +\sqrt{\frac{M+N}{M N}}\,
t^{a}_{l \bar{k}}  \, 
 \de_{\si \bar{\tau}}\, 
 J^{(\bar{\rho} j)} \, J^{(\bar{\tau} l)}\, J^{(\si \bar{k})}
 \nonu \\
 && -2 \, \sqrt{\frac{M+N}{M N}}\,
t^{a}_{k \bar{l}}  \, 
 \de^{j \bar{l}}\, 
 J^{u(1)}  \, J^{(\bar{\rho} k)}\, J^{u(1)}
 -2 \,
t^{\al}_{\rho \bar{\si}} \, \de^{\rho \bar{\rho}} \, \de^{j \bar{l}} \,
t^a_{k \bar{l}}\, J^{u(1)}\, J^{(\bar{\si} k)}\, J^{\al}
+ \frac{2}{M}\,  J^{u(1)}\, J^{(\bar{\rho} j)}\, J^{a}\nonu \\
&& + (i \, f +d)^{a b c} \,
t^c_{k \bar{j}} \, \de^{j \bar{j}} \,
J^{u(1)}\, J^{(\bar{\rho} k)}\, J^b+N \,
t^{a}_{k \bar{j}}  \, 
 \de^{j \bar{j}}\, 
 J^{u(1)}  \, \pa \, J^{(\bar{\rho} k)}
 ) \, a_7\nonu \\
 && + (
t^{a}_{m \bar{j}}  \, 
 \de^{j \bar{j}}\, \de_{\si \bar{\tau}}\, \de_{k \bar{j}}\, 
 J^{(\bar{\rho} m)}  \, J^{(\si \bar{j})} \, J^{(\bar{\tau} k)}
 + t^{a}_{m \bar{j}}  \, 
 \de^{j \bar{j}}\, \de_{\si \bar{\tau}}\, \de_{k \bar{j}}\, 
 J^{(\bar{\rho} m)}  \, J^{(\bar{\tau} k)}\,  J^{(\si \bar{j})}
 \nonu \\
 && - 2 \, 
\sqrt{\frac{M+N}{M N}}\,
J^{a}  \, J^{(\bar{\rho} j)}\, J^{u(1)}
 -2 \, t^{\al}_{\rho \bar{\tau}}  \, 
  \de^{\rho \bar{\rho}}\,
  J^{a}  \,  J^{(\bar{\tau} j)}\, J^{\al}
  + 2 \,
t^{b}_{k \bar{j}}  \, 
 \de^{j \bar{j}}\, 
 J^{a}  \, J^{(\bar{\rho} k)}\, J^b
 \nonu \\
 && +
  (M+N)\, J^a \,  \pa \, J^{(\bar{\rho} j)}
 )\, a_8
 +(\sqrt{\frac{M+N}{M N}}\,
d^{a b c}
\de^{\tau \bar{\rho}} \, \de^{j \bar{k}} \,
J^{(\bar{\tau} k)} \, J^b \, J^c
\nonu \\
&&- d^{a b c}
t^b_{k \bar{j}} \, \de^{j \bar{j}} \,
J^{u(1)} \, J^{(\bar{\rho} k)}  \, J^c
-
 d^{a b c}
t^c_{k \bar{j}} \, \de^{j \bar{j}} \,
J^{u(1)} \, J^b\, J^{(\bar{\rho} k)}  
)\, a_9 \nonu \\
&& +(-  i\, f^{a b c}
t^b_{k \bar{j}} \, \de^{j \bar{j}} \,
\pa \, J^{(\bar{\rho} k)}  \, J^c
-
 i\, f^{a b c}
t^c_{k \bar{j}} \, \de^{j \bar{j}} \,
\pa \,J^b\,  J^{(\bar{\rho} k)}  
) \, a_{11}
\nonu \\
&& + (
-\sqrt{\frac{M+N}{M N}}\,
t^a_{l \bar{k}} \, \de^{j \bar{k}} \,
J^{(\bar{\rho} l)} \, \pa \, J^{u(1)} -
t^{\al}_{\rho \bar{\tau}} \, \de^{\rho \bar{\rho}} \, \de^{j \bar{k}} \,
t^a_{l \bar{k}}\, J^{(\bar{\tau} l)}\, \pa \, J^{\al}
+ \frac{1}{M}\,  J^{(\bar{\rho} j)}\, \pa \, J^a
\nonu \\
&& + \frac{1}{2}\,
 (i \, f +d)^{a b c} \,
t^c_{l \bar{j}} \, \de^{j \bar{j}} \,
J^{(\bar{\rho} l)}\, \pa \, J^b
+
\frac{N}{2}\,
t^a_{k \bar{j}}\, \de^{j \bar{j}} \,
\pa^2 \, J^{(\bar{\rho} k)}
)\, a_{12}
\nonu \\
&&+ (
-\sqrt{\frac{M+N}{M N}}\,
t^a_{k \bar{l}} \, \de^{j \bar{l}} \,
\pa \, J^{(\bar{\rho} k)}  \, J^{u(1)}-
t^{\al}_{\rho \bar{\si}} \, \de^{\rho \bar{\rho}} \, \de^{j \bar{l}} \,
t^a_{k \bar{l}}\, \pa \, J^{(\bar{\rho} k)} \, J^{\al}
+ \frac{1}{M}\,  \pa \, J^{(\bar{\rho} j)} \, J^a
\nonu \\
&& +
\frac{1}{2}\,
 (i \, f +d)^{a b c} \,
t^c_{k \bar{j}} \, \de^{j \bar{j}} \,
\pa \, J^{(\bar{\rho} k)}\, J^b
)\, a_{13} -
t^a_{k \bar{j}}\, \de^{j \bar{j}} \, \pa^2 \,
J^{(\bar{\rho} k)} 
\, a_{16}
\nonu \\
&& + (
-\frac{2}{M}\,
t^a_{k \bar{j}} \, \de^{j \bar{j}} \,
J^{(\bar{\rho} k)}\, \, J^b\, J^b-
\frac{1}{2}\,
 (i \, f +d)^{a b e} \,
d^{e c d}\,  
t^b_{k \bar{j}} \, \de^{j \bar{j}} \,
 J^{(\bar{\rho} k)}\,J^c \,  J^d
-\frac{2}{M}\,
t^b_{k \bar{j}} \, \de^{j \bar{j}} \,
J^{(\bar{\rho} k)}\, \, J^b\, J^a
\nonu \\
&& -
\frac{1}{2}\,
 (i \, f +d)^{a d e} \,
d^{e b c}\,  
t^b_{k \bar{j}} \, \de^{j \bar{j}} \,
J^{(\bar{\rho} k)}\,J^c \,  J^d
-\frac{2}{M}\,
t^b_{k \bar{j}} \, \de^{j \bar{j}} \,
J^{(\bar{\rho} k)}\, \, J^a\, J^b
\nonu \\
&&
-\frac{1}{2}\,
 (i \, f +d)^{a c e} \,
d^{e b d}\,  
t^b_{k \bar{j}} \, \de^{j \bar{j}} \,
J^{(\bar{\rho} k)}\,J^c \,  J^d
-\frac{2}{M}\,
t^c_{k \bar{j}} \, \de^{j \bar{j}} \,
J^a\, J^{(\bar{\rho} k)}\, \, J^c\nonu \\
&&
-\frac{1}{2}\,
 (i \, f +d)^{a b e} \,
d^{e c d}\,  
t^c_{k \bar{j}} \, \de^{j \bar{j}} \,
J^b\, J^{(\bar{\rho} k)}\,  J^d
-\frac{2}{M}\,
t^b_{k \bar{j}} \, \de^{j \bar{j}} \,
J^b\, J^{(\bar{\rho} k)}\, \, J^a
\nonu \\
&& -\frac{1}{2}\,
 (i \, f +d)^{a d e} \,
d^{e b c}\,  
t^c_{k \bar{j}} \, \de^{j \bar{j}} \,
J^b\, J^{(\bar{\rho} k)}\,  J^d
-\frac{2}{M}\,
t^a_{k \bar{j}} \, \de^{j \bar{j}} \,
J^b\, J^{(\bar{\rho} k)}\, \, J^b
\nonu \\
&&
-\frac{1}{2}\,
 (i \, f +d)^{a c e} \,
d^{e b d}\,  
t^c_{k \bar{j}} \, \de^{j \bar{j}} \,
J^b\, J^{(\bar{\rho} k)}\,  J^d
-\frac{2}{M}\,
t^c_{k \bar{j}} \, \de^{j \bar{j}} \,
J^a\, J^c\, J^{(\bar{\rho} k)}
\nonu \\
&& -\frac{1}{2}\,
 (i \, f +d)^{a b e} \,
d^{e c d}\,  
t^d_{k \bar{j}} \, \de^{j \bar{j}} \,
J^b\, J^c\, J^{(\bar{\rho} k)}
-\frac{2}{M}\,
t^a_{k \bar{j}} \, \de^{j \bar{j}} \,
J^b \, J^b\, J^{(\bar{\rho} k)}
\nonu \\
&& -\frac{1}{2}\,
 (i \, f +d)^{a d e} \,
d^{e b c}\,  
t^d_{k \bar{j}} \, \de^{j \bar{j}} \,
J^b\, J^c\, J^{(\bar{\rho} k)}
-\frac{2}{M}\,
t^b_{k \bar{j}} \, \de^{j \bar{j}} \,
J^b \, J^a \, J^{(\bar{\rho} k)}
\nonu \\
&& -\frac{1}{2}\,
 (i \, f +d)^{a c e} \,
d^{e b d}\,  
t^d_{k \bar{j}} \, \de^{j \bar{j}} \,
J^b\, J^c\, J^{(\bar{\rho} k)}
) \, a_{17},
\nonu
\eea
where the coefficients are in (\ref{avalues}).

\section{Some OPEs in terms of coset fields in section $6$}

We present the coset realizations for the following OPEs
\bea
G^{+,a}(z) \, G^{-,b}(w) &= &
\frac{1}{(z-w)^3}\, k N \,\de^{a b} 
\nonu \\
& + & 
\frac{1}{(z-w)^2} \, \Bigg[ \frac{(k+M+N)}{M} \, K +
  \frac{N}{2} \, (i f - d)^{a b c} \, J^c +\frac{k}{2} \,
  ( i f + d)^{a b c} \, J_f^c \Bigg](w) 
\nonu \\
&+& \frac{1}{(z-w)}\,
\Bigg[  (t^b \, t^a)_{l\bar{i}}\,
  \de_{\rho \bar{\nu}} \, J^{(\rho \bar{i})}\, J^{(\bar{\nu} l)}
 - k \,  (t^a \, t^b)_{j\bar{k}}\,
 \de_{\tau \bar{\si}} \,
 \psi^{(\tau \bar{k})}\, \pa \, \psi^{(\bar{\si} j)}
 \nonu \\
 &-& \sqrt{\frac{M+N}{M N}} \,
 (t^a \, t^b)_{j\bar{k}}\,
 \de_{\tau \bar{\si}} \, J^{u(1)}\,
 \psi^{(\tau \bar{k})} \, \psi^{(\bar{\si} j)}
-t^{\al}_{\si \bar{\tau}}\, (t^a \, t^b)_{j\bar{k}}\,
  J^{\al}\,
 \psi^{(\tau \bar{k})} \, \psi^{(\bar{\si} j)}
 \nonu \\
 &+ &   (t^a \, t^c\, t^b)_{j\bar{k}}\,
\de_{\tau \bar{\si}} \,
 \psi^{(\tau \bar{k})} \, J^c\, \psi^{(\bar{\si} j)}
 \Bigg](w) +\cdots.
\nonu
\eea
The first order pole can be simplified and is given by
(\ref{ope2}).

The OPEs with the spin-$2$ currents are 
\bea
G^{+,a}(z) \, K^b(w) &=&
-\frac{1}{(z-w)^2}\, \frac{2(k^2-1)(2k+M+N)}{k(2k+M)}\,
(t^a \, t^b)_{j \bar{i}}\, \de_{\rho \bar{\si}} \, J^{(\rho \bar{i})}\,
\psi^{(\bar{\si} j)}
\nonu \\
&+& \frac{1}{(z-w)}\, \Bigg[
 -\frac{2(k^2-1)(2k+M+N)}{k(2k+M)}\,
 (t^a \, t^b)_{j \bar{i}}\, \de_{\rho \bar{\si}} \, \pa \, (
 J^{(\rho \bar{i})}\,
\psi^{(\bar{\si} j)})
\nonu \\
&+&
2(k+N) \,  (t^a \, t^b)_{j \bar{i}}\, \de_{\rho \bar{\si}} \,
\pa \, 
 J^{(\rho \bar{i})}\,
 \psi^{(\bar{\si} j)} -\frac{2}{k M}\, (k+M+N)\, J^b \, G^{+,a}
 \nonu \\
 &-& \frac{2}{k}\, (k+N)\, \sqrt{\frac{M+N}{M N}}\,
 (t^a \, t^b)_{j \bar{l}}\, \de_{\rho \bar{\si}} \,
 J^{u(1)}\,  J^{(\rho \bar{l})}\,
 \psi^{(\bar{\si} j)}
 \nonu \\
 &-& (i \, f -\frac{2k+M+2N}{2k+M})^{b c d}\,
  (t^a \, t^d)_{j \bar{l}} \,\de_{\rho \bar{\si}}\,
 J^{c}\,  J^{(\rho \bar{l})}\,
 \psi^{(\bar{\si} j)}
 \nonu \\
 &-& 2 \, t^{\al}_{\si \bar{\si}}\,
 (t^a \, t^b)_{j \bar{l}} \,
 J^{\al}\,  J^{(\si \bar{l})}\,
 \psi^{(\bar{\si} j)}
\Bigg](w) + \cdots,
\nonu
\eea
and 
\bea
G^{-,a}(z) \, K^b(w) &=&
-\frac{1}{(z-w)^2}\, \frac{2(k^2-1)(2k+M+N)}{k(2k+M)}\,
(t^b \, t^a)_{k \bar{i}}\, \de_{\rho \bar{\si}} \,
J^{(\bar{\si} k)}\,
\psi^{(\rho \bar{i})}
\nonu \\
&+& \frac{1}{(z-w)}\, \Bigg[
 -\frac{2(k^2-1)(2k+M+N)}{k(2k+M)}\,
 (t^b \, t^a)_{k \bar{i}}\, \de_{\rho \bar{\si}} \,
\pa \, (J^{(\bar{\si} k)}\,
\psi^{(\rho \bar{i})})
\nonu \\
&+&
2(k+N) \,  (t^b \, t^a)_{k \bar{i}}\, \de_{\rho \bar{\si}} \,
\pa \, 
 J^{(\bar{\si} k)}\,
 \psi^{(\rho \bar{i})} +
 \frac{2}{k M}\, (k+M+N)\, J^b \, G^{-,a}
 \nonu \\
 &+& \frac{2}{k}\, (k+N)\, \sqrt{\frac{M+N}{M N}}\,
 (t^b \, t^a)_{k \bar{l}}\, \de_{\rho \bar{\si}} \,
 J^{u(1)}\,  J^{(\bar{\si} k)}\,
 \psi^{(\rho \bar{i})}
 \nonu \\
 &-& (i \, f +\frac{2k+M+2N}{2k+M})^{b c d}\,
  (t^d \, t^a)_{k \bar{i}} \,\de_{\rho \bar{\si}}\,
 J^{c}\,  J^{(\bar{\si} k)}\,
 \psi^{(\rho \bar{i} )}
 \nonu \\
 &+& 2 \, t^{\al}_{\rho \bar{\si}}\,
 (t^b \, t^a)_{k \bar{i}} \,
 J^{\al}\,  J^{(\bar{\si} k)}\,
 \psi^{(\rho \bar{i} )}
\Bigg](w) + \cdots.
\nonu
\eea
They are simplified and are described in (\ref{gkOPE}) and
(\ref{GKope}).

\section{ The OPEs between the nonsinglet spin-$2$ operators in section $6$ }

In the computation of the OPE between the nonsinglet spin-$2$ current
(\ref{nonspin2})
and itself, we should use the following OPEs. 

$\bullet$ The OPEs between the first term and the remaining terms 
\bea
&& K^a(z) \, J^b\, J^{u(1)}_f(w) = \frac{1}{(z-w)}\,
i \, f^{a b c}\, J^{u(1)}_f \, K^c(w)+ \cdots,
\nonu \\
&& K^a(z) \, (i \, f+ d)^{c b d} \, J^c\, J^{d}_f(w) =
-\frac{1}{(z-w)}\,
(i \, + d)^{c b d} \, i\, f^{c a e} \, J^{d}_f \, K^e(w)
+ \cdots,
\nonu \\
&& K^a(z) \, d^{b c d} \, J^c\, J^{d}(w) =
\frac{1}{(z-w)^2}\, M \, d^{a b c}\, K^c \nonu \\
&& +
\frac{1}{(z-w)}\, \Big[ i \, f^{a c e} \, d^{b c d}\,
  K^e \, J^d + i \, f^{a d e}\, d^{b  c d}\, J^c \, K^e
  \Big](w) + \cdots,
\nonu \\
&& K^a(z) \, J^b\, J^{u(1)}(w) =
\frac{1}{(z-w)}\, i\, f^{a b c} \, J^{u(1)}\, K^c(w)
+\cdots.
\nonu
\eea

$\bullet$  The OPEs between the second term and the remaining terms 
\bea
&& t^a_{j \bar{i}} \, \de_{\rho \bar{\si}} \,
\psi^{(\rho \bar{i})}\,
\pa \psi^{(\bar{\si} j)}(z) \,
t^b_{k \bar{l}} \, \de_{\mu \bar{\nu}} \,
\psi^{(\mu \bar{l})}\,
\pa \psi^{(\bar{\nu} k)} =-\frac{1}{(z-w)^4}\,
N \, \de^{a b} \nonu \\
&& +\frac{1}{(z-w)^2}\, \Big[-\frac{2}{M}\, \de^{a b}\,
  \de_{k \bar{i}} \, \de_{\rho \bar{\nu}}\, \psi^{(\rho \bar{i})}\,
  \pa \psi^{(\bar{\nu} k)} - d^{a b c} \, t^c_{k \bar{i}}\, \de_{\rho \bar{\nu}}
  \, \psi^{(\rho \bar{i})}\,
\pa \psi^{(\bar{\nu} k)}\Big](w)
\nonu \\
&&+ \frac{1}{(z-w)}\, \Big[
-\frac{1}{M}\, \de^{a b}\,
\de_{k \bar{i}} \, \de_{\rho \bar{\nu}}\,
\pa \, \psi^{(\rho \bar{i})}\,
  \pa \psi^{(\bar{\nu} k)}
  -\frac{1}{2}\, (i \, f+d)^{b a c}\,  t^c_{k \bar{i}}\,
  \de_{\rho \bar{\nu}}
  \, \pa \, \psi^{(\rho \bar{i})}\,
\pa \, \psi^{(\bar{\nu} k)}
\nonu \\
&& - \frac{1}{M}\, \de^{a b}\,
\de_{k \bar{i}} \, \de_{\rho \bar{\nu}}\,
 \psi^{(\rho \bar{i})}\,
 \pa^2 \psi^{(\bar{\nu} k)}
 -\frac{1}{2}\, (i \, f+d)^{ a b c}\,  t^c_{k \bar{i}}\,
  \de_{\rho \bar{\nu}}
   \, \psi^{(\rho \bar{i})}\,
\pa^2\,  \psi^{(\bar{\nu} k)}
 \Big](w) + \cdots,
\nonu \\
&& t^a_{j \bar{i}} \, \de_{\rho \bar{\si}} \,
\psi^{(\rho \bar{i})}\,
\pa \psi^{(\bar{\si} j)}(z) \,
J^{u(1)}
\, J_f^b(w) =
\frac{1}{(z-w)^3}\, N \, \de^{a b}\, J^{u(1)}(w)
\nonu \\
&&+
\frac{1}{(z-w)^2}\, \Big[
  \frac{1}{M}\, \de^{a b} \, J^{u(1)}\, J_f^{u(1)}
  -\frac{1}{2}\, (i \, f+ d)^{b a c}\, J^{u(1)}\,
  J_f^{c}
  \Big](w)\nonu \\
&&+ \frac{1}{(z-w)}\, \Big[
 \frac{1}{M}\, \de^{a b} \, J^{u(1)}\, \pa \, J_f^{u(1)}
  -\frac{1}{2}\, (i \, f+ d)^{b a c}\, J^{u(1)}\,
  \pa \, J_f^{c}
  \nonu \\
  && +
  \frac{1}{2}\,
  (i \, f+ d)^{a b c}\, t^c_{j \bar{k}}\, \de_{\rho
    \bar{\si}}\, J^{u(1)}\,
  \psi^{(\rho \bar{k})}\,
\pa \psi^{(\bar{\si} j)}
\nonu \\
 && -
  \frac{1}{2}\,
  (i \, f+ d)^{ b a c}\, t^c_{l \bar{i}}\, \de_{\rho
    \bar{\si}}\, J^{u(1)}\,
  \psi^{(\rho \bar{i})}\,
\pa \psi^{(\bar{\si} l)}
\Big](w)+ \cdots,
\nonu \\
&& t^a_{j \bar{i}} \, \de_{\rho \bar{\si}} \,
\psi^{(\rho \bar{i})}\,
\pa \psi^{(\bar{\si} j)}(z) \,
 t^{\al}_{\mu \bar{\nu}} \, t^b_{k \bar{l}} \, J^{\al} \,
 \psi^{(\mu \bar{l})}\,
 \psi^{(\bar{\nu} k)}(w)=
 \nonu \\
 && -\frac{1}{(z-w)^2}\, \Big[ \frac{1}{M}\, \de^{b a}\, \de_{k \bar{i}}+
   \frac{1}{2} \, (i \, f + d)^{b a c}\, t^c_{k \bar{i}}\Big]
  t^{\al}_{\rho \bar{\nu}}  \, J^{\al} \,
 \psi^{(\rho \bar{i})}\,
 \psi^{(\bar{\nu} k)}(w)
 \nonu \\
 &&+
 \frac{1}{(z-w)}\, \Big[
 -\frac{1}{M}\, \de^{b a}\, \de_{k \bar{i}}
  t^{\al}_{\rho \bar{\nu}}  \, J^{\al} \,
 \pa \, \psi^{(\rho \bar{i})}\,
 \psi^{(\bar{\nu} k)}
 - \frac{1}{2} \, (i \, f + d)^{b a c}\, t^c_{k \bar{i}}\,
  t^{\al}_{\rho \bar{\nu}}  \, J^{\al} \,
\pa \, \psi^{(\rho \bar{i})}\,
 \psi^{(\bar{\nu} k)}
 \nonu \\
 && -\frac{1}{M}\, \de^{b a}\, \de_{k \bar{i}}
  t^{\al}_{\rho \bar{\nu}}  \, J^{\al} \,
  \psi^{(\rho \bar{i})}\,
 \pa \, \psi^{(\bar{\nu} k)}
 - \frac{1}{2} \, (i \, f + d)^{a b c}\, t^c_{k \bar{i}}\,
t^{\al}_{\rho \bar{\nu}}  \, J^{\al} \,
 \psi^{(\rho \bar{i})}\,
\pa \,  \psi^{(\bar{\nu} k)}
 \Big](w) 
 +\cdots,
 \nonu \\
&& t^a_{j \bar{i}} \, \de_{\rho \bar{\si}} \,
\psi^{(\rho \bar{i})}\,
\pa \psi^{(\bar{\si} j)}(z) \, J^b \, J_f^{u(1)}(w) =
\frac{1}{(z-w)^2}\, J^b \, J^a_f(w) +
\frac{1}{(z-w)}\, J^b \, \pa \, J^a_f(w) + \cdots,
\nonu \\
&& t^a_{j \bar{i}} \, \de_{\rho \bar{\si}} \,
\psi^{(\rho \bar{i})}\,
\pa \psi^{(\bar{\si} j)}(z) \, (i \, f+d)^{c b d}\, J^c \, J^d_f(w)
= \frac{1}{(z-w)^3}\, N \, (i\, f +d)^{c b a}\, J^c
\nonu \\
&&+\frac{1}{(z-w)^2}\, \Big[
\frac{1}{M}\, \de^{d a}\, \de_{l \bar{i}}+\frac{1}{2}\,
(i\, f +d)^{d a e}\, t^e_{l \bar{i}}
\Big]\, \de_{\rho \bar{\si}}\, (i\, f +d)^{c b d}\,
J^c\, \psi^{(\rho \bar{i})}\, \psi^{(\bar{\si} l)}(w)
\nonu \\
&& +
  \frac{1}{(z-w)} \Big[
\frac{1}{M}\, \de^{d a}\, \de_{l \bar{i}}\,
(i\, f +d)^{c b d}\, t^e_{l \bar{i}}
\, \de_{\rho \bar{\si}}\,
J^c\, \pa \, (
\psi^{(\rho \bar{i})}\, \psi^{(\bar{\si} l)})
\nonu \\
&& +\frac{1}{2} \, (i\, f +d)^{ d a e }\, t^e_{l \bar{i}}
\, \de_{\rho \bar{\si}}\,
(i \, f +d)^{c b d}\, J^c \,
 \pa \, (
\psi^{(\rho \bar{i})}\, \psi^{(\bar{\si} l)})
\nonu \\
&& +
\frac{1}{2} \, (i\, f +d)^{ a d e }\,
t^e_{j \bar{k}}
\, \de_{\rho \bar{\si}}\,
(i \, f +d)^{c b d}\, \de_{\rho \bar{\si}} \, J^c \,
\psi^{(\rho \bar{k})}\, \pa \, \psi^{(\bar{\si} j)}
\nonu \\
&& -
\frac{1}{2} \, (i\, f +d)^{ d a e }\,
t^e_{l \bar{i}}
\, \de_{\rho \bar{\si}}\,
(i \, f +d)^{c b d} \, J^c \,
\psi^{(\rho \bar{i})}\, \pa \, \psi^{(\bar{\si} l)}
\Big](w) +\cdots,
  \nonu \\
  && t^a_{j \bar{i}} \, \de_{\rho \bar{\si}} \,
\psi^{(\rho \bar{i})}\,
\pa \psi^{(\bar{\si} j)}(z) \,\pa \, J_f^b(w)=
\frac{1}{(z-w)^4}\, 3N \, \de^{a b}
\nonu \\
&& + \frac{1}{(z-w)^3} \Big[ \frac{2}{M} \, \de^{b a}
  \, J_f^{u(1)}- (i \, f + d)^{b a c}\, J_f^c \Big](w)
\nonu \\
&& +\frac{1}{(z-w)^2}\, \Big[ \frac{2}{M}\, \de^{b a}\,
  \pa \, J_f^{u(1)} - (i \, f + d )^{b a c}
  \, \pa \, J_f^c + i \, f^{a b c}\, t^c_{j \bar{k}}\,
  \de_{\rho \bar{\si}}\, \psi^{(\rho \bar{k})}\, \pa \,
  \psi^{(\bar{\si} j)}
  \Big](w)
\nonu \\
&& + \frac{1}{(z-w)}\, \Big[
\frac{1}{M}\, \de^{b a}\,
\pa^2 \, J_f^{u(1)} - \frac{1}{2} \,
(i \, f + d )^{b a c}
\, \pa^2 \, J_f^c
+ i \, f^{a b c}\, t^c_{j \bar{k}}\,
\de_{\rho \bar{\si}}\, \pa \, (
\psi^{(\rho \bar{k})}\, \pa \,
\psi^{(\bar{\si} j)})
  \Big](w) + \cdots.
\nonu
\eea

$\bullet$  The OPEs between the third term and the remaining terms 
\bea
&& J^{u(1)}\, J_f^a(z) \, J^{u(1)}\, J_f^b(w)
=
\frac{1}{(z-w)^4}\, k \, N \, \de^{a b}+
\frac{1}{(z-w)^3}\, k \, i \, f^{a b c}\, J_f^c
\nonu \\
&& + \frac{1}{(z-w)^2}\, \Big[ k \,
  J_f^a \, J_f^b + N \, \de^{a b} \, J^{u(1)}\,
  J^{u(1)} \Big](w)\nonu \\
&& + \frac{1}{(z-w)}\, \Big[
 k \,\pa \,
  J_f^a \, J_f^b + N \, \de^{a b} \, J^{u(1)}\,
  \pa \, J^{u(1)} + i \, f^{a b c}\,
  J^{u(1)}\, J^{u(1)}\, J_f^c
  \Big](w)+
\cdots,
\nonu \\
&& J^{u(1)}\, J_f^a(z)\,
 t^{\al}_{\mu \bar{\nu}} \, t^b_{k \bar{l}} \, J^{\al} \,
 \psi^{(\mu \bar{l})}\,
 \psi^{(\bar{\nu} k)}(w)=\frac{1}{(z-w)}\,
 \Big[-\frac{1}{M}\, \de^{a b}\,
   \de_{j \bar{k}}\, t^{\al}_{\rho \bar{\si}}\,
   J^{\al} (( J^{u(1)}\, \psi^{(\rho \bar{k})})
   \psi^{(\bar{\si} j)})
   \nonu \\
   && -\frac{1}{2}\, (i \, f+ d)^{b a c}\,
   t^c_{j \bar{k}}\, t^{\al}_{\rho \bar{\si}}\,
   J^{\al} (( J^{u(1)}\, \psi^{(\rho \bar{k})})
   \psi^{(\bar{\si} j)})
   +
   \frac{1}{M}\, \de^{a b}\,
   \de_{l \bar{i}}\, t^{\al}_{\rho \bar{\si}}\,
   J^{\al} \,  \psi^{(\rho \bar{i})}
   \, J^{u(1)} \, \psi^{(\bar{\si} l)}
   \nonu \\
   &&+
   \frac{1}{2}\, (i \, f+ d)^{a b c}\,
   t^c_{l \bar{i}}\, t^{\al}_{\rho \bar{\si}}\,
   J^{\al} \, \psi^{(\rho \bar{i})}\, J^{u(1)}\,
   \psi^{(\bar{\si} l)} \Big](w) +\cdots,
 \nonu \\
&& J^{u(1)}\, J_f^a(z) \, (i \, f +d)^{c b d}\, J^c \,
 J_f^d(w) =
 \frac{1}{(z-w)^2}\, N \, \de^{d a}\, (i \, f +d)^{
 c b d} \, J^c \, J^{u(1)}(w)
 \nonu \\
 && + \frac{1}{(z-w)}\, \Big[
 N \, \de^{d a}\, (i \, f +d)^{
 c b d} \, J^c \, \pa \, J^{u(1)}
 - i \, f^{d a e}\, (i \, f +d)^{c b d}\,
 J^c \, J^{u(1)}\, J_f^e \Big](w)
 + \cdots,
 \nonu \\
 && J^{u(1)}\, J_f^a(z) \, \pa  \,
 J_f^b(w) =
 \frac{1}{(z-w)^3}\, 2 N \, \de^{a b}\, J^{u(1)} +
 \frac{1}{(z-w)^2}\, \Big[
 2 N \, \de^{a b}\, \pa \, J^{u(1)} - i \, f^{a b c}\, J^{u(1)}
 \,
 J_f^c \Big](w)
 \nonu \\
 && + \frac{1}{(z-w)}\, \Big[
 N \, \de^{a b}\, \pa^2 \, J^{u(1)} - i \, f^{a b c}\, \pa (J^{u(1)}
 \,
 J_f^c) \Big](w)+\cdots,
 \nonu \\
 &&J^{u(1)}\, J_f^a(z)   \,
 J^b \, J^{u(1)}(w) = \frac{1}{(z-w)^2}\, k \, J^b \, J^a_f(w)
 \frac{1}{(z-w)}\, k \, J^b \, \pa \, J^a_f(w) + \cdots.
 \nonu
\eea

$\bullet$  The OPEs between the fourth term and the remaining terms 
\bea
 && t^{\al}_{\rho \bar{\si}} \, t^a_{j \bar{i}} \, J^{\al} \,
 \psi^{(\rho \bar{i})}\,
 \psi^{(\bar{\si} j)}(z) \,
  t^{\beta}_{\mu \bar{\nu}} \, t^b_{k \bar{l}} \, J^{\beta} \,
 \psi^{(\mu \bar{l})}\,
 \psi^{(\bar{\nu} k)}(w) =
 \frac{1}{(z-w)^4}\, k \, (N^2-1)\, \de^{a b}
 \nonu \\
 && -\frac{1}{(z-w)^3} \, k \, (N -\frac{1}{N})\, i \, f^{a b c}\,
 J_f^c(w) \nonu \\
 && +\frac{1}{(z-w)^2}\, \Big[ k \, (\de_{\rho \bar{\nu}} \, \de_{\mu
     \bar{\si}} -\frac{1}{N} \,\de_{\rho \bar{\si}} \, \de_{\mu
     \bar{\nu}} )\, t^a_{j \bar{i}}\, t^b_{k \bar{l}}\, (\psi^{(\rho \bar{i})}\,
   \psi^{(\bar{\si} j)})(\psi^{(\mu \bar{l})}\, \psi^{(\bar{\nu} k)})
   \nonu \\
   &&  -N \, (\frac{2}{M}\, \de^{a b}\, \de_{k \bar{i}}
 \, t^{\ga}_{\rho
  \bar{\nu}} 
   + d^{a b c}\, t^{\ga}_{\rho
  \bar{\nu}} \, t^c_{k \bar{i}})\, J^{\ga}\, \psi^{(\rho \bar{i})}\, \psi^{(\bar{\nu} k)}
   + \de^{a b}\, J^{\al}\, J^{\al} \Big](w)
 \nonu \\
 &&+ \frac{1}{(z-w)}\, \Big[
   k \, (\de_{\rho \bar{\nu}} \, \de_{\mu
     \bar{\si}} -\frac{1}{N} \,\de_{\rho \bar{\si}} \, \de_{\mu
     \bar{\nu}} ) \,  t^a_{j \bar{i}}\, t^b_{k \bar{l}}\, (\pa \,
   (\psi^{(\rho \bar{i})}\,
   \psi^{(\bar{\si} j)}))(\psi^{(\mu \bar{l})}\, \psi^{(\bar{\nu} k)})
   \nonu \\
   && -  i\, f^{\beta \al \ga}\, t^{\al}_{\rho \bar{\si}}\, t^{\beta}_{
     \mu \bar{\nu}} \, t^a_{j \bar{i}}\, t^b_{k \bar{l}}\,
   (J^{\ga}\,\psi^{(\rho \bar{i})}\,
   \psi^{(\bar{\si} j)} )(\psi^{(\mu \bar{l})}\, \psi^{(\bar{\nu} k)})
   + \de^{a b}\, J^{\al}\, \pa \, J^{\al}  \nonu \\
   && +(\frac{1}{N}\, \de^{\al \beta}\, \de_{\rho \bar{\nu}} +
   \frac{1}{2} \, (i \, f+ d)^{\al \beta \ga} \, t^{\ga}_{\rho \bar{\nu}})
   (\frac{1}{M}\, \de^{b a}\, \de_{k \bar{i}} +
   \frac{1}{2} \, (i \, f+ d)^{b a c} \, t^{c}_{k \bar{i}})
   \, J^{\beta}\, J^{\al}\, \psi^{(\rho \bar{i})}\, \psi^{(\bar{\nu} k)}
   \nonu \\
   && -
(\frac{1}{N}\, \de^{\al \beta}\, \de_{\mu \bar{\si}} +
   \frac{1}{2} \, (i \, f+ d)^{\beta \al \ga} \, t^{\ga}_{\mu \bar{\si}})
   (\frac{1}{M}\, \de^{b a}\, \de_{j \bar{l}} +
   \frac{1}{2} \, (i \, f+ d)^{ a b c} \, t^{c}_{j \bar{l}})
   \, J^{\beta}\, J^{\al}\, \psi^{(\mu \bar{l})}\, \psi^{(\bar{\si} j)}
   \Big](w) \nonu \\
 && +\cdots,
 \nonu \\
  && t^{\al}_{\rho \bar{\si}} \, t^a_{j \bar{i}} \, J^{\al} \,
 \psi^{(\rho \bar{i})}\,
 \psi^{(\bar{\si} j)}(z) \, (i \, f +d)^{c b d} \, J^c \, J_f^d(w) =
 \nonu \\
 && \frac{1}{(z-w)}\,
 (i \, f+ d)^{ c b d} \, i \, f^{a d e} \, t^{\al}_{\rho \bar{\si}}\,
 t^{e}_{j \bar{k}}
   \, J^{c}\, J^{\al}\, \psi^{(\rho \bar{k})}\, \psi^{(\bar{\si} j)}
   + \cdots,
   \nonu \\
&&  t^{\al}_{\rho \bar{\si}} \, t^a_{j \bar{i}} \, J^{\al} \,
 \psi^{(\rho \bar{i})}\,
 \psi^{(\bar{\si} j)}(z) \, \pa \, J_f^b(w) =
 \frac{1}{(z-w)^2}\, i \, f^{a b c}\, t^{\al}_{\rho \bar{\si}}\,
 t^c_{j \bar{k}}\, J^{\al}\, \psi^{(\rho \bar{k})}\,
 \psi^{(\bar{\si} j)}
 \nonu \\
 &&+
 \frac{1}{(z-w)}\, i \, f^{a b c}\, t^{\al}_{\rho \bar{\si}}\,
 t^c_{j \bar{k}}\, \pa \, (J^{\al}\, \psi^{(\rho \bar{k})}\,
 \psi^{(\bar{\si} j)}) + \cdots.
 \nonu
   \eea

$\bullet$  The OPEs between the fifth term and the remaining terms 
\bea
&& J^a \, J^{u(1)}_f(z) \,J^b \, J^{u(1)}_f(w) =
\frac{1}{(z-w)^4}\, k \, M \, N\, \de^{a b} +
\frac{1}{(z-w)^3}\, M \, N \, i \,f^{a b c}\, J^c(w) 
\nonu \\
&& + \frac{1}{(z-w)^2}\, \Big[ k \, \de^{a b}\, J_f^{u(1)} \,
  J_f^{u(1)} + i \, f^{a b c}\, M \, N \, \pa \, J^c+
  M \, N\, J^b \, J^a \Big](w) 
\nonu \\
&&+ \frac{1}{(z-w)} \, \Big[
 k \, \de^{a b}\, \pa \, J_f^{u(1)} \,
  J_f^{u(1)} + i \, f^{a b c}\, ((J^c \, J_f^{u(1)})\, J_f^{u(1)})+
  M \, N\, J^b \, \pa \, J^a
  \Big](w) + \cdots,
\nonu \\
&& J^a \, J^{u(1)}_f(z) \, (i \, f+ d)^{c b d}\, J^c \, J^{d}_f(w)=
\frac{1}{(z-w)^2}\, k \, (i \, f +d)^{a b c}\, J_f^c \, J_f^{u(1)}(w)
\nonu \\
&&+ \frac{1}{(z-w)}\, \Big[
 k \, (i \, f +d)^{a b c}\, J_f^c \, \pa \, J_f^{u(1)}
 -(i \, f +d)^{c b d}\, i\, f^{c a e}\, J^d_f \, J^e\, J_f^{u(1)}
 \Big](w) +\cdots,
\nonu \\
&& J^a \, J^{u(1)}_f(z) \, d^{b c d}\, J^c \, J^d(w)=
\frac{1}{(z-w)^2}\, (2k +M)\, d^{a b c}\, J_f^{u(1)}\, J^c(w)
\nonu \\
&&+ \frac{1}{(z-w)} \Big[
(2k +M)\, d^{a b c}\, \pa \, (J_f^{u(1)}\, J^c)
  -(2k +M)\, d^{a b c}\, J_f^{u(1)}\, \pa \, J^c
  \nonu \\
  && + i \, f^{a b c} \, d^{c d e}\, J_f^{u(1)}\, J^d \, J^e \Big](w) +\cdots,
\nonu \\
&& J^a \, J^{u(1)}_f(z) \, J^b \, J^{u(1)}(w)=
\frac{1}{(z-w)^2}\, k \, \de^{a b}\, J^{u(1)}\, J_f^{u(1)}(w)
\nonu \\
&& + \frac{1}{(z-w)}\, \Big[
  k \, \de^{a b}\, J^{u(1)}\, \pa \, J_f^{u(1)}+
  i \, f^{a b c}\, J^{u(1)}\, J^c \,J_f^{u(1)} 
  \Big](w) + \cdots.
\nonu
\eea

$\bullet$  The OPEs between the sixth term and the remaining terms 
\bea
&& (i \, f+ d)^{f a c}\, J^f \, J^{c}_f(z)\,
(i \, f+ d)^{d b e}\, J^d \, J^{e}_f(w)\,
= \frac{1}{(z-w)^4}\, k \, N\, (i \, f + d)^{f a c}(i \, f +d)^{f b c}
\nonu \\
&& + \frac{1}{(z-w)^3} \, \Big[
k \, i \, f^{c e g} \, (i \, f + d)^{f a c}(i \, f +d)^{f b e}\, J_f^g
- N \,i \, f^{d f g} \, (i \, f + d)^{f a c}(i \, f +d)^{d b c}\, J^g
\Big](w)\nonu \\
&& + \frac{1}{(z-w)^2}\, \Big[
  k\, (i \, f + d)^{f a c}(i \, f +d)^{f b e}\, J_f^c\, J_f^e
  -N\, i \, f^{d f g}\,  (i \, f + d)^{f a c}(i \, f +d)^{d b c}\,
  \pa \,J^g 
  \nonu \\
  && -
  f^{d h f} \, f^{e c g}\,
  (i \, f + d)^{h a c}(i \, f +d)^{d b e}\, J^f\, J_f^g
  + N \,
  (i \, f + d)^{f a c}(i \, f +d)^{d b c}\, J^d\, J^f
  \Big](w)
\nonu \\
&&+ \frac{1}{(z-w)} \, \Big[
 k\, (i \, f + d)^{f a c}(i \, f +d)^{f b e}\, \pa \, J_f^c\, J_f^e
 + N \,
  (i \, f + d)^{f a c}(i \, f +d)^{d b c}\, J^d\, \pa \, J^f
 \nonu \\
 && - i\, f^{d g f} \, (i \, f + d)^{g a c}(i \, f +d)^{d b e}\,
 ((J^f \, J_f^c) J_f^e)
 - i \, f^{e c g}\,  (i \, f + d)^{h a c}(i \, f +d)^{d b e}\,
 J^d \, J^h \, J_f^g
 \Big](w)\nonu \\
&& + \cdots,
\nonu \\
&& (i \, f+ d)^{f a c}\, J^f \, J^{c}_f(z)\,
\pa \, J_f^b(w)=
\frac{1}{(z-w)^3}\, 2 \, N \, \de^{b c}\, (i \, f+ d)^{d a c }\, J^d(w)
\nonu \\
&& +\frac{1}{(z-w)^2}\, \Big[ 2 \, N \, \de^{b c}\, (i \, f+d)^{d a c}\,
  \pa \, J^d - i \, f^{b c g}\, (i \, f+d)^{d a c}\, J^d \, J_f^g
  \Big](w)
\nonu \\
&& + \frac{1}{(z-w)}\,
 \Big[ N \, \de^{b c}\, (i \, f+d)^{d a c}\,
   \pa^2 \, J^d - i \, f^{b c g}\, (i \, f+d)^{d a c}\,
   \pa \, (J^d \, J_f^g)
  \Big](w) + \cdots,
 \nonu \\
 && (i \, f+ d)^{f a c}\, J^f \, J^{c}_f(z)\,
 d^{b d e}\, J^d \, J^e(w)=
 \frac{1}{(z-w)^2}\, (2k+M)\, d^{d b c} \, (i \, f+d)^{d a e}\,
 J^c \, J_f^e(w) \nonu \\
 && +\frac{1}{(z-w)}\, \Big[
 (2k+M)\, d^{d b c} \, (i \, f+d)^{d a e}\,
   \pa \, (J^c \, J_f^e)-
    (2k+M)\, d^{d b c} \, (i \, f+d)^{d a e}\,
   \pa \, J^c \, J_f^e\nonu \\
   && + i \, f^{f b c}\, d^{c d e}\, (i \, f +d)^{f a g}
   \, J^d \, J^e \, J_f^g
   \Big](w) +\cdots,
 \nonu \\
 && (i \, f+ d)^{f a c}\, J^f \, J^{c}_f(z)\,
 J^b \, J^{u(1)}(w) =
 \frac{1}{(z-w)^2}\, k \, (i \, f +d)^{b a c}\, J^{u(1)}\,
 J_f^c \nonu \\
 && +\frac{1}{(z-w)}\,
 \Big[
 k \, (i \, f +d)^{b a c}\, \pa \, (J^{u(1)}\,
 J_f^c) -
  k \, (i \, f +d)^{b a c}\, \pa \, J^{u(1)}\,
  J_f^c \nonu \\
  && +
  i \, f^{e b d}\, (i\, f +d)^{e a c}\, J^d \, J^{u(1)}\,
  J_f^c
   \Big](w)+\cdots.
 \nonu
\eea

$\bullet$  The OPEs between the seventh term and the remaining terms 
\bea
&& \pa \, J_f^a(z)\, \pa \, J_f^b(w) =
-\frac{1}{(z-w)^4}\, 6 \, N\, \de^{a b}
- \frac{1}{(z-w)^3}\, 2 \, i \, f^{a b c}\, J_f^c(w) 
\nonu \\
&& - \frac{1}{(z-w)^2}\,  i \, f^{a b c}\, \pa \, J_f^c(w)+
\cdots.
\nonu
\eea

$\bullet$  The OPEs between the eighth term and the remaining terms 
\bea
&& d^{a c d}\, J^c\, J^d(z) \, J^b \, J^{u(1)}(w) =
\frac{1}{(z-w)^2}\, (2k+M)\, d^{b a c}\, J^c \, J^{u(1)}(w) \nonu \\
&& +
\frac{1}{(z-w)}\, \Big[ (2k+M)\, d^{b a c}\, \pa \, J^c \, J^{u(1)}-
  i \, f^{b a c}\, d^{c d e}\, J^d \, J^e \, J^{u(1)} \Big](w) + \cdots.
\nonu
\eea

$\bullet$  The OPE between the last term and itself 
\bea
&&
J^a \, J^{u(1)}(z) \, J^b \, J^{u(1)}(w) =
\frac{1}{(z-w)^4}\, k^2 \, \de^{a b}+\frac{1}{(z-w)^3}\, k \, i \, f^{a b c}
\, J^c(w) 
\nonu \\
&& +\frac{1}{(z-w)^2}\, \Big[k \, \de^{a b}\, J^{u(1)}\, J^{u(1)}+
 k \, i \, f^{a b c}
\, \pa \, J^c+ k \, J^b \, J^a
\Big](w)\nonu \\
&& + \frac{1}{(z-w)}\, \Big[
  k \, \de^{a b}\, \pa \, J^{u(1)}\, J^{u(1)}
  + i \, f^{a b c}\, ((J^c \, J^{u(1)}) J^{u(1)})
  + k \, J^b \, \pa \, J^a \Big](w) +\cdots.
\nonu
\eea
The other OPEs where the order of the operators in the left hand side is
reversed can be calculated by using the above OPEs with the appropriate
contributions from the higher order poles.

\section{
 The OPEs between the singlet spin-$2$ operators in section $7$ 
}

In order to calculate the OPE in (\ref{spin2spin2ope}),
we need to compute the following OPEs
\bea
&& \de_{\rho \bar{\si}} \de_{j \bar{i}} \,
J^{(\rho \bar{i})}  J^{(\bar{\si} j)}(z) \,
\de_{\mu \bar{\nu}} \de_{l \bar{k}} \,
J^{(\mu \bar{k})}  J^{(\bar{\nu} l)}(w)
=  \frac{1}{(z-w)^4}\,
k N M \, (k-N-M)\nonu \\
&& +  \frac{1}{(z-w)^2}\, \Bigg[
2 (k+N+M)^2\, T_{boson} 
- 2(M+k)\, (k+M+N)\, J^a \, J^a
\nonu \\
&& + (\frac{M}{2(k+N)})^2 \, 4 \, (N+k)\,
J^{\al}\, J^{\al}
+ (\frac{M+N}{2k})^2\, 4 \, k\, J^{u(1)}\,
J^{u(1)}\nonu \\
&& + k(M+N)\, M N\, \sqrt{\frac{M+N}{M N}}
\pa \, J^{u(1)} \Bigg](w)
+  {\cal O}(\frac{1}{(z-w)})
+ \cdots,
\nonu \\
&& J^{\al}\, J^{\al}(z) \,
J^{\beta}\, J^{\beta}(z)  = 
\frac{1}{(z-w)^4} \, 2k\, (N+k)\, (N^2-1)\nonu \\
&& + 
\frac{1}{(z-w)^2}\, 4 \, (N+k)\, J^{\beta}\, J^{\beta}
 +  {\cal O}(\frac{1}{(z-w)})
+ \cdots,
\nonu \\
&& J^{\al} \, J_f^{\al}(z)
\,
J^{\beta} \, J_f^{\beta}(w)  = 
\frac{1}{(z-w)^4}\, M k \, (N^2-1)+
\frac{1}{(z-w)^2}\, \Bigg[
  k \, J_f^{\al}\, J_f^{\al}\nonu \\
  && -  2 N \, J^{\al}\, J_f^{\al}
  + M \, J^{\al}\, J^{\al}
  \Bigg](w) + {\cal O}(\frac{1}{(z-w)})
 + \cdots,
\nonu \\
&& J^{u(1)} J^{u(1)}_f(z) \,
J^{u(1)} J^{u(1)}_f(w)  = 
\frac{1}{(z-w)^4}\, k M N \nonu \\
&& + 
\frac{1}{(z-w)^2}\, \Bigg[
  k\, J_f^{u(1)}\, J_f^{u(1)}
  + M N \, J^{u(1)}\, J^{u(1)} 
  \Bigg](w)
 +  {\cal O}(\frac{1}{(z-w)})
+ \cdots,
\nonu \\
&& J^a \, J^a_f(z) \,
J^b \, J^b_f(w)  = 
\frac{1}{(z-w)^4}\, k N (M^2-1)
\nonu \\
&& + \frac{1}{(z-w)^2}\,
\Bigg[ k J_f^a \,J_f^a- 2 M \, J^a \, J^a_f +
  N \, J^a \, J^a \Bigg](w)
+  {\cal O}(\frac{1}{(z-w)})
+ \cdots,
\nonu \\
&& \de_{\rho \bar{\si}} \, \de_{j \bar{i}} \,
\pa \, \psi^{(\rho \bar{i})} \, \psi^{(\bar{\si} j)}(z)
\,
\de_{\mu \bar{\nu}} \, \de_{l \bar{k}} \,
\pa \, \psi^{(\mu \bar{k})} \, \psi^{(\bar{\nu} l)}(w)
=  -\frac{1}{(z-w)^4}
\, N M
\nonu \\
&& +  \frac{1}{(z-w)^2}\, 2 \,
\de_{\mu \bar{\nu}} \, \de_{l \bar{k}} \,
\pa \, \psi^{(\mu \bar{k})} \, \psi^{(\bar{\nu} l)}(w)
+  {\cal O}(\frac{1}{(z-w)})
+ \cdots,
\nonu \\
&& J^a \, J^a(z) \,
 J^b \, J^b(w)  = 
\frac{1}{(z-w)^4}\,2k \, (M+k)\, (M^2-1)
\nonu \\
&& +  \frac{1}{(z-w)^2}\, 4 \,(M+k)\, J^a \, J^a
+ {\cal O}(\frac{1}{(z-w)})
 +  \cdots,
\nonu \\
&& \de_{\rho \bar{\si}} \, \de_{j \bar{i}} \,
 \psi^{(\rho \bar{i})} \, \pa\, \psi^{(\bar{\si} j)}(z)
 \,
 \de_{\mu \bar{\nu}} \, \de_{l \bar{k}} \,
 \psi^{(\mu \bar{k})} \, \pa\, \psi^{(\bar{\nu} l)}(w)  = 
  -\frac{1}{(z-w)^4}\, N M \nonu \\
  && -  \frac{1}{(z-w)^2}\, 2 \,
\de_{\mu \bar{\nu}} \, \de_{l \bar{k}} \,
\psi^{(\mu \bar{k})} \, \pa\, \psi^{(\bar{\nu} l)}(w)
 +  {\cal O}(\frac{1}{(z-w)})
+ \cdots,
\nonu \\
&& J^{u(1)} \, J^{u(1)}(z) \,
 J^{u(1)} \, J^{u(1)}(w)  = 
\frac{1}{(z-w)^4}\, 2k^2+
\frac{1}{(z-w)^2}\, 4k \, J^{u(1)} \, J^{u(1)}
\nonu \\
&& +  {\cal O}(\frac{1}{(z-w)})
 +  \cdots,
\nonu \\
&& \pa \, J^{u(1)}(z) \,
 \pa \, J^{u(1)}(w)  = 
-\frac{1}{(z-w)^4}\, 6 k
+ {\cal O}(\frac{1}{(z-w)})
+ \cdots.
\label{variousope}
\eea
We only present the OPEs up to the second order pole in
(\ref{variousope}) where
the bosonic stress energy tensor $T_{boson}$
is given by the footnote \ref{bosonT}.
Other OPEs between the operators of $W^{-(2),0}$ can be performed
similarly.

\section{ Some of the defining relations
in section $8$}

For convenience, we present the previous relations in \cite{AK1411}.

The OPEs between the supersymmetry generators of spin-$\frac{3}{2}$
are given by
\bea
\hat{G}_{11}(z) \, \hat{G}_{11}(w) & = & 
\frac{1}{(z-w)} \, \frac{4}{(N+k+2)} \left[ -\hat{A}_{+} \hat{B}_{-} 
\right](w) + \cdots,
\nonu \\
\hat{G}_{11}(z) \, \hat{G}_{12}(w) & = &
\frac{1}{(z-w)^2} \left[ 4 i \, \gamma_A  \hat{A}_{+}  \right](w)
\nonu \\
& + &  
\frac{1}{(z-w)} \left[ 2 i \, \gamma_A  \pa \hat{A}_{+} +\frac{4}{(N+k+2)}
\hat{A}_{+} \hat{B}_3 
\right](w) +\cdots,
\nonu \\
\hat{G}_{11}(z) \, \hat{G}_{21}(w) & = &
\frac{1}{(z-w)^2} \left[ -4 i \, \gamma_B  \hat{B}_{-}  \right](w)
\nonu \\
&+ &  
\frac{1}{(z-w)} \left[ -2 i \, \gamma_B  \pa  \hat{B}_{-}
  +\frac{4}{(N+k+2)}
\hat{A}_3  \hat{B}_{-}
\right](w) +  \cdots,
\nonu \\
\hat{G}_{11}(z) \, \hat{G}_{22}(w) & = &
\frac{1}{(z-w)^3} \, \frac{2}{3}c_{\mbox{Wolf}} +
\frac{1}{(z-w)^2} \left[ 4 i \left( \gamma_A  \hat{A}_3 - \gamma_B  \hat{B}_3 \right) \right](w)
+\frac{1}{(z-w)}  \left[ 2 \hat{T} \right.
\nonu \\
 & + & \left. 2 i \pa \left( \gamma_A  \hat{A}_3 - \gamma_B  \hat{B}_3 \right)+
 \frac{2}{(k+N+2)}\left( 
  \hat{A}_i \, \hat{A}_i +\hat{B}_i \, \hat{B}_i 
+2 \hat{A}_3 \, \hat{B}_3    \right) \right] (w) 
+  \cdots,
\nonu \\
\hat{G}_{12}(z) \, \hat{G}_{12}(w) & = & 
\frac{1}{(z-w)} \, \frac{4}{(N+k+2)} \left[ \hat{A}_{+} 
\hat{B}_{+} \right](w) + \cdots,
\nonu \\
\hat{G}_{12}(z) \, \hat{G}_{21}(w) & = &
\frac{1}{(z-w)^3} \, \frac{2}{3}c_{\mbox{Wolf}} +
\frac{1}{(z-w)^2} \left[ 4 i \left( \gamma_A  \hat{A}_3 + \gamma_B  \hat{B}_3 \right) \right](w)
+\frac{1}{(z-w)} \left[ 2 \hat{T} \right.
\nonu \\
& + &  \left.  2 i \pa \left( \gamma_A  \hat{A}_3 + \gamma_B  \hat{B}_3 \right) +
 \frac{2}{(k+N+2)}\left( 
    \hat{A}_i \, \hat{A}_i +\hat{B}_i \, \hat{B}_i 
-2 \hat{A}_3 \, \hat{B}_3    \right) \right](w) 
+  \cdots,
\nonu \\
\hat{G}_{12}(z) \, \hat{G}_{22}(w) & = &
\frac{1}{(z-w)^2} \left[ -4i\, \gamma_B  \hat{B}_{+} \right](w)
\nonu \\
&+& 
\frac{1}{(z-w)} \left[ -2 i\, \gamma_B \pa  \hat{B}_{+} 
 + \frac{4}{(N+k+2)} \hat{A}_3  \hat{B}_{+} 
 \right](w) +  \cdots,
\nonu \\
\hat{G}_{21}(z) \, \hat{G}_{21}(w) & = &  
\frac{1}{(z-w)} \frac{4}{(N+k+2)} \left[ \hat{A}_{-}
\hat{B}_{-} 
\right](w) +\cdots,
\nonu \\
\hat{G}_{21}(z) \, \hat{G}_{22}(w) & = &
\frac{1}{(z-w)^2} \left[ 4i\, \gamma_A  \hat{A}_{-} \right](w) +  
\frac{1}{(z-w)} \left[ 2 i\, \gamma_A  \pa  \hat{A}_{-}
  + \frac{4}{(N+k+2)} \hat{A}_{-}
 \hat{B}_3 \right](w) \nonu \\
& + & \cdots,
\nonu \\
\hat{G}_{22}(z) \, \hat{G}_{22}(w) & = &  
\frac{1}{(z-w)} \frac{4}{(N+k+2)} \left[ -\hat{A}_{-} \hat{B}_{+}  
\right](w) +\cdots,
\label{ggopenonlinear}
\eea
where the two parameters are given by
$\gamma_A \equiv \frac{N}{N+k+2}$, $\gamma_B \equiv 
\frac{k}{N+k+2}$, and we introduce the spin-$1$ currents 
$\hat{A}_{\pm}(z) \equiv \hat{A}_1 \pm i \hat{A}_2(z)$ and 
$\hat{B}_{\pm}(z) \equiv \hat{B}_1 \pm i \hat{B}_2(z)$
and the central term above is given by
$
c_{\mbox{Wolf}}   =  
\frac{6 k N}{(2+k+N)}$.

The higher spin-$\frac{3}{2}$ currents can be obtained
from the following OPEs
\bea 
\hat{G}_{21} (z) \, T^{(1)} (w) &=&
\frac{1}{(z-w)} \left[ \hat{G}_{21}  + 2 T_{+}^{(\frac{3}{2})} \equiv
 \hat{G}'_{21} \right](w)+\cdots,
\nonu \\
\hat{G}_{12} (z) \, T^{(1)} (w) &=&
\frac{1}{(z-w)} \left[ -\hat{G}_{12}  + 2 T_{-}^{(\frac{3}{2})}
\equiv
 \hat{G}'_{12}
  \right](w)+\cdots,
\nonu \\
\hat{G}_{11} (z) \, T^{(1)} (w) &=&
\frac{1}{(z-w)} \left[ \hat{G}_{11}  + 2 U^{(\frac{3}{2})}
\equiv
 \hat{G}'_{11}
  \right](w)+\cdots, 
\nonu \\
\hat{G}_{22} (z) \, T^{(1)} (w) &=&
\frac{1}{(z-w)} \left[- \hat{G}_{22}  + 2 V^{(\frac{3}{2})}
\equiv
 \hat{G}'_{22}
  \right](w)+\cdots.
\label{spin3halfgenerating}
\eea

The higher spin-$2$ currents are determined by
\bea
\hat{G}_{11}(z) \, \hat{G}'_{12} (w)
&=&
\frac{1}{(z-w)} \, \left[ - 2 U_{-}^{(2)} -\frac{4}{(N+k+2)} 
 \hat{A}_{+}  B_3  \right](w) +\cdots,
\nonu \\
\hat{G}_{21}(z) \,  \hat{G}'_{22} (w)
&=&
\frac{1}{(z-w)} \, \left[  2 V_{+}^{(2)} -\frac{4}{(N+k+2)} 
 \hat{A}_{-}  B_3  \right](w) +\cdots,
\nonu \\
\hat{G}_{11}(z) \, \hat{G}'_{21} (w)
&=&
\frac{1}{(z-w)} \, \left[ - 2 U_{+}^{(2)} -\frac{4}{(N+k+2)} 
\hat{A}_3 \hat{B}_{-}  \right](w) +\cdots,
\nonu \\
\hat{G}_{12}(z) \, \hat{G}'_{22} (w)
&=&
\frac{1}{(z-w)} \, \left[ 2 V_{-}^{(2)} -\frac{4}{(N+k+2)} 
 \hat{A}_3 \hat{B}_{+}  \right](w) +\cdots,
\nonu \\
 \hat{G}_{12}(z) \, \hat{G}'_{21} (w)
&=&
\frac{1}{(z-w)^2} \, 2 T^{(1)} (w) +
\frac{1}{(z-w)} \, \left[ -2 T^{(2)} +\pa T^{(1)} +  \frac{2(k+N)}{(k+N+2kN)} 
\hat{T}  \right.
\nonu \\
&+& \left.   \frac{2}{(N+k+2)} \left( \hat{A}_i \hat{A}_i +\hat{B}_i \hat{B}_i
-2 \hat{A}_3 \hat{B}_3 \right)
 \right](w) +\cdots,
 \nonu \\
\hat{G}_{11}(z) \, \hat{G}'_{22} (w)
&=&
\frac{1}{(z-w)^2} \, 2 T^{(1)} (w) +
\frac{1}{(z-w)}\, \left[ 2 W^{(2)} +\pa T^{(1)} -2 \hat{T}  \right.
\nonu \\
&-& \left.   \frac{2}{(N+k+2)} \left( \hat{A}_i \hat{A}_i +\hat{B}_i \hat{B}_i
+2 \hat{A}_3 \hat{B}_3 \right)
 \right](w) +\cdots,
\nonu \\
\hat{G}_{21}(z) \, \hat{G}'_{12} (w)
&=&
\frac{1}{(z-w)^2} \, 2 T^{(1)} (w) +
\frac{1}{(z-w)} \, 
\left[ 2 T^{(2)} +\pa T^{(1)} -  \frac{2(k+N)}{(k+N+2kN)} \hat{T}  \right.
\nonu \\
&-& \left.   \frac{2}{(N+k+2)} \left( \hat{A}_i \hat{A}_i +\hat{B}_i \hat{B}_i
-2 \hat{A}_3 \hat{B}_3 \right)
 \right](w) +\cdots,
 \nonu \\
\hat{G}_{22}(z) \, \hat{G}'_{11} (w)
&=&
\frac{1}{(z-w)^2} \, 2 T^{(1)} (w) +
\frac{1}{(z-w)} \, \left[ -2 W^{(2)} +\pa T^{(1)} + 2 \hat{T}  \right.
\nonu \\
&+& \left.   \frac{2}{(N+k+2)} \left( \hat{A}_i \hat{A}_i +\hat{B}_i \hat{B}_i
+2 \hat{A}_3 \hat{B}_3 \right)
 \right](w) +\cdots.
\label{spin2generating1} 
\eea
Here $\hat{G}'_{mn}$ is defined in (\ref{spin3halfgenerating}).

Similarly, the higher spin-$\frac{5}{2}$ currents can be
obtained from 
\bea
\hat{G}_{21} (z) \, U^{(2)}_{-} (w) 
&=&
\frac{1}{(z-w)^2}\, \left[ \frac{(N+2k)}{(N+k+2)} \hat{G}_{11} + 
\frac{2(N+2k+1)}{(N+k+2)} U^{(\frac{3}{2})} \right](w)
\nonu \\
&+& 
\frac{1}{(z-w)} \, \left[ U^{ (\frac{5}{2}) } 
+\frac{1}{3} \pa \, (\mbox{pole-2})
\right](w) + \cdots,
\nonu \\
\hat{G}_{21} (z) \, V^{(2)}_{-} (w) 
&=&
\frac{1}{(z-w)^2}  \, \left[ - \frac{(2N+k)}{(N+k+2)} \hat{G}_{22} + 
\frac{2(2N+k+1)}{(N+k+2)} 
V^{(\frac{3}{2})} \right](w)
\nonu \\
&+& 
\frac{1}{(z-w)} \, \left[ V^{ (\frac{5}{2}) } 
+\frac{1}{3} \pa \,  (\mbox{pole-2})
\right] (w)+ \cdots,
\nonu \\
\hat{G}_{21} (z) \, W^{(2)} (w) 
&=&
\frac{1}{(z-w)^2}  \, \left[ \frac{(N+2k+1)}{(N+k+2)} 
\hat{G}_{21} + \frac{(k-N)}{(N+k+2)} T_{+}^{(\frac{3}{2})} \right](w)
\nonu \\
&+& 
\frac{1}{(z-w)} \, \left[ W_{+}^{(\frac{5}{2})} 
+\frac{1}{3} \pa \,  (\mbox{pole-2})
\right] (w)+ \cdots,
\nonu \\
\hat{G}_{12} (z) \, W^{(2)} (w) 
&=&
\frac{1}{(z-w)^2}  \, \left[ \frac{(N+2k+1)}{(N+k+2)} 
\hat{G}_{12} + \frac{(N-k)}{(N+k+2)} T_{-}^{(\frac{3}{2})} \right](w)
\nonu \\
&+& 
\frac{1}{(z-w)} \, \left[ W_{-}^{(\frac{5}{2})} 
+\frac{1}{3} \pa \,  (\mbox{pole-2})
\right] (w)+ \cdots.
\label{spin5halfgenerating}
\eea
In next Appendix,
we express the higher spin-$\frac{5}{2}$ currents for fixed $N=5$ with $M=2$.

\section{ The higher spin-$\frac{5}{2}$ currents
for fixed $N=5$ and $M=2$ in section $8$}

We present the higher spin-$\frac{5}{2}$ currents as follows:
\bea
U^{(\frac{5}{2})} & =&
 \left(\frac{1}{k+7}\right)^{3/2}\, \Bigg[
i \, f^{3 1 d} \Big(\frac{1}{2}
\, i \, f^{3 d c}\, V^{+(\frac{5}{2}),c}+
\frac{(2 k+7)}{6 k}\, i \, f^{3 d c}\, \pa \,
G^{-,c}\Big)
\nonu \\
& + & i\,  f^{3 2 d} \Big(\frac{1}{2}  \, f^{3 d c}\, V^{+(\frac{5}{2}),c}
+\frac{  (2 k+7)}{6 k}
  \, f^{3 d c}\, \pa \,
G^{-,c}\Big)
\nonu \\
& + & i\,  f^{3 1 d} \Big(-\frac{1}{2} 
\, i \, f^{3 d c}\, V^{-(\frac{5}{2}),c}
-\frac{7 }{6 k}
\, i \, f^{3 d c}\, \pa \,
G^{-,c}\Big)
\nonu \\
&+& i\,  f^{3 2 d} \Big(-\frac{1}{2}  \, f^{3 d c}\, V^{-(\frac{5}{2}),c}
-\frac{7  }{6 k}\,
 f^{3 d c}\, \pa \,
G^{-,c}\Big)
- \frac{(k+7)}{2 k}\, f^{3 1 d}\,  f^{3 d e}\,
G^{+,e} \, K
\nonu \\
& + & \frac{i (k+7)}{2 k}\,
  f^{3 2 d}\,  f^{3 d e}\,
G^{+,e} \, K
-\frac{(k+7)}{2 k}\,
  f^{3 1 d}\,  f^{3 d e}\,
 G^{-,e} \, K
 -
  i \, f^{3 1 d}\, J^d\,
 G^{-,3} 
 \nonu \\
 &+& \frac{i (k+7)}{2 k}\,
   f^{3 2 d}\,  f^{3 d e}\,
  G^{-,e} \, K
  -f^{3 2 d}\, J^d\,
 G^{-,3} 
 +
 i \, f^{3 1 d}\, G^{+,d}\, J_f^3
 \nonu \\
 & + &  f^{3 2 d}\, G^{+,d}\, J_f^3
  +\frac{ (2 k+7)}{k}\,
  \de^{3 3}\, J^1 \, G^+
  -\frac{i  (2 k+7)}{k}\,
   \de^{3 3}\, J^2 \, G^+
   \nonu \\
   &+&
   i\, f^{3 1 d}\, G^{-,d}\, J_f^3
   + f^{3 2 d}\, G^{-,d}\, J_f^3
+ \frac{(k+7)}{k}\,
    \de^{3 3}\, J^1 \, G^- -\frac{i (k+7)}{k}\,
      \de^{3 3}\, J^2 \, G^- \, \Bigg],
      \nonu \\
      V^{(\frac{5}{2})} &=&\left(\frac{1}{k+7}\right)^{3/2}
      \, \Bigg[
-i \, f^{3 1 d} \, \frac{1}{2} 
\,  f^{d e f} \, f^{3 f g } \, J^e \, G^{+,g}      
\nonu \\
&+& i \, f^{3 2 d}\, \Big( -\frac{1}{2}
\, f^{d e f}\, i \, f^{3 f g } \, J^e \, G^{+,g}
-\frac{ \left(k^2-7\right)}{2 k}\,
 f^{3 d e}\, \pa \, G^{+,e}
\Big)
\nonu \\
&+&  i \, f^{3 1 d}\, \Big(
\frac{1}{2} \,
i \, f^{3 d e}\, V^{-(\frac{5}{2}),e}
- f^{d e f} \, f^{3 f g } \, J^e \, G^{-,g}
-  \frac{(k+7)}{2 k}\,
i\, f^{3 d e}\, J^e \, G^{-}
\Big) \nonu \\
& + & i \, f^{3 2 d}\, \Big(
-\frac{1}{2} \, f^{3 d e}\, V^{-(\frac{5}{2}),e}
-i \,
 f^{d e f} \, f^{3 f g } \, J^e \, G^{-,g}
\nonu \\
&
-&
\frac{  \left(k^2+10 k-14\right)}{6 k}\,
 f^{3 d e}\, \pa \, G^{-,e}
+\frac{ (k+7)}{2 k}\,
 f^{3 d e}\, J^e\, G^{-}
\Big)
\nonu \\
& +& \frac{i (k+7)}{2 k}\,
 f^{3 2 d} \, f^{3 d e}\,G^{+,e}\, K
+ 
i \, f^{3 1 d}\, J^d \, G^{-,3}- f^{3 2 d}\, J^d \, G^{-,3}
\nonu \\
&- &
G^+\, J_f^1
-2 \,
t^1_{j \bar{i}}\, \de_{k \bar{l}}\, \de_{\rho \bar{\mu}}\,
\de_{\tau \bar{\si}}\, \psi^{(\bar{\si} j)}\, \psi^{(\rho \bar{i})}\,
\psi^{(\bar{\mu} k)} \, J^{(\tau \bar{l})}
 + k 
\pa \, G^{+,1}-2\,
G^-\, J_f^1\nonu \\
& - & 2 \,
t^1_{i \bar{j}}\, \de_{l \bar{k}}\, \de_{\mu \bar{\rho}}\, \de_{\si \bar{\tau}}\, 
 \psi^{(\si \bar{j})}\, \psi^{(\bar{\rho} i)}\,
 \psi^{(\mu \bar{k})} \, J^{(\bar{\tau} l)}
 +\frac{ \left(k^2+10 k-14\right)}{3 k}\,
 \pa \, G^{-,1}
 \nonu \\
 & + &  \sqrt{\frac{7}{5}} \,
  f^{3 2 c}\, J^{u(1)}\, G^{+,c}
 +  \frac{1}{5}  \sqrt{2} (5 k+1)\,
  f^{3 2 d}\, t^d_{j \bar{i}}\, \de_{\rho \bar{\si}}\,
 \pa \, J^{(\rho \bar{i})}\, \psi^{(\bar{\si} j)}
 \nonu \\
 & -&
  \sqrt{2} \,
  f^{3 2 c}\, t^{\al}_{\rho \bar{\si}}\, t^c_{j \bar{i}}\,
 J^{(\rho \bar{i})}\, \psi^{(\bar{\si} j)}\, J^{\al}+
 i \sqrt{2}\,
 t^3_{l \bar{k}}\, \de_{\tau \bar{\mu}}\, (( \psi^{(\bar{\mu} l)}\,
 J^{(\tau \bar{k})})\, J_f^2)
 \nonu \\
 &- & \frac{\sqrt{\frac{7}{5}} (k+5)
  }{k}\,
 i \, f^{3 1 c}\, J^{u(1)}\, G^{+,c}
  -  \frac{1}{5} \sqrt{2}  (5 k+1)\,
 i \, f^{3 1 c} \, t^c_{j \bar{i}}\, \de_{\rho \bar{\si}}\,
 \pa \, J^{(\rho \bar{i})}\, \psi^{(\bar{\si} j)}
 \nonu \\
 & + & \sqrt{2}\,
 i \, f^{3 1 c}\, t^{\al}_{\rho \bar{\si}}\, t^c_{j \bar{i}}\, J^{(\rho \bar{i})}\,
 \psi^{(\bar{\si} j)}\, J^{\al}
 +  2 \sqrt{2} \,
 t^{\al}_{\rho \bar{\si}}\, t^1_{j \bar{i}}\,
 t^{\al}_{\nu \bar{\mu}}\, t^3_{k \bar{l}}\,
 \psi^{(\bar{\si} j)}\,\psi^{(\rho \bar{i})}\,
 \psi^{(\nu \bar{l})}\,J^{(\bar{\mu} k)}
 \nonu \\
 & - & \frac{1}{\sqrt{2}}\,
  f^{3 c e} \, f^{c 1 d}\, G^{+,e}\, J_f^d
 - \frac{ \sqrt{35}}{k}\,
  f^{3 2 c}\, J^{u(1)}\, G^{-,c}
 +\frac{\sqrt{35}}{k}\,
 i \, f^{3 1 c}\, J^{u(1)}\, G^{-,c}
 \nonu \\
 &-&\frac{1}{5} \sqrt{2} \,
 i \, f^{3 1 c}\, t^c_{ \bar{j} i}\, \de_{\si \bar{\rho}}\,
 \pa \, J^{(\bar{\rho} i)}\, \psi^{(\si \bar{j})} + 
 \frac{12}{5} \sqrt{2} \,
 t^3_{ \bar{l} k}\, \de_{\mu \bar{\tau}}\, ((\psi^{(\mu \bar{l})}\, J^{(\bar{\tau} k)}
 )\, J_f^1) \, \Bigg],
 \nonu \\
 W^{(\frac{5}{2})}_{+} & = & \left(\frac{1}{k+7}\right)^{3/2}\,
 \Bigg[
\frac{ (k-2) (k+1) }{2 k}\,
 f^{1 2 d}\, \de^{3 d}\, \pa \, G^+
 +  \frac{1}{10}   (10 k+37)\,
 f^{1 2 d}\, \de^{3 d}\, \pa \, G^-
\nonu \\
& - & \frac{ (k+7)}{7 k}\,
 f^{1 2 d}\, \de^{3 d}\,  G^+\, K
-\frac{1}{10}  (k+7)\,
 f^{1 2 d}\, \de^{3 d}\,  G^-\, K
+ 
i \, f^{b 3 c}\, G^{+,b}\, J_f^c
\nonu\\
&-&
2 \,
t^3_{j \bar{i}}\, \de_{k \bar{l}}\, \de_{\rho \bar{\mu}}\,
\de_{\tau \bar{\si}}\, \psi^{(\bar{\si} j)}\, \psi^{(\rho \bar{i})}\,
\psi^{(\bar{\mu} k)} \, J^{(\tau \bar{l})}
-
G^{+,3}\, J_f^{u(1)}
-
\frac{ \left(k^2+13 k+7\right)}{3 k}\,
\pa \, G^{+,3}
\nonu \\
& + &
2 \,
i \, f^{b 3 c}\, G^{-,b}\, J_f^c
- 2 \,
\de_{l \bar{k}}\, \de_{\mu \bar{\rho}}\,
 \de_{\si \bar{\tau}}\, t^3_{ \bar{i} j}\,
 \psi^{(\si \bar{j})}\,\psi^{(\bar{\rho} i)}\,
 \psi^{(\mu \bar{k})}\,J^{(\bar{\tau} l)}+
 G^{-,3}\, J_f^{u(1)}
 \nonu \\
 &-&\frac{ \left(2 k^2+8 k-7\right)}{3 k}\,
 \pa \, G^{-,3}+
 \frac{1}{70} \sqrt{2} \, (35 k+2)\,
 \de^{3 3}\, \de_{\rho \bar{\si}}\, \de_{j\bar{i}}\,\pa \, J^{(\rho \bar{i})}
 \, \psi^{(\bar{\si}) j}
 \nonu \\
 &- &
 \frac{10}{7}  \sqrt{2} \,
 t^{\al}_{\rho \bar{\si}}
 \, t^3_{j \bar{i}}\, t^{\al}_{\mu \bar{\nu}}\, t^3_{l \bar{k}}\,
\, \psi^{(\bar{\si} j)}\, \psi^{(\rho \bar{i})}\,
\psi^{(\bar{\nu} l)} \, J^{(\mu \bar{k})} 
- \frac{ (7 k+10)}{\sqrt{35} k}\,
\de^{3 3}\, J^{u(1)}\, G^+
\nonu \\
& - & \sqrt{2}\,
\de^{3 3}\, \de_{j \bar{i}}\, t^{\al}_{\rho \bar{\si}}\, J^{(\rho \bar{i})}\, \psi^{(\bar{\si}
  j)} \, J^{\al}
+ \frac{1}{35} \sqrt{2}\, (35 k+12)\,
\de^{3 3}\, \de_{i \bar{j}}\, \de_{\si \bar{\rho}}\,
\pa \, J^{(\bar{\rho} i)}\, \psi^{(\si \bar{j})}
\nonu \\
&+ & \sqrt{2}\,
\de^{3 3}\, \de_{i \bar{j}}\, t^{\al}_{\si \bar{\rho}}\,
J^{(\bar{\rho} i)}\, \psi^{(\si \bar{j})} \, J^{\al}
+  \frac{10}{7} \sqrt{2} \,
t^{\al}_{\si \bar{\rho}}\, t^3_{i \bar{j}}\,
 t^{\al}_{\nu \bar{\mu}}\, t^3_{k \bar{l}}\,
 \psi^{(\si \bar{j})}\,\psi^{(\bar{\rho} i)}\,
 \psi^{(\nu \bar{l})}\,J^{(\bar{\mu} k)}
 \nonu \\
 & + &
 \frac{ (7 k+60)}{35 \sqrt{2}}\,
 \de^{3 3}\, G^-\, J_f^{u(1)}
 -  \frac{1}{\sqrt{2}}\,
  f^{3 c d} \, f^{3 d e}\, J^c \, G^{+,e}+
 \sqrt{2}\,
 J^3 \, G^{+,3}+
 \sqrt{2} \,
 J^3 \, G^{-,3}
 \nonu \\
 & - & 
 V^{+(\frac{5}{2}),3}+
 V^{-(\frac{5}{2}),3} -\frac{(k+7)}{k}\,
 G^{+,3}\, K -  \frac{(2 k+7)}{k}\,
 J^3 \, G^+
 \nonu \\
 &-& \frac{(k+7)}{k}\, G^{-,3}\, K
 -\frac{7 }{k}\,
 J^3 \, G^-
 -i \sqrt{2} \,
 J^1 \, G^{-,2}
 +  \frac{5}{7}  \, f^{3 1 d} \, G^{+,d}\, J_f^2
 \nonu \\
 & + & \frac{12}{7}  
 \, f^{3 1 d} \, G^{-,d}\, J_f^2+ 
 \frac{5}{14} \,
  f^{e 1 c}  \, f^{2 c d}  \, f^{3 e f} \,
 G^{+,f}\, J_f^d+
 \frac{6}{7}  \,
  f^{e 1 c} \, f^{2 c d} \, f^{3 e f} \,
 G^{-,f}\, J_f^d
 \nonu \\
 & - & \frac{i }{\sqrt{2}}\,
   f^{e 1 c}  \, f^{2 c d} \,
 J^e \, G^{-,d} \Bigg],
 \nonu \\
 W^{(\frac{5}{2})}_{-} & = &
 \left(\frac{1}{k+7}\right)^{3/2}\, \Bigg[
 \frac{ \left(k^2+3 k-6\right)}{k}\,
 f^{1 2 d}\, \de^{3 d}\, \pa \, G^+
+ \frac{1}{10}   (5 k+2)\,
 f^{1 2 d}\, \de^{3 d}\, \pa \, G^-
\nonu \\
&- &\frac{6  (k+7)}{7 k}\,
 f^{1 2 d}\, \de^{3 d}\,  G^+\, K
-\frac{1}{10}  (k+7)\,
 f^{1 2 d}\, \de^{3 d}\,  G^-\, K
+\,2 \,
i \, f^{b 3 c}\, G^{+,b}\, J_f^c
\nonu\\
&-& 2
\,
t^3_{j \bar{i}}\, \de_{k \bar{l}}\, \de_{\rho \bar{\mu}}\,
\de_{\tau \bar{\si}}\, \psi^{(\bar{\si} j)}\, \psi^{(\rho \bar{i})}\,
\psi^{(\bar{\mu} k)} \, J^{(\tau \bar{l})}
-
G^{+,3}\, J_f^{u(1)}
-\frac{\left(2 k^2+8 k-7\right)}{3 k}
\,
\pa \, G^{+,3}\nonu \\
& + &
\,
i \, f^{b 3 c}\, G^{-,b}\, J_f^c
- 2 \,
\de_{l \bar{k}}\, \de_{\mu \bar{\rho}}\,
 \de_{\si \bar{\tau}}\, t^3_{ i \bar{j} }\,
 \psi^{(\si \bar{j})}\,\psi^{(\bar{\rho} i)}\,
 \psi^{(\mu \bar{k})}\,J^{(\bar{\tau} l)}+
 G^{-,3}\, J_f^{u(1)}
 \nonu \\
 &-&\frac{ \left(k^2+13 k+7\right)}{3 k}\,
 \pa \, G^{-,3}
 +  \frac{1}{35} \sqrt{2}  (35 k+12)
 \,
 \de^{3 3}\, \de_{\rho \bar{\si}}\, \de_{j\bar{i}}\,\pa \, J^{(\rho \bar{i})}
 \, \psi^{(\bar{\si}) j}
 \nonu \\
 &+ &\frac{10}{7} \sqrt{2} 
 \,
 t^{\al}_{\rho \bar{\si}}
 \, t^3_{j \bar{i}}\, t^{\al}_{\mu \bar{\nu}}\, t^3_{l \bar{k}}\,
\, \psi^{(\bar{\si} j)}\, \psi^{(\rho \bar{i})}\,
\psi^{(\bar{\nu} l)} \, J^{(\mu \bar{k})} 
- \frac{(7 k+60)}{\sqrt{35} k}\,
\de^{3 3}\, J^{u(1)}\, G^+
\nonu \\
& - & \sqrt{2}\,
\de^{3 3}\, \de_{j \bar{i}}\, t^{\al}_{\rho \bar{\si}}\,
J^{(\rho \bar{i})}\, \psi^{(\bar{\si}
  j)} \, J^{\al}
+\frac{1}{35} \sqrt{2} (35 k+2)\,
\de^{3 3}\, \de_{i \bar{j}}\, \de_{\si \bar{\rho}}\,
\pa \, J^{(\bar{\rho} i)}\, \psi^{(\si \bar{j})}
\nonu \\
&+ & \sqrt{2} \,
\de^{3 3}\, \de_{i \bar{j}}\, t^{\al}_{\si \bar{\rho}}\,
J^{(\bar{\rho} i)}\, \psi^{(\si \bar{j})} \, J^{\al}
-  \frac{10}{7} \sqrt{2}\,
t^{\al}_{\si \bar{\rho}}\, t^3_{i \bar{j}}\,
 t^{\al}_{\nu \bar{\mu}}\, t^3_{k \bar{l}}\,
 \psi^{(\si \bar{j})}\,\psi^{(\bar{\rho} i)}\,
 \psi^{(\nu \bar{l})}\,J^{(\bar{\mu} k)}
 \nonu \\
 & + &
\frac{ (7 k+10)}{35 \sqrt{2}}\,
 \de^{3 3}\, G^-\, J_f^{u(1)}
 - 
\sqrt{2}\,
J^3 \, G^{+,3}
+\frac{1}{\sqrt{2}}\,
 f^{3 c d}\, f^{3 d e}\, J^c \, G^{-,e}
\nonu \\
& - & \sqrt{2} \,
J^3 \, G^{-,3}
+ 
 V^{+(\frac{5}{2}),3}-
 V^{-(\frac{5}{2}),3} +\frac{(k+7)}{k}\,
 G^{+,3}\, K \nonu \\
 & + & \frac{7 }{k}\,
 J^3 \, G^+
 + \frac{(k+7)}{k}\, G^{-,3}\, K
 +\frac{(2 k+7)}{k}\,
 J^3 \, G^-
 \nonu\\
 & - & i \sqrt{2} \,
 J^1\, G^{+,2}
 -  \frac{12}{7}  \,
  \, f^{3 1 d} \, G^{+,d}\, J_f^2
 -\frac{5}{7}  \, f^{3 1 d} \, G^{-,d}\, J_f^2\nonu \\
 & - &\frac{6}{7}  \, f^{e 1 c}  \, f^{2 c d}  \, f^{3 e f} \,
 G^{+,f}\, J_f^d- 
 \frac{5}{14}   \, f^{e 1 c}  \, f^{2 c d}  \, f^{3 e f} \,
 G^{-,f}\, J_f^d
 \nonu \\
 & - &\frac{i }{\sqrt{2}}\,
   f^{e 1 c}  \, f^{2 c d} \,
  J^e \, G^{+,d} \Bigg].
  \label{5halfm2case}
\eea
The general $N$  dependence on these currents can be
determined by calculating Appendix (\ref{spin5halfgenerating}) explicitly.
Then the spin-$\frac{5}{2}$ currents can be obtained from
$V^{\pm (\frac{5}{2}),a}$ and other composite operators.


\end{document}